\documentclass[a4paper,11pt]{article}
\pdfoutput=1
\usepackage{jheppub}

\arxivnumber{1912.11923}

\newcommand{\U}{{\rm U}}

\newcommand\scalemath[2]{\scalebox{#1}{\mbox{\ensuremath{\displaystyle #2}}}}

\newcommand*{\rdmathspace}[1][h]{%
\if h#1%
\thinmuskip=2mu\medmuskip=1mu\thickmuskip=3mu\fi%
\if m#1%
\thinmuskip=2.3mu\medmuskip=2mu\thickmuskip=3.5mu\fi%
\if s#1%
 \thinmuskip=3mu minus 0.7mu\medmuskip=4mu minus 2mu\thickmuskip=5mu plus 3.5mu minus 1.5mu\fi%
\if d#1%
\thinmuskip=3mu\medmuskip=4mu\thickmuskip=5mu plus 5mu\fi}

\providecommand\inspire[1]{\href{https://inspirehep.net/search?p=find+#1}{{\tiny IN}{\footnotesize SPIRE}}}

\providecommand\erratum[4][ibid.\ ]{\emph{Erratum #1}{\bf #2} (#3) #4}

\providecommand{\jhep}[3] {\ifnum#2>2009%
\href{https://doi.org/10.1007/JHEP#1(#2)#3}{\emph{JHEP} {\bf #1} (#2) #3}%
\else%
\href{https://doi.org/10.1088/1126-6708/#2/#1/#3}{\emph{JHEP} {\bf #1} (#2) #3}%
\fi}
\def\issueFromCounter.#1#2#3#4#5#6.{#2#3}
\providecommand{\jstat}[2]{\PackageWarningNoLine{\jname}{The macro \protect\jstat\space is obsolete!\MessageBreak Please typeset JSTAT as any other journal}%
  \href{https://doi.org/10.1088/1742-5468/#1/\issueFromCounter.#2./#2}{\emph{J.\ Stat.\ Mech.\ }(#1) #2}} 
\providecommand{\hepth}[1]{\href{https://arxiv.org/abs/hep-th/#1}{\tt hep-th/#1}}

\providecommand{\heplat}[1]{\href{https://arxiv.org/abs/hep-lat/#1}{\tt hep-lat/#1}}

\providecommand{\condmat}[1]{\href{https://arxiv.org/abs/cond-mat/#1}{\tt cond-mat/#1}}

\providecommand{\arXivid}[1]{\href{https://arxiv.org/abs/#1}{\tt arXiv:#1}}
\providecommand{\Math}[2]{%
\if!#1!%
\href{https://arxiv.org/abs/math/#2}{\tt math/#2}%
\else%
\href{https://arxiv.org/abs/math.#1/#2}{\tt math.#1/#2}%
\fi}

\usepackage{mathtools}
\usepackage{color}
\usepackage[all]{xy}
\usepackage{simplewick}
\usepackage{array}
\usepackage{tikz}
\usepackage{tkz-euclide}
\usepackage{amsthm}
\usepackage{enumitem}
\usepackage{bbm}
\usepackage{bbold}

\newcommand{\Tr}{{\rm Tr}}
\newcommand{\tr}{{\rm tr}}
\newcommand{\orho}{\overline{\rho}}

\newcommand{\vev}[1]{\langle #1\rangle}

\usetikzlibrary{shapes.misc}

\tikzset{cross/.style={cross out, draw=black, fill=none, minimum size=2*(#1-\pgflinewidth), inner sep=0pt, outer sep=0pt}, cross/.default={2pt}}

\def\Tr{\textrm{Tr}}

\def\firstcircle{(0,0) circle (1.5cm)}
\def\secondcircle{(1.75,0) circle (1.5cm)}
\def\thirdcircle{(0.85,1.75) circle (1.5cm)}
\def\fourthcircle{(8.5,0) circle (1.5cm)}
\def\fifthcircle{(10.25,0) circle (1.5cm)}
\def\sixthcircle{(9.35,1.75) circle (1.5cm)}

\usetikzlibrary{shapes.misc}

\title{Spectral fluctuations in the Sachdev-Ye-Kitaev model}

\author{Yiyang Jia}
\author{and Jacobus J.M.\ Verbaarschot}

\affiliation{Department of Physics and Astronomy, Stony Brook University,\\
 Stony Brook, New York 11794, U.S.A.}

\emailAdd{yiyang.jia@stonybrook.edu}
\emailAdd{jacobus.verbaarschot@stonybrook.edu}

\abstract{We present a detailed quantitative analysis of spectral
  correlations in the Sachdev-Ye-Kitaev (SYK) model. We find that the
  deviations from universal Random Matrix Theory (RMT) behavior are
  due to a small number of long-wavelength fluctuations (of the order
  of the number of Majorana fermions $N$) from one realization of the
  ensemble to the next one. These modes can be parameterized
  effectively in terms of Q-Hermite orthogonal polynomials, the main
  contribution being due to scale fluctuations for which we give a
  simple analytical estimate.  Our numerical results for $N=32$ show
  that only the lowest eight polynomials are needed to eliminate the
  nonuniversal part of the spectral fluctuations.  The covariance
  matrix of the coefficients of this expansion can be obtained
  analytically from low-order double-trace moments. We evaluate the
  covariance matrix of the first six moments and find that it agrees
  with the numerics.  We also analyze the spectral correlations in
  terms of a nonlinear $\sigma$-model, which is derived through a
  Fierz transformation, and evaluate the one and two-point spectral
  correlation functions to two-loop order. The wide correlator is
  given by the sum of the universal RMT result and corrections whose
  lowest-order term corresponds to scale fluctuations.  However, the
  loop expansion of the $\sigma$-model results in an ill-behaved
  expansion of the resolvent, and it gives universal RMT fluctuations
  not only for $q=4$ or higher even $q$-body interactions, but also
  for the $q=2$ SYK model albeit with a much smaller Thouless energy
  while the correct result in this case should have been Poisson
  statistics.  In our numerical studies we analyze the number variance
  and spectral form factor for $N=32$ and $q=4$.  We show that the
  quadratic deviation of the number variance for large energies
  appears as a peak for small times in the spectral form factor. After
  eliminating the long-wavelength fluctuations, we find quantitative
  agreement with RMT up to an exponentially large number of level
  spacings for the number variance or exponentially short times in the
  case of the spectral form factor.  }

\begin{document}
\maketitle\flushbottom

\section{Introduction}
Starting with the seminal talk by Kitaev~\cite{kitaev2015}, the Sachdev-Ye-Kitaev (SYK) model~\cite{sachdev1993}
has
attracted a great deal of attention in recent years, in particular as a model
for two-dimensional gravity. The low-energy limit of the SYK model is
given by the Schwarzian action which can also be obtained from Jackiw-Teitelboim
gravity~\cite{maldacena2016,Iliesiu-ml-2019xuh}.
In this limit,  the SYK model
is dual to  a black hole~\cite{Sachdev_2010},
and because of this an initial state has to thermalize which is only possible if
its dynamics are chaotic. In fact the SYK model turned out to be maximally chaotic~\cite{shenker2013,maldacena2016,kitaev2015,Cotler-ml-2019egt}, which was
 shown by the calculation of
 Out-of-Time-Order Correlators (OTOC)~\cite{kitaev2015,maldacena2016,Borgonovi-ml-2019mrk}.

 A different measure of chaos in quantum systems is the extent to which
correlations of eigenvalues are given by random matrix theory. This goes back
to the Bohigas-Giannoni-Schmidt conjecture~\cite{bohigas1984} stating that
if the corresponding classical system is chaotic, the  eigenvalue
correlations of the quantum system are given by  random
matrix theory. In the SYK model this can be investigated by the exact
diagonalization of the SYK Hamiltonian.
It was found that level statistics
are given by the random matrix theory
with the same anti-unitary symmetries as the ($N$ mod 8)-dependent anti-unitary
symmetries of the SYK model~\cite{You-ml-2016ldz,Garcia-Garcia-ml-2016mno,Cotler2016}.
This has been understood
analytically in terms of a two-replica nonperturbative saddle point of the
so called $\Sigma G$ formulation of the SYK model~\cite{Saad-ml-2018bqo}.
However, there are also deviations from random
matrix theory at many level spacings  or small times which follow from
a nonlinear $\sigma$-model formulation~\cite{Verbaarschot1984,Altland-ml-2017eao}
or from a moment calculation of the SYK model~\cite{Garcia-Garcia-ml-2018ruf,Gharibyan-ml-2018jrp}.

The so-called complex SYK model~\cite{sachdev1993} was first introduced in nuclear
physics~\cite{french1971,bohigas1971,mon1975} to reflect the four-body nature of the
nuclear interaction (known as a two-body interaction in the nuclear physics literature)
as well as the exponential increase of the level density and the random matrix
behavior of nuclear level correlations.
The great advance that was made in~\cite{sachdev1993} is to formulate the SYK model as a path
integral which isolates $N$ as a prefactor of the action, making it possible to evaluate the Green's
functions of the model by mean field theory. This analysis revealed one of the most striking properties
of the SYK model, namely that its ground state entropy is nonzero and extensive, making it a model for
non-Fermi liquids and  black hole physics alike~\cite{Sachdev_2010,sachdev2015}.
This approach also showed that the level density of the SYK model increases as
$\exp\sqrt{E-E_0}$~\cite{Cotler2016,bagrets2016,Garcia-Garcia-ml-2017pzl,Stanford-ml-2017thb} exactly as the
phenomenologically successful Bethe formula~\cite{bethe1936} for the nuclear level density, which was  actually not
realized in the early nuclear physics literature.
The randomness of
the SYK model is not expected to be important, and it has been shown for several
non-random SYK-like models, specifically tensor models,
have a similar melonic mean
field behavior~\cite{Witten-ml-2016iux,Klebanov-ml-2016xxf,Kim-ml-2019upg}.

The SYK model has become a paradigm of quantum many-body physics.
It
has been used to understand thermalization~\cite{Kourkoulou-ml-2017zaj,Almheiri-ml-2019jqq},
eigenstate thermalization~\cite{Sonner-ml-2017hxc} and decay of the
thermofield double state~\cite{delCampo-ml-2017ftn}. The chaotic-integrable
phase transition has been studied in a
mass-deformed SYK~\cite{Nosaka-ml-2018iat,Garcia-Garcia-ml-2017bkg}.
Coupled SYK models have been used to get a deeper understanding of wormholes~\cite{Maldacena-ml-2018lmt,Garcia-Garcia-ml-2019poj,Okuyama-ml-2019xvg,Penington-ml-2019kki,Maldacena-ml-2019ufo}
and black hole microstates~\cite{deBoer-ml-2019kyr}.
Lattices of coupled SYK models describe phase transitions involving non-Fermi
liquids~\cite{Kruchkov-ml-2019idx,Altland-ml-2019czw,Karcher-ml-2019ogt}.
  The complex SYK model has a conserved charge (the total number of fermions)
  and its effects were recently analyzed in the $\Sigma G$ formulation~\cite{Gu-ml-2019jub}. It has also been used to construct a model for quantum batteries~\cite{Rossini-ml-2019nfu}.
  The duality been the SYK model and Jackiw-Teitelboim
  gravity has been further explored in random matrix theories with the spectral
  density of the SYK model at low energies~\cite{Saad-ml-2019lba,Stanford-ml-2019vob,Saad-ml-2019pqd,Garcia-Garcia-ml-2019zds,Gross-ml-2019uxi}.

There are  different observables to study level correlations. The best
known one is the spacing distribution of neighboring levels or the ratio
of the maximum and the minimum of two consecutive spacings~\cite{oganesyan2007}. The disadvantage of
these
measures is that they include both two-point and higher-point correlations. In this paper
we will focus on
the number variance, which is the variance of the number of eigenvalues
in a interval that contains $n$ eigenvalues on average, and the spectral
form factor which is the Fourier transform of the pair correlation function.
Note that the number variance is an integral transform of both the pair
correlation function (by definition) and the spectral form factor.

The average spectral density is not universal, and to analyze
the universality of spectral correlators,  one has to eliminate this
non-universal part which appears in two different places.
First, one has to subtract the
disconnected part of the two-point correlator, and second, one has to
unfold the spectrum by a smooth transformation resulting in a spectral density
that is  uniform. Note that the number variance is already defined in terms
of level numbers and no further unfolding is needed (although in practice it is often
convenient to do so).  The number
variance and the spectral form factor are complementary observables,
each with their  own advantages and disadvantages.

\enlargethispage*{\baselineskip}\relax

The goal of this paper is to determine quantitatively the agreement of level
correlations in the SYK model with random matrix theory. To do this we
distinguish between level correlations within one specific realization of the
SYK model and level correlations due to fluctuations from one realization
to the next. The latter are expected to be large. Since the SYK model
is determined by only ${N \choose q}$ independent random variables, the relative
error in a observable is of order $1/{N \choose q}^{1/2}$. Such fluctuations
in the average level density result in a contribution to the number variance of
$\sim n^2 /{N \choose q}$ which for $q=4$ becomes important when $n \sim N^2$, which is not
large in comparison to the total number of levels of $2^{N/2}/2$. This effect
gives a constant contribution to the pair correlation function, and a function proportional to
$\delta(\tau)$ to the spectral form factor, which becomes the
dominant contribution for $\tau < {N\choose q}^{-1}$. It was already realized
many years ago that these scale fluctuations can be eliminated by rescaling
the eigenvalues according to the width of the spectral density for each configuration~\cite{French1973,Flores_2001}. Recently, it was shown that the main long-range spectral fluctuations  in the SYK model are of this nature~\cite{Altland-ml-2017eao,Garcia-Garcia-ml-2018ruf,Gharibyan-ml-2018jrp}.
However, these are not the only long-range fluctuations: they are just the first term in a ``multipole''
expansion of the smoothened spectral density for each realization.
In this paper we systematically study such long-range fluctuations.  Since the spectral
density is close to the weight function of the Q-Hermite polynomials~\cite{Cotler2016,Garcia-Garcia-ml-2017pzl,Klebanov-ml-2019jup}, it
is natural to expand the deviations in terms of Q-Hermite polynomials. It
turns out that we only need a small number of  polynomials, suggesting a separation of scales between long-wavelength fluctuations of the spectral
density and the short-wavelength fluctuations of the universal RMT spectral correlations.
The long-wavelength fluctuations are determined by  low-order moments, and we present an analytical calculation
of the covariance matrix of the first six moments.

We also analyze the deviations from random matrix theory in terms of the replica limit
of a spectral determinant~\cite{Verbaarschot1984,benet2003,Srednicki-ml-2002,Altland-ml-2017eao}.
We  evaluate the wide correlator to
two-loop order. The random matrix contribution to this correlator is given by the massless
part of the propagator, whereas the deviations are due to massive modes. For $q=4$ these
results are in agreement with~\cite{Altland-ml-2017eao} and the numerical results of~\cite{Garcia-Garcia-ml-2016mno}
(for values of $N$ in the universality class of the Gaussian Unitary Ensemble (GUE)).
However, the expressions for $q=2$ are qualitatively the same albeit with a Thouless energy
that scales as $N$ rather than $N^2$ for $q=4$. It is clear though, that for $q=2$, with energies given
by sums of single-particle energies, spectral
correlations are given by Poisson statistics. Presently, the mechanism that nullifies the contribution from
the zero modes is not clear.
Both the $\Sigma G$ formulation~\cite{Saad-ml-2018bqo} and
the spectral determinant~\cite{Verbaarschot1984,Altland-ml-2017eao} of the SYK model make use of the replica trick. Although the replica trick may result in an incorrect answer~\cite{Verbaarschot1985}, we expect that at the mean field level only replica
diagonal solutions contribute to one-point functions~\cite{Wang_2019}
while solutions that couple the two replicas, but are otherwise diagonal, appear
in the calculation of the two-point function~\cite{Verbaarschot1984,Altland-ml-2017eao,Saad-ml-2018bqo,Arefeva-ml-2018att,Arefeva-ml-2018vfp,Okuyama-ml-2019xvg}.

Our strategy to eliminate the collective fluctuations of the spectral density is discussed in section~\ref{sec:specDenAndUnfolding} after introducing the SYK model in section~\ref{sec:SYKdef}.
In section~\ref{sec:col} we give a simple argument to determine the leading correction to the number variance.
The $\sigma$-model formulation of the SYK model is discussed in section~\ref{sec:sigmaModel}.
We calculate the one-point function and
the two-point function to two-loop order. We also obtain a very efficient expansion for the  expansion
of the resolvent in terms of powers of the resolvent of a semicircle rescaled to the actual width of the
spectrum of the SYK model. This expansion is obtained by a resummation of the expansion in semicicular
resolvents obtained in~\cite{Altland-ml-2017eao}.
The covariance matrix is obtained in section~\ref{sec:covariance}, where we give explicit results for the first six double-trace moments.
Numerical results for the number variance and spectral form factor
are presented in section~\ref{sec:numberVarAndSpecForm}. In this section we also discuss the effects of ensemble averaging and the fluctuations relative to the ensemble average, the latter contribute to the number variance and the spectral form factor. The structure of high-order double-trace moments and the convergence to the
random matrix result is discussed in section~\ref{sec:doubleTrRMTvalidity}.
Concluding remarks are
made in section~\ref{sec:conclusion}.
In appendix~\ref{app:A} we derive  the $\sigma$-model for the spectral determinant from a Fierz transformation. Several
combinatorial formulas are given in
appendix~\ref{app:one}. The replica limit of the one-point and two-point functions of the GUE are calculated in appendix~\ref{app:GUEreplicaLimit}. Examples for the calculation of double-trace moments are worked out in appendix~\ref{app:mom3} and explicit results for low-order
moments are given in appendix~\ref{sec:lowOrderMoments}.

\section{The Sachdev-Ye-Kitaev (SYK) model} \label{sec:SYKdef}
The SYK model is a model of $N$ interacting Majorana fermions with a $q$-body Hamiltonian given by
\begin{equation}
H = \sum_\alpha J_\alpha \Gamma_\alpha,
\end{equation}
with $\alpha$ being a multi-index set of $q$ integer elements:
\begin{equation}
\alpha :=\{i_1,i_2,\ldots, i_q\}, \quad 1\leq i_1<i_2<\cdots<i_q\leq N,
\end{equation}
and hence $\alpha$ can have $\binom{N}{q}$ configurations. Furthermore,
\begin{equation}\label{eqn:GammaAlphaDef}
\Gamma_\alpha := (\sqrt{-1})^{\frac{q(q-1)}{2}} \gamma_{i_1}\gamma_{i_2}\cdots\gamma_{i_q}, \qquad  \{\gamma_i, \gamma_j\} = 2\delta_{ij},
\end{equation}
and the $J_\alpha$ are independently Gaussian distributed random
variables with variance $v^2$ given~by
\begin{equation}\label{eqn:singleParticleVariance}
v^2 = \frac{J^2(q-1)!}{ 2^q N^{q-1}}.
\end{equation}
This choice results in a many-body variance (the following bracket $\vev{\cdots}$ denotes ensemble average over all $J_\alpha$)
\begin{equation}\label{eqn:ManyBodyVariance}
\sigma^2 = 2^{-N/2}\vev{\Tr H^2} =\binom{N}{q} v^2,
\end{equation}
and a ground state energy that scales linearly with $N$, see equation~\eqref{eqn:groundStateEnergy}. In this paper we only consider
even values of $q$, especially $q=4$.
The Majorana fermions are represented as Dirac $\gamma$ matrices which is an effective way to obtain
a Hamiltonian that can be diagonalized numerically.

\section{Spectral density}\label{sec:specDenAndUnfolding}

The average spectral density of the SYK model is well approximated~\cite{Garcia-Garcia-ml-2017pzl,Klebanov-ml-2019jup}
by the weight function of the Q-Hermite polynomials (in units
where the many-body variance $\sigma^2$ is one)~\cite{Viennot-1987}:
\begin{equation}
\vev{\rho_{\text{SYK}}(x)}\approx\rho_{\rm QH}(x) = \frac {\Gamma_{\eta^2}(\frac 12)}{\pi\sqrt{1+\eta}} \sqrt{1-\frac 14 (1-\eta)x^2}
    \prod_{k=1}^\infty\left [1-\frac{x^2(1-\eta)\eta^k}{(1+\eta^k)^2}\right],
        \end{equation}
        where
      \begin{equation}\label{eqn:etaFirstappearance}
      \eta =   2^{-N/2}{N\choose q}^{-1} \sum_\beta  \Tr \Gamma_\alpha \Gamma_\beta\Gamma_\alpha \Gamma_\beta={N\choose q}^{-1}  \sum_{k=0}^q (-1)^{q-k} {q\choose k} {N-q \choose q-k},
      \end{equation}
      and
      $\Gamma_{u}(x)$ is defined by
      \begin{equation}
\Gamma_{u}(s) =  (1-u)^{1-s}\frac{\prod_{j=0}^\infty (1-u^{j+1})}{\prod_{j=0}^\infty (1-u^{j+s})}.
      \end{equation}
In  physical units the energy $E$ is related to dimensionless energy $x$ by
\begin{equation}
x= \frac{E}{\sigma},
\end{equation}
and  the ground state energy is given by
      \begin{equation}\label{eqn:groundStateEnergy}
      E_0^2 = \frac {4\sigma^2}{1-\eta},
      \end{equation}
     where $\sigma^2$ was given in~\eqref{eqn:ManyBodyVariance}. Note that  $E_0$ is extensive in $N$. This results in the dimensionful spectral density
\begin{equation}
\rho_{\rm QH}(E) = \frac {\Gamma_{\eta^2}(\frac 12)}{\pi \sigma \sqrt{1+\eta}}
     \sqrt{1-(E/E_0)^2}
   \prod_{k=1}^\infty\left [1-\frac{4 E^2\eta^k}{E_0^2(1+\eta^k)^2}\right ].
        \end{equation}
    In this paper we will only use the dimensionless energy $x$.

        The average spectral density of the SYK model can be expanded  in
        terms of the orthogonal polynomials corresponding to this weight
        function,
        \begin{equation}
       \langle  \rho_{\rm SYK}(x)\rangle = \rho_{QH}(x)\left ( 1+\sum_{k=1}^ \infty \vev{c_k}
    H^\eta_{k}\left (x\right )\right).
\label{rhoavexp}
    \end{equation}
    The average spectral density is even in $x$  so that all
    odd coefficients vanish after averaging. Since
    the second and fourth moments of the SYK model coincide with those of
    the Q-Hermite spectral density we
    also have
    \begin{equation}
    \vev{c_2}=0,\qquad \vev{c_4} =0.
    \end{equation}
    Both $\rho_{\rm SYK}$ and $\rho_{\rm QH}$ are normalized to unity.

 The Q-Hermite polynomials satisfy the recursion relation~\cite{Viennot-1987}
 \begin{equation}
 H_{n+1}^\eta(x) = x H_n^\eta(x) - \sum_{k=0}^{n-1}\eta^k H_{n-1}^\eta(x)
 \end{equation}
 with
 \begin{equation}
 H_0^\eta(x) = 1 \qquad {\rm and } \qquad H_1^\eta(x) = x.
 \end{equation}
 The orthogonality relations are given by
 \begin{equation}
 \int_{-\frac{2}{\sqrt{1-\eta}}}^{\frac{2}{\sqrt{1-\eta}}} dx \rho_{QH}(x) H_n^\eta\left (x \right )
 H_m^\eta\left (x \right )=\delta_{nm}n_\eta!,
 \end{equation}
 where $n_\eta!$ is the Q-factorial defined as
 \begin{equation}
 n_\eta! = \prod_{k=1}^{n-1}\left(\sum_{s=0}^k\eta^s\right).
 \end{equation}
 The expansion~(\ref{rhoavexp}) is a generalization of the Gram-Charlier expansion --- for $\eta=1$ it becomes the
 Gram-Charlier expansion. It is
 only positive-definite when the expansion coefficients are sufficiently
 small. For the SYK model with $N=32$ and $q=4$ we find that $|\vev{c_k}| < 0.01$
 and it decreases for larger values of $k$. The expansion converges very well in this case.

\begin{figure}
\centering
 \includegraphics[trim = 0 20 0 0, clip, width=6cm]{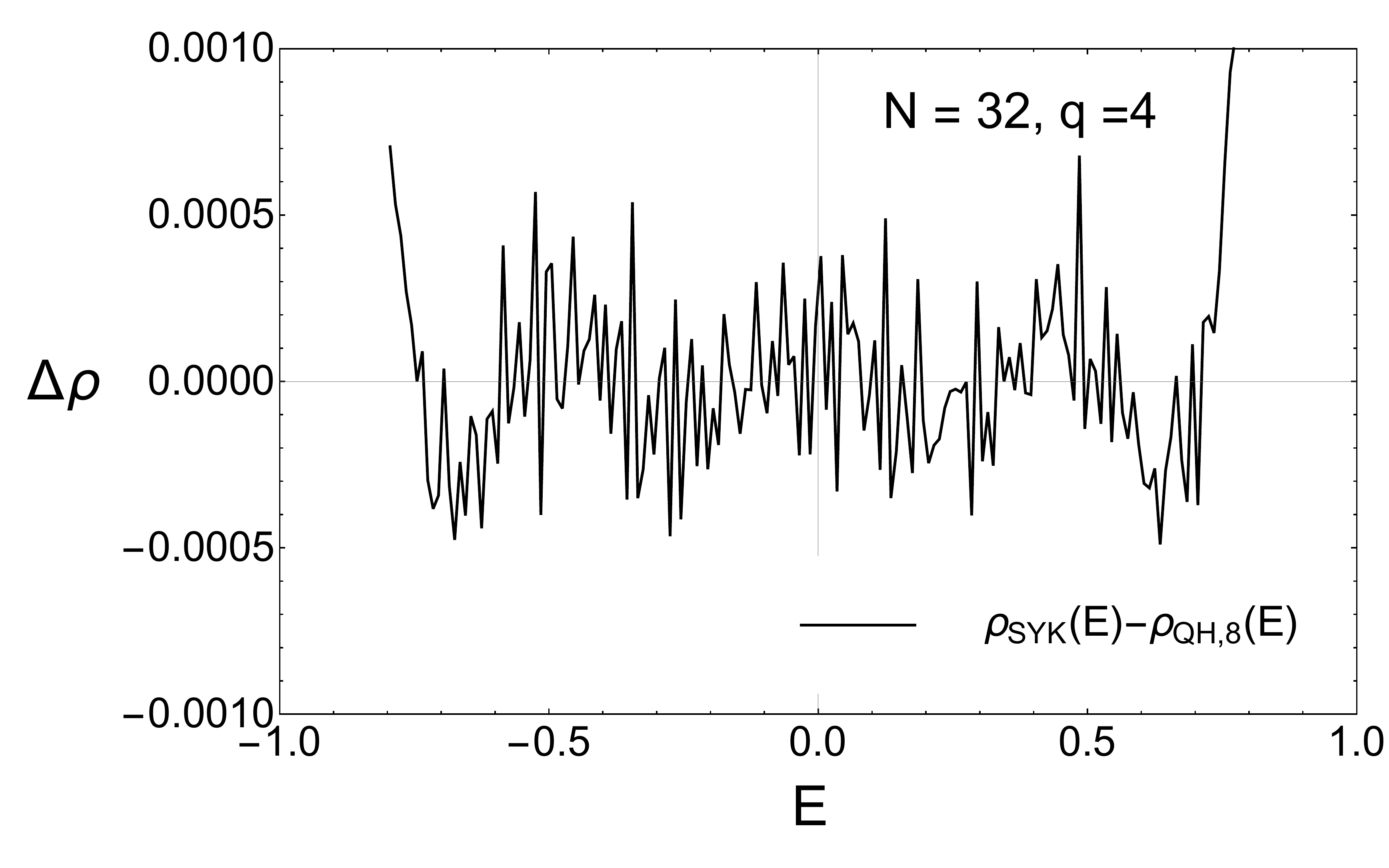}\quad
 \includegraphics[trim = 0 20 0 0, clip, width=6cm]{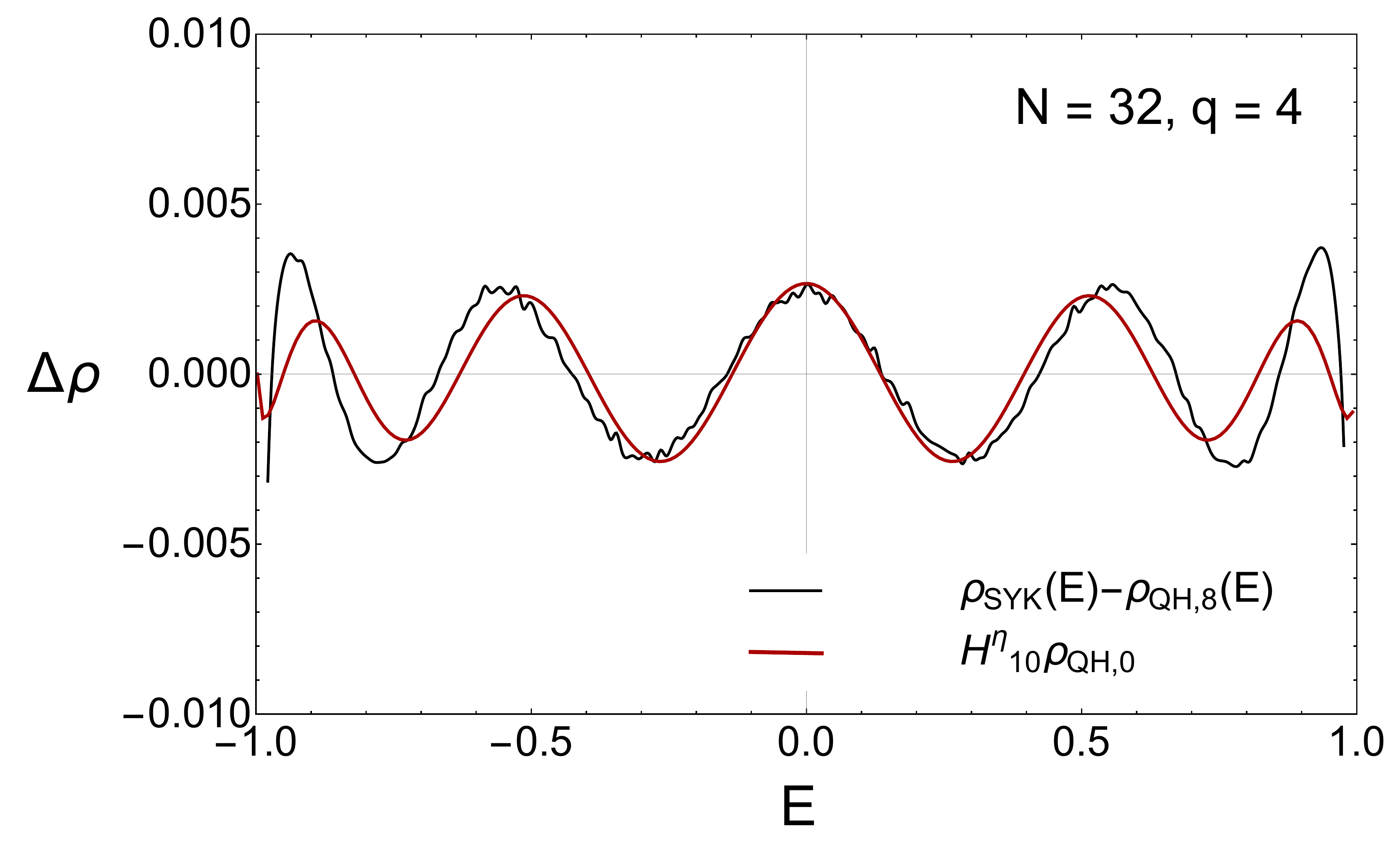}
 \caption{Difference of the of the ensemble average of the spectral density
   of the SYK Hamiltonian and the eighth order Q-Hermite approximation (black curve). In the
   left figure, the coefficients are calculated by minimization, and in
   the right figure by calculating scalar products. The red curve in the
   right figure is a fit of the contribution of the tenth order
   Q-Hermite polynomial.\label{fit-average}}
\end{figure}

 The expansion coefficients can be calculated in two ways. First, from the
scalar product of the numerically calculated average spectral density
and the Q-Hermite orthogonal polynomials, and second, from minimizing
the $L_2$ norm of the difference between the numerical result and the expansion
up to a given order. The two should give the same results. However,
an expansion in orthogonal polynomials does not converge well near
the end points of the spectrum, and a much better fit is obtained
by excluding a small fraction of the spectrum in this region. In that case, the coefficients
cannot be obtained by taking scalar
products, but they still can be obtained by minimization.
 We
illustrate this in figure~\ref{fit-average} for the average spectral density
of an ensemble of 400 SYK Hamiltonians for $N=32$ and $q=4$. What is plotted
is the difference of the average spectral density and the eighth order
Q-Hermite approximation. In the right figure the coefficients are obtained
by calculating the scalar products and the difference (black curve) is
close to the contribution of the tenth order Q-Hermite polynomial (red curve).
In the left figure, the coefficients are obtained by minimization of $\chi^2$
value of the difference not taking into account 2.5\% of the total number
of eigenvalues at each end of the spectrum. This gives a fit that is a factor
10 better with an accuracy of 1 part in 10000.

We can also expand the spectral density of the SYK Hamiltonian for \emph{each configuration}
in terms
of Q-Hermite polynomials
        \begin{equation}
        \rho_{\rm SYK}(x) = \rho_{\rm QH}(x)\left[ 1+\sum_{k=1}^ \infty c_k
    H^\eta_{k}\left (x \right )\right].
        \end{equation}
      Now, the $c_k$ are stochastic variables whose averages give the
    expansion coefficients $\vev{c_k}$ of the average spectral density
    in equation~(\ref{rhoavexp}). The fluctuations of $c_k$ correspond to spectral
    fluctuations with wavelength scale of $E_0/k$. We expect that these fluctuations
    are non-universal for small values of $k$, but for large values of $k$,
    they will correspond to universal random matrix correlations. The aim
    of this paper is to study at which scale this transition to universal random
    matrix behavior takes place.

    Since the coefficients $c_k$ are determined by ${N\choose q} $ independent stochastic variables,
    the relative error of the  $c_k$ as well as of $\rho(x)$ is of order  $1/{N\choose q}^{1/2} $.
    We thus expect that the variance of the number of levels in an interval containing $\bar n$ levels on average behaves as
    \begin{equation}
    \Sigma^2(\bar n) \sim \bar n^2 {N\choose q}^{-1}
    \end{equation}
    in agreement with the explicit calculation of $\vev{\Tr H^2 \Tr H^2}$ for the SYK model~\cite{Altland-ml-2017eao,Gharibyan-ml-2018jrp}.
    The covariance matrix $\langle c_k c_l\rangle$ can be calculated analytically for small values for $k$ and $l$
    by calculation double-trace moments using  methods similar to those first introduced for the calculation of the single-trace moments~\cite{mon1975,Garcia-Garcia-ml-2018kvh,Berkooz-ml-2018qkz,Berkooz-ml-2018jqr}, see
    section~\ref{sec:covariance}.

  \section{Collective fluctuations of eigenvalues}
  \label{sec:col}

  As was noticed in earlier work,
  the main contribution to the number variances comes from overall rescaling
of the eigenvalues from one configuration to the next~\cite{French1973,Flores_2001,Altland-ml-2017eao,Garcia-Garcia-ml-2018ruf,Gharibyan-ml-2018jrp}. Such fluctuations
can be written as
\begin{equation}
x_n \to x_n(1+\xi)
\end{equation}
where $\xi$ is a stochastic variable with a zero average and a finite variance.
  The corresponding spectral density fluctuates as
  \begin{equation}
  \rho(x) =\sum_k \delta(x-x_k(1+\xi))= \sum_k \frac 1{1+\xi}\delta(\frac x{1+\xi}-x_k) \approx \frac  1{1+\xi} \overline {\rho} \left ( \frac x{1+\xi}\right ),
  \label{rho-scale}
  \end{equation}
 where $\overline {\rho }$ is the ensemble-averaged spectral density function so that $ \overline {\rho }(z) = \langle \rho(z)\rangle$.
  This results in a contribution to the two-point correlation function
  \begin{equation}
  \langle  \rho_2(x,y) \rangle = \langle  \rho(x) \rho(y) \rangle -\langle  \rho(x) \rangle \langle  \rho(y) \rangle
  =\langle \xi^2\rangle \langle\rho(x)\rangle \langle\rho(y) \rangle +O(\xi^4),
\label{rho2-scale}
\end{equation}
where $\langle \xi^2 \rangle$ denotes averaging over the square of the stochastic variable $\xi$.
Since we consider spectral correlations on a scale $|x-y|\ll 1$, terms involving the first and second derivative of $\rho(x)$ can be ignored.
The expectation value of $\xi^2$ can be obtained from the normalized (by Hilbert space dimension) and rescaled (by variance)
double-trace moment $\tilde M_{2,2}$, with $\tilde M_{m,n}$ defined as
\begin{equation}
\tilde{M}_{m,n}:=\frac{\vev{ \Tr H^m\Tr H^n}}{2^N \sigma^{m+n}},
\end{equation}
where $\sigma$ is given by~\eqref{eqn:ManyBodyVariance}. As a special case we introduce the notation for (normalized and rescaled) single-trace moments
\begin{equation}
\tilde{M}_{n} :=\tilde{M}_{0,n}.
\end{equation}
The $\tilde M_{2, 2}$ in terms of scale fluctuations is given by
{\rdmathspace[h]
\begin{align}
\tilde M_{2,2}-\tilde{M}_{2}^2 & \approx
\int dx dy \left[ \left \langle \frac{x^2 y^2}{(1+\xi)^2}
  \orho\left(\!\frac x{1+\xi}\right) \orho\left(\!\frac{y}{1+\xi}\right)\!\right\rangle
  -\left\langle\frac{x^2}{1+\xi}
  \orho\left(\!\frac x{1+\xi}\right)\!\right \rangle
  \left \langle \frac{y^2}{1+\xi}\orho\left(\!\frac{y}{1+\xi}\right)\!\right\rangle \right ]\nonumber\\
 & \approx
\left\langle (1+\xi)^4 \right \rangle
-\left\langle (1+\xi)^2 \right \rangle^2  \nonumber\\
 & \approx  4\left\langle \xi^2\right \rangle .
\end{align}}\relax
On the other hand, this quantity can be calculated directly in the SYK model~\cite{Gharibyan-ml-2018jrp}:
\begin{equation}
\tilde M_{2,2}-\tilde{M}_{2}^2= 2{N \choose q}^{-1}.
\end{equation}
We thus find
\begin{equation}\label{eqn:scaleFluc}
\langle \xi^2\rangle = \frac 12 {N \choose q}^{-1}.
\end{equation}
This results in the number variance
\begin{align}
\Sigma ^2(\bar n)  & =   \int_{x-\Delta/2}^{x+\Delta/2} \int_{x-\Delta/2}^{x+\Delta/2}
\langle \xi^2\rangle \langle \rho(y)\rangle \langle \rho(z) \rangle dy dz\nonumber\\
 & =  \langle \xi^2\rangle \bar n^2
\end{align}
with the average number of levels in the interval given by
\begin{equation}
\bar n = \int_{x-\Delta/2}^{x+\Delta/2} dy \langle \rho(y)\rangle.
\end{equation}

We can also calculate the contribution of  scale fluctuations to arbitrary double-trace  moments:
\begin{equation}
\tilde{M}_{2m,2n}-\tilde{M}_{2m} \tilde{M}_{2n}
\approx 2mn {N \choose q}^{-1}
\tilde{M}_{2m} \tilde{M}_{2n}
\label{eqn:mommn}
\end{equation}
This follows from the elementary calculation
\begin{align}
\tilde{M}_{2m,2n}-\tilde{M}_{2m} \tilde{M}_{2n}
 & \approx  \left [\langle (1+\xi)^{2m} (1+\xi)^{2n} \rangle
  -\langle (1+\xi)^{2m}\rangle\langle (1+\xi)^{2n} \rangle\right ]\tilde{M}_{2m} \tilde{M}_{2n}
\nonumber \\
 & \approx  4mn\langle \xi^2\rangle \tilde{M}_{2m} \tilde{M}_{2n} \nonumber\\
 & =  2mn{N\choose q}^{-1} \tilde{M}_{2m} \tilde{M}_{2n} .
\label{eqn:doubleTrFromScaleFluc}
\end{align}
We will see below that
in the SYK model, this result is equal to the average double-trace moment with two cross
contractions, see equation~\eqref{eqn:2crossFactorization}.

\section{Nonlinear \texorpdfstring{\boldmath $\sigma$}{sigma}-model}
\label{sec:sigmaModel}

  Using standard random matrix techniques it is possible to derive a
  nonlinear $\sigma$-model for the SYK model. For the complex SYK model
  this was done already in the early eighties~\cite{Verbaarschot1984} but because
  of the coupling between the massive and massless modes, it was hard
  to analyze the $\sigma$-model reliably. The SYK model with Majorana
  fermions is simpler and Altland and Bagrets obtained~\cite{Altland-ml-2017eao} the following nonlinear $\sigma$-model
 for the $\beta =2$ universality class
  (see appendix~\ref{app:A} for
  a derivation),
  \begin{equation}
  Z =\int D a_\mu e^{-\frac 12 \sum{}^{'}_\mu T_\mu^{-1} \tr a_\mu^2
     -{\rm Tr} \log\left( z + \frac \sigma{ \sqrt D} \sum{}^{'}_\mu a_\mu X^U_\mu\right)} .
  \label{twogen}
  \end{equation}
\looseness=-1     The coefficient of $a_\mu^2$ is not positive definite, but the integrals
    can be made convergent by an appropriate rotation of the integration
    contour in the complex plane~\cite{Altland-ml-2017eao}. However, we evaluate the
    integral perturbatively in a loop expansion about its saddle point when
    the choice of the integration contour is irrelevant. The $X_\mu$'s are all the linearly independent products of Dirac matrices in $N$ dimensions and they form a basis for the vector space of all \mbox{$2^{N/2}\times 2^{N/2}$} matrices.
    Because the Hamiltonian commutes with the Dirac chirality matrix $\gamma_c$, the Hamiltonian splits into two blocks, and the
      partition function is for one of the two blocks.
      Therefore,   the sum over $\mu$ is, as indicated by the prime,  over the upper block of the  $X_\mu$ with even $|\mu|$ (the length of the multi-index $\mu$), and only ranges from $0\le |\mu|\le N/2$ (with only half the generators for $|\mu|=N/2$, see appendix~\ref{app:A}). We use the normalization convention that $X_\mu^2 =1$ and denote the upper block by $X_\mu^U$.
       As an example, for $N=4$ we~have
\begin{equation*}
\begin{split}
|\mu|=0: \{X_{\mu}\} &=\{\mathbb{1}\}, \\
|\mu|=1: \{X_{\mu}\} &=\{\gamma_i|i=1,2,3,4\},\\
|\mu|=2: \{X_{\mu}\} &=\{i\gamma_{i_1}\gamma_{i_2}|1\leq i_1<i_2\leq 4\},\\
|\mu|=3: \{X_{\mu}\} &=\{i\gamma_{i_1}\gamma_{i_2}\gamma_{i_3}|1\leq i_1<i_2<i_3\leq 4\},\\
|\mu|=4: \{X_{\mu}\} &=\{\gamma_1\gamma_2\gamma_3\gamma_4\},
\end{split}
\end{equation*}
and in this case the $\sum{}^{'}_\mu$ would be a sum over the set $\{\mathbb{1}, i\gamma_1\gamma_2,i\gamma_1\gamma_3, i\gamma_1\gamma_4\}$.
      In the analytical calculation we sum over all $\mu$ and correct for
      that by including the appropriate combinatorial factor.
  Note that contrary to~\cite{Altland-ml-2017eao} we use Hermitian generators
  ${X_\mu}^\dagger = X_\mu$.
The matrix $a_\mu$ is an $2n\times 2n$ matrix in replica space
and
\begin{equation}
z = (\underbrace{x+i\epsilon, \cdots, x+i\epsilon}_{n},
\underbrace{y-i\epsilon,\cdots ,y-i\epsilon}_{n}).
\label{z2}
\end{equation}
The matrix block corresponding to $x$ or $y$ will be denoted by $11$ or $22$, respectively.
The trace over the $2n$-dimensional replica space ($n$-dimensional in case
of the one-point function)
is denoted by $\tr$, while the combined trace over the Clifford algebra and
the replica space is denoted by $\Tr$.
The coefficient
  $T_\mu$ is a combinatorial factor due to the commutation of the $\gamma$ matrices:
  \begin{equation}\label{eqn:TmuDef}
  T_{\mu} = \frac{ 1 }{{N \choose q}}
  \sum_{j=0}^q (-1)^j{|\mu|\choose j}{N-|\mu|\choose q-j}
  \end{equation}
  with $|\mu|$ the cardinality of index set $\mu$.
  It is a Fierz coefficient of the Fierz transformation that arises in the
  derivation of the $\sigma$-model, see appendix~\ref{app:A}:
  \begin{equation}
  T_\mu = \frac 1D{N\choose q}^{-1} \sum_\alpha \Tr\, X^U_\mu \Gamma^U_\alpha X^U_\mu \Gamma^U_\alpha=2^{-\frac N2}{N\choose q}^{-1} \sum_\alpha \Tr\, X_\mu \Gamma_\alpha X_\mu \Gamma_\alpha
  \end{equation}
  with $\Gamma_\alpha$ the $q$-body operator defined in equation~\eqref{eqn:GammaAlphaDef} and $D = 2^{N/2}/2$.

  The pair correlation function is given by
  \begin{align}
  C(x,y)  & = \lim_{n\to0} \frac 1{n^2} \frac 1{D^2}\frac d{dx}\frac d{dy} Z(x,y)\nonumber\\
   & =  \lim_{n\to 0}\frac 1{n^2} \frac1{D} \frac 1{\sigma^2}
  \left \langle\tr P_{11}  a_0 \; \tr P_{22}  a_0 \right \rangle,
\end{align}
where $P_{pp}$ are the projections onto the $11$ block and $22$ block for $p=1$ and $p=2$, respectively. To obtain this expression we have written the derivatives
with respect to $x$ and $y$ as a derivative with respect to $a_0$.

The generating function for the one-point function has the same form as in the case
of the two-point function,~(\ref{twogen}), but now $a_\mu$ is an $n\times n$ matrix and
$z = \text{diag}(\underbrace{x, \cdots, x}_{n})$. The resolvent is given by
  \begin{align}
  G(z) & =
  -\frac 1n \frac 1D \Tr \frac d{dz} \log Z \nonumber\\
   & =
   -\lim_{n\to 0}\frac 1{n} \frac 1{\sigma \sqrt D}
  \left \langle\tr\;  a_0  \right \rangle,
\label{g-rep}
  \end{align}
    where the right-hand side is obtained after a partial integration by expressing the $z$-derivative as
    a derivative with respect to $a_0$.

    The aim of this section is to show that the $\sigma$ model~(\ref{twogen}) reduces to    the universal $\sigma$ model of the Gaussian Unitary Ensemble
    which is given in appendix~\ref{app:replicaGUETwoPoint}. The model~(\ref{twogen})
    spontaneouls breaks the $\U(n,n)$ symmetry of the replica space of
    the advanced and retarded sectors to $\U(n)\times \U(n)$. Because of that
    we know that the low energy effective Lagrangian
is based on this pattern of spontaneous symmetry breaking and
    is given by the one
    of the GUE\@. However, this argument
    does not determine the prefactor of this universal
    term, and neither do we know the effect of the coupling between the
    Goldstone modes and massive modes. As was already realized in 1984~\cite{Verbaarschot1984}, these couplings give large corrections which
    ultimately should cancel in order to arrive at the universal GUE result.
    However, to this date it has not been shown that such cancellations indeed
    do happen. The  goal of this paper is more modest. We show by a perturbative
    calculation that the $\sigma$ model~(\ref{twogen}) gives rise to
    long range eigenvalue correlations scaling as $1/(x-y)^2$, as in case of  the GUE,
    and  are consistent with low order moments which can be calculated independently.
    We also address, the issue, that for
    the $q=2$ case, which is integrable and should have Poisson statistics,
    the corrections terms should nullify the universal term. Also in this case
    we were not able to  solve this problem, but we do find
    a significant quantitative difference between the spectral correlator for
    $q=2$ and $q=4$.

\subsection{One-point function}\label{subsec:onepoint}

To better understand the convergence properties of the nonlinear sigma model, it is instructive to work out
the one-point function.
The generating function for the one-point function has the same form as in the case
of the two-point function,~(\ref{twogen}), but now $a_\mu$ is an $n\times n$ matrix and
$z = (\underbrace{x, \cdots, x}_{n})$.
  The resolvent is given by
  \begin{equation}
  G(z)= \int D a_\mu \frac {\sqrt D}{\sigma} (-a_0) e^{-\frac 12  \sum{}^{'}_\mu T_\mu^{-1}
 \tr   a_\mu^2   -{\rm Tr}  \log\left( z + \frac \sigma{ \sqrt D} \sum{}^{'}_\mu a_\mu X^U_\mu\right)}
,
  \end{equation}
  where $a_0$ is the diagonal element of the replica matrix.
    We evaluate the integral by a loop expansion about the saddle point.
  The saddle point equation is given by
  \begin{equation}
  T_{\mu}^{-1} a_\mu +\Tr  \frac {\sigma  X^U_\mu/\sqrt D } {z+\sigma \sum{}^{'}_\mu X^U_\mu a_\mu/\sqrt D}=0,
  \label{saddle-one}
  \end{equation}
  where the trace $\Tr$ is over the gamma matrices.
  By inspection one realizes
  that a solution is given by~\cite{Verbaarschot1984}
  \begin{equation}
  a_\mu =\bar a \delta_{\mu 0},
  \end{equation}
  with
\begin{equation}
  \bar a + \frac{\sigma \sqrt D }{z + \sigma \bar a/\sqrt D}=0.
  \end{equation}
For $\mu\neq 0$ this follows from the fact that $\Tr X^U_\mu =0$.
     The saddle point result for the resolvent is equal to
    \begin{equation}
    \bar G(z) = - \frac {\bar a } {\sigma \sqrt D}.
\label{res}
    \end{equation}
    The resolvent thus satisfies the saddle point equation
    \begin{equation}
    \bar G + \frac 1{z + \sigma^2 \bar G}=0,
    \end{equation}
    and is given by
    \begin{equation}
    \bar G=  \frac z{2\sigma^2} - \frac 1{2\sigma^2} \sqrt{z^2-4\sigma^2}.
\end{equation}
    We expand $a_\mu$ about its saddle point
    \begin{equation}
    a_\mu = \bar a  +\alpha_\mu.
    \end{equation}
    Using the saddle point equation for $\overline{ a}$,
    we expand the logarithm as
  \begin{align}
  -\Tr \log (z+\frac \sigma {\sqrt{D}} \sum_\mu{}^{'} a_\mu X^U_\mu)
 & =  \Tr \sum_{k=1}^\infty \frac 1k \left(\frac {\bar a}{D} \sum{}^{'}_\mu \alpha_\mu X^U_\mu\right)^k\\
   & =  \Tr \sum_{k=1}^\infty \frac 1k \left(\frac {-\bar G\sigma }{\sqrt D} \sum{}^{'}_\mu \alpha_\mu X^U_\mu\right)^k.
  \end{align}
  This results in the  propagator
  \begin{equation}
  \langle \alpha_\mu^{kl} \alpha_\nu^{mn}\rangle
  = \delta_{\mu \nu}\delta_{lm}\delta_{kn}
  \frac { T_{\mu}}{1- T_{\mu} \left (\; {\bar G\sigma } \right )^2},
\label{prop-one}
  \end{equation}
  while the vertices of the loop expansion are given by
\begin{equation}
  \Tr \sum_{k=3}^\infty \frac 1k \left(\frac {-\bar G\sigma }{\sqrt D} \sum_\mu{}^{'} \alpha_\mu X^U_\mu\right)^k.
  \end{equation}

  We now calculate the resolvent to two-loop order in this expansion. The one-loop contribution,
  $\langle \Tr a_0 \Tr (\sum{}^{'}_\mu\alpha_\mu X^U_\mu)^3 \rangle $,
  vanishes in the replica limit. The first nonvanishing contribution,
    $\langle \Tr a_0 \Tr (\sum{}^{'}_\mu\alpha_\mu X^U_\mu)^5 \rangle $, is given
 by the diagram
 \begin{equation}
\contraction[1ex]{\tr\,} {\alpha}{\,\tr\,}{\alpha}
\contraction[2ex]{\tr\, \alpha \,\tr\, \alpha}{\alpha}{\alpha }{\alpha}
\contraction[3ex]{\tr\, \alpha \,\tr\, \alpha\alpha}{\alpha}{\alpha}{\alpha}
\tr\, \alpha \, \tr\, \alpha\alpha\alpha\alpha\alpha.
\end{equation}
 Writing out the indices and the traces over the replica space we obtain
\begin{equation}
\begin{tikzpicture}[baseline=(current  bounding  box.center)]
\node at (0,0) {$\sum\limits_{k,l=1}^n \sum\limits_{\mu \nu}
  \alpha_0^{kk}\alpha_0^{kk}\alpha_\mu^{kl}\alpha_\nu^{lk}\alpha_\mu^{kl}\alpha_\nu^{lk}.$};
 \draw (-0.9,0.2)--(-0.9,0.5)--(-0.25,0.5)--(-0.25,0.2);
 \draw (0.3,0.2)--(0.3,0.5)--(1.35,0.5)--(1.35,0.2);
  \draw (0.85,0.2)--(0.85,0.7)--(1.8,0.7)--(1.8,0.2);
\end{tikzpicture}
\end{equation}
From the progator~(\ref{prop-one}) it is immediately clear that the
only nonvanishing contractions are those with replica indices $k=l$. Since
the propagtor is replica symmetric, the sum over $k$ gives an overall
factor $n$ which is cancelled by the $1/n$ factor of the replica limit~(\ref{g-rep}).
Combining this with the factors $X_\mu^U$ we obtain
  \begin{equation}
    \sum_\mu{}^{'} \sum_\nu{}^{'} \frac{\bar G^5\sigma^4}{D^2(1-\bar G^2)}\frac{ T_\mu}{1-T_\mu
    \left (  {\bar G \sigma}\right )^2}
    \frac{T_\nu}{1-T_\nu \left(\bar G \sigma \right )^2}
    \frac 1D  \Tr X^U_\mu X^U_\nu X^U_\mu X^U_\nu.
    \label{dia1}
     \end{equation}
     The trace can be evaluated  by observing that $X^U_\mu$  and $X^U_\nu$ commute
     or anti-commute depending on the number of gamma matrices they have in
     common, while $T_\mu$ only depends on the number of indices in $\mu$.
     This results in
     \begin{align}
&  \sum_m{}^{'} \sum_n{}^{'} {N\choose m}\frac{\bar G^5\sigma^4}{D^2(1-\bar G^2)}
\frac{ T_m}{1-T_m    \left (  {\bar G \sigma}\right )^2}
    \frac{T_n}{1-T_n \left(\bar G \sigma \right )^2}
\sum _{p=0}^m (-1)^p {m\choose p}{N-m \choose n-p}
    \nonumber \\  & =
     \frac {\sigma^4 \eta}{z^5} + \frac {\sigma^6(6\eta+2\eta^2)}{z^7} +O(1/z^9).
     \end{align}
     The coefficients of the $1/z$ expansion can be obtained by using
     various combinatorial identities, see appendix~\ref{app:one}.
 The normalized fourth moment is thus given by
     \begin{equation}
     \frac {M_4}{\sigma^4} = 2+\eta.
     \end{equation}
     Next we consider the contribution
     \begin{equation}
\lim_{n\to 0} \frac 1n \frac 1D
     \left \langle\frac 1{12} \bar G(z) \tr \alpha_0 {\Tr}
     \left ( \frac{ \bar  G(z)\sigma}{\sqrt D} \sum_\mu{}^{'} \alpha_\mu X^U_\mu\right )^3
     {\Tr}  \left ( \frac {\bar  G(z)\sigma}{\sqrt D} \sum_\mu{}^{'} \alpha_\mu X^U_\mu\right )^4 \right \rangle,
     \end{equation}
which permits two different contraction patterns.
For the diagram
\begin{equation}
\contraction[1ex]{\tr\,}{ \alpha} {\,\tr\, \alpha\alpha\alpha \,\tr\,}{\alpha}
\contraction[2ex]{\tr\, \alpha \,\tr\,}{\alpha} {\alpha\alpha \,\tr\, \alpha}{\alpha}
  \contraction[3ex]{\tr\, \alpha \,\tr\, \alpha}{\alpha} {\alpha\,\tr\, \alpha\alpha}{\alpha}
    \contraction[4ex]{\tr\, \alpha \,\tr\, \alpha\alpha}{\alpha}{\,\tr\, \alpha\alpha\alpha}{\alpha}
\tr\, \alpha \,\tr\, \alpha\alpha\alpha \,\tr \,\alpha\alpha\alpha\alpha
\label{dia2}
\end{equation}
we find after taking the replica limit
\begin{align}
  \frac 1{D^4}
\frac {\bar G^7\sigma^6}{1 -\sigma^2\bar G^2}
\sum_{\mu\nu\rho}{}^{'}
\frac{T_\mu}{1-T_\mu
  \left (  {\bar G \sigma}\right )^2}
\frac{ T_\nu}{1-T_\nu
  \left (  {\bar G \sigma}\right )^2}
\frac{ T_\rho}{1-T_\rho
    \left (  {\bar G \sigma}\right )^2}
         \Tr X^U_\mu X^U_\nu X^U_\rho \Tr X^U_\mu X^U_\nu X^U_\rho.\label{dia21}
\end{align}
The traces are only nonzero if $X^U_\rho$ contains the
gamma matrices that
         $X^U_\mu$ and $X^U_\nu$ do not have in common.          For $m_1$ gamma matrices
         in $X^U_\mu$, $m_2$ gamma matrices in $X^U_\nu$ and $s$ common gamma
         matrices this gives the combinatorial factor
         \begin{equation}
         D^2 {N \choose m_1} {m_1\choose s} {N-m_1 \choose  m_2-s}(-1)^s,
         \end{equation}
         where the phase factor is due to the fact that only $i^s X^U_\mu X^U_\nu$
         is Hermitian.      The diagram~(\ref{dia21})  is thus equal to
         \begin{align}
&         \frac 1{D^2}
         \frac {\bar G^7\sigma^6}{1 -\sigma^2 \bar G^2}
       \frac 14  \sum_{m_1=0}^N \sum_{m_2=0}^N \sum_{s=0}^{m_1}
\frac{ T_{m_1}}{1-T_{m_1}
  \left (  {\bar G \sigma}\right )^2}
\frac{T_{m_2}}{1-T_{m_2}
  \left (  {\bar G \sigma}\right )^2}
\frac{T_{m_1+m_2 -2s}}{1-T_{m_1+m_2 -2s}
    \left (  {\bar G \sigma}\right )^2}
\nonumber \\ &\times
{N \choose m_1} {m_1\choose s} {N-m_1 \choose  m_2-s}(-1)^s.
     \end{align}
The sums for the leading order term in the $1/z$ expansion of the resolvent
can be evaluated as the contribution $T_6$~\cite{Garcia-Garcia-ml-2018kvh},\footnote{The notation $T_6$ in~\cite{Garcia-Garcia-ml-2018kvh} refers to the single-trace chord diagrams with three chords all intersecting with each other. It is a bit unfortunate that the notation $T_6$ clashes with our notation of $T_\mu$ defined in~\eqref{eqn:TmuDef}.} see appendix~\ref{app:one}.
The combinatorial factor $1/4$ is to account for the prime on the sums in equation~(\ref{dia21}).
Note that the sum over $m_1$ and $m_2$ only runs up to $N/2$ while the
sum over odd $m_1$ and $m_2$ vanishes anyway.
We thus  obtain the large $z$ expansion of the correction
\begin{equation}
\frac {T_6}{z^{7}}+O(1/z^9).
\end{equation}
The second diagram
     \begin{equation}
\contraction[1ex]{\tr\,} {\alpha}{\,\tr\,}{\alpha}
\contraction[2ex]{\tr\, \alpha \,\tr\, \alpha}{\alpha}{\alpha \,\tr\, }{\alpha}
\contraction[3ex]{\tr\, \alpha \,\tr\, \alpha\alpha}{\alpha}{\,\tr\, \alpha\alpha}{\alpha}
\contraction[4ex]{\tr\, \alpha \,\tr\, \alpha\alpha\alpha \,\tr\, \alpha}{\alpha}{\alpha}{\alpha}
\tr\, \alpha \,\tr\, \alpha\alpha\alpha \,\tr\, \alpha\alpha\alpha\alpha
\end{equation}
 is equal to
\begin{align}
&\frac 1{D^4}
\frac {\bar G^7\sigma^6}{1 -\sigma^2 \bar G^2}
\sum_{\mu\nu}{}^{'}
\frac{T_\mu^2}{(1-T_\mu
  \left (  {\bar G \sigma}\right )^2)^2}
\frac{ T_\nu}{1-T_\nu
  \left (  {\bar G \sigma}\right )^2}
         \Tr X^U_\mu X^U_\mu  \Tr X^U_\mu X^U_\nu X^U_\mu X^U_\nu
         \nonumber\\
          & =  \frac {\sigma^6 \eta^2}{z^7} + O(1/z^9),
         \end{align}
         where we have used combinatorial identities to obtain the coefficient of $1/z^7$.

         To summarize, starting from the generating
         function $Z$, we have obtained the normalized sixth moment
         \begin{equation}
         \frac{M_6}{\sigma^6} = 5 + 6\eta +3\eta^2 +T_6
         \end{equation}
         in agreement with an explicit moment calculation of the
         SYK model~\cite{Garcia-Garcia-ml-2016mno}.

         If we continue this expansion we will recover the full $1/z$ expansion of
         the resolvent of the SYK model. This expansion is not well behaved
         for $z$ close to the support of the spectrum and will have to be resummed to obtain nonperturbative results, see section~\ref{sec:resolventCutIssue} for
         more discussion of this issue.

\subsection{Loop expansion of the two-point function}\label{sec:loopTwoPoint}
The saddle-point equation for the two-point function  has the same form as equation~(\ref{saddle-one}) but now $z$ is a vector of length $2n$, see equation~(\ref{z2}), and $a_\mu$ is a $2n\times 2n $ matrix. The solution with $a_\mu^{12} = 0$ is the same as for
the one-point function with $z$ replaced by $x$ and $y$ for the 11-sector and the 22-sector, respectively,
\begin{equation}
  \bar a^{11} + \frac{\sigma \sqrt D }{x + \sigma \bar a^{11}/\sqrt D} = 0,\qquad
  \bar a^{22} + \frac{\sigma \sqrt D }{y + \sigma \bar a^{22}/\sqrt D}=0.
  \end{equation}
  The saddle point value of the resolvent is thus given by
\begin{equation}
  \bar G = \left ( \begin{array}{cc} \bar G(x) \mathbb{I}_n & 0 \\ 0 & \bar G(y) \mathbb{I}_n \end{array} \right ),
  \end{equation}
  where $\bar G(x)$ and $\bar G(y)$ are related to $\bar a_{11}$ and $\bar a_{22}$ according to
  equation~(\ref{res}), in this order.

  The propagator and vertices follow from the expansion of the logarithm in the action
  \begin{align}
 -\Tr \log (z+\frac \sigma {\sqrt{D}} \sum_\mu{}^{'} a_\mu X^U_\mu)
   & =  \Tr \sum_{k=1}^\infty \frac 1k \left(\frac {\bar a}{D} \sum_\mu{}^{'} \alpha_\mu X^U_\mu\right)^k\\
  & =  \Tr \sum_{k=1}^\infty \frac 1k \left (\frac {-\bar G\sigma }{\sqrt D}
 \sum_\mu{}^{'} \alpha_\mu X^U_\mu \right)^k.
  \end{align}
The propagator is thus given by
  \begin{equation}
  \langle \alpha_\mu^{pq} \alpha_\nu^{qp}\rangle = \delta_{\mu \nu}
  \frac { T_{\mu}}{1- T_{\mu}\sigma^2 \bar G(x_p)\bar G(x_q) }.
  \end{equation}

  We first calculate the one-loop contribution to the two-point correlator
  \begin{equation}
  C(x,y)
   =   \left \langle \frac 1n \frac 1D \frac {\sqrt D}\sigma\tr P_{11} \alpha_0\frac 1n \frac 1D \frac {\sqrt D}\sigma\tr P_{22}\alpha_0\right \rangle.
  \end{equation}
  As is the case for the GUE (see appendix~\ref{app:replicaGUETwoPoint}) we have two contributions.
  The first contribution is given by
  \begin{align}
&  \frac 1n \frac 1D
  \left \langle
  \frac {\sqrt D}\sigma\tr P_{11} \alpha_0\frac 1n \frac 1D \frac {\sqrt D}\sigma\tr P_{22}\alpha_0 \frac 14
   \Tr \left (\frac {-\bar G\sigma }{\sqrt D} \sum_\mu{}^{'} \alpha_\mu X^U_\mu \right)^4
  \right \rangle\nonumber \\
 & =   \frac 1{n^2} \frac 1{D^3} \sigma^2
\bar G^2(x)\bar G^2(y)
  \left \langle
   \tr \alpha^{11}_0 \tr \alpha^{22}_0
 \sum_\mu{}^{'} \Tr \alpha^{11}_0 \alpha_\mu^{12} X^U_\mu \alpha_0^{22} \alpha_\mu^{21} X^U_\mu
  \right \rangle\nonumber \\
 & =   \frac{\sigma^2}{D^2}
  \sum_\mu{}^{'} \frac{\bar G^2(x)\bar G^2(y)T_{\mu}}{(1-T_{\mu} \sigma^2 \bar G(x)
    \bar G(y))
    (1- \sigma^2 \bar G^2(x))(1- \sigma^2 \bar G^2(y))}\nonumber \\
 & = \left( \sigma^4 \frac 1{x^3y^3}+2\sigma^6 \frac 1{x^4y^4}\right )
 {N\choose q}^{-1}+\frac{\sigma^6}{D^2}\frac 1{x^4y^4} \sum_\mu{}^{'} T_{\mu}^3 +\cdots\,
\label{dia-two1}
 \end{align}
  where we have used that
  \begin{equation}
  \sum_\mu{}^{'} T_\mu = 0, \qquad
\frac 1{D^2}  \sum_\mu{}^{'} T_\mu^2 = \frac 1{D^2}  \frac 14 \sum_{|\mu|=0}^N {N\choose |\mu|}T_\mu^2  =   {N\choose q}^{-1},
  \end{equation}
  and that the sum over $\mu$ in equation~(\ref{dia-two1}), as indicated by the primes,
  only runs  over the even values of $\mu < N/2$ resulting in the combinatorial factor of $ 1/4$.

    The second one-loop contribution to the two-point function is given by
 \begin{align}
&  \frac 1n \frac 1D
  \left \langle
  \frac {\sqrt D}\sigma\tr P_{11} \alpha_0\frac 1n \frac 1D \frac {\sqrt D}\sigma\tr P_{22}\alpha_0 \frac 1{18}
  \Tr^2 \left (\frac {-\bar G\sigma }{\sqrt D} \sum_\mu{}^{'} \alpha_\mu X^U_\mu \right)^3
  \right \rangle\nonumber \\
 & =   \frac 1{n^2} \frac 1{D^4} \sigma^4
\bar G^3(x)\bar G^3(y)
  \left \langle
   \tr \alpha^{11}_0 \tr \alpha^{22}_0
 \sum_\mu{}^{'} \Tr \alpha^{11}_0 \alpha_\mu^{12} X^U_\mu  \alpha_\mu^{21} X^U_\mu
\sum_\nu{}^{'} \Tr \alpha^{22}_0 \alpha_\nu^{21} X^U_\nu  \alpha_\nu^{12} X^U_\nu
 \right \rangle\nonumber \\
 & =    \frac{\sigma^4}{D^2}
\sum_\mu{}^{'} \frac{\bar G^3(x)\bar G^3(y)T_{\mu}^2}{(1-T_{\mu}\sigma^2 \bar G(x)\bar G(y))^2(1- \sigma^2 \bar G^2(x))(1- \sigma^2 \bar G^2(y))}\nonumber \\
  & =  \left( \sigma^4\frac 1{x^3y^3}-2\sigma^6\frac 1{x^4y^4}\right ) {{N\choose q} }^{-1}+ 2\frac{\sigma^6}{D^2}\frac 1{x^4y^4} \sum_\mu{}^{'} T_{\mu}^3 +\cdots\
\end{align}
after taking the replica limit.

\begin{figure}
\centering
\includegraphics[width=8cm]{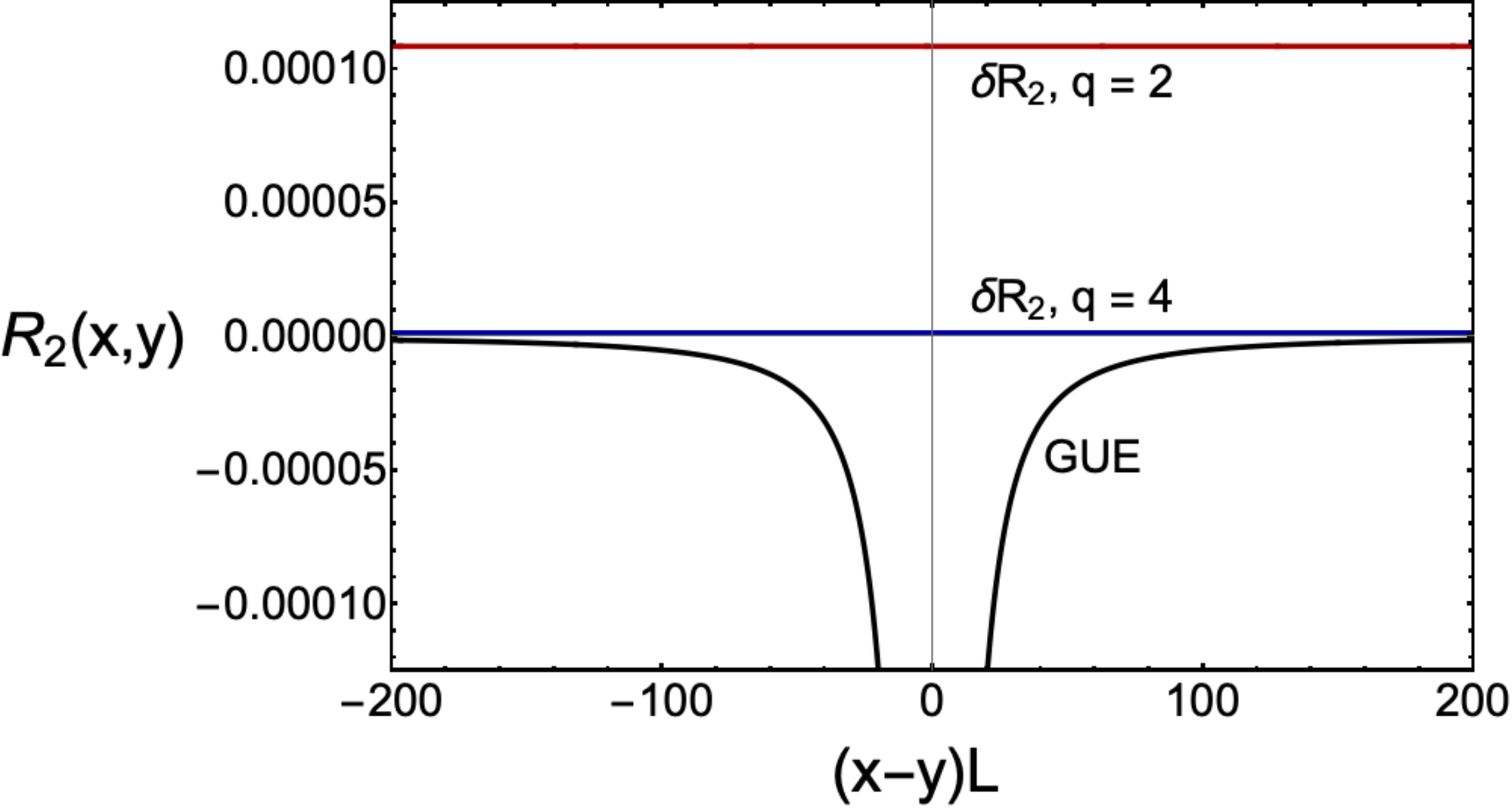}
 \caption{Perturbative calculation of the two-point correlator. The
   GUE result is the $\mu=0$ term in~(\ref{cor3a}), while the
   correction $\delta R_2$ for $q=2$ (red) and $q=4$ (blue) is due to
   the $\mu\ne 0$ terms.  The ratio of the $q=2$ and $q=4$ correction
   scales as $N^2$, and in the figure its value is
   142.\label{fig:corr}}
\end{figure}

  The sum of the two contributions is equal to
  \begin{align}
& \frac{\sigma^2}{D^2}
  \sum_\mu{}^{'} \frac{\bar G^2(x)\bar G^2(y)T_{\mu}}{(1-T_{\mu}\sigma^2 \bar G(x)\bar G(y))^2(1- \sigma^2 \bar G(x)\bar G(x))(1- \sigma^2 \bar G(y)\bar G(y))}\nonumber\\
   & =  2 \sigma^4\frac 1{x^3y^3} {{N\choose q} }^{-1}+ 3\frac{\sigma^6}{D^2}\frac 1{x^4y^4} \sum_\mu{}^{'} T_{\mu}^3 +\cdots\,
\label{cor3a}
  \end{align}
  The $\mu = 0$ term is the large $N$ limit of the two-point correlator for the GUE (see appendix~\ref{app:replicaGUETwoPoint}). It is
  given by
  \begin{align}
\frac{1}{D^2}\frac{ \sigma^2 \bar G^2(x)\bar G^2(y)}{(1-\sigma^2 \bar G(x)\bar G(y))^2(1- \sigma^2 \bar G^2(x))(1- \sigma^2 \bar G^2(y))}.
 \end{align}
The corresponding spectral correlation function follows from the discontinuity across the real axis
\begin{align}
\langle \rho(x) \rho(y)\rangle_c  & = -\frac 1{4\pi^2}\langle(G(x+i\epsilon)-G(x-i\epsilon)) (G(y+i\epsilon)-G(y-i\epsilon)) \rangle
\nonumber \\  & = -\frac 1{2\pi^2} \frac 1{(x-y)^2 2^{N}/4}. \end{align}
This is exactly the asymptotic behavior of
\begin{equation}
-\frac {\sin^2 [\pi(x-y)D]}{(  D\pi(x-y))^2}
\approx  -\frac 1{2\pi^2} \frac 1{D^2 (x-y)^2},
  \end{equation}
  where $D$ is the total number of eigenvalues, $D = 2^{N/2}/2$.
  In figure~\ref{fig:corr} we show the GUE result $\mu=0$ and the sum of the correction terms $\mu\ne 0$) both for
  $q=2$ and $ q=4$.  The correction of  the $q=2$ result is much larger,
  by a factor of order $N^2$, but is not qualitatively
  different from that of the $q=4$ result. We observe that in both cases we have a scale separation between
  the GUE result and the correction due to the massive modes.
  Since other correction terms are of higher order in $1/D$, it is
puzzling how
  the correction
  terms  for $q=2$ can nullify the GUE contribution in order to get Poisson statistics.
  \relax
  The only way out seems to be that interaction terms between the zero
modes ($|\mu| =0$ or $|\mu| =N$ terms) and the massive modes ($|\mu| \ne 0, N$ terms) are important for $q=2$ while their contributions cancel for $q=4$. The details of
this cancellation are not clear, in particular because the lowest order contribution of such terms was large
in the complex SYK model~\cite{Verbaarschot1984}.
Since $T_\mu \sim 1 - 2|\mu|q/N + \cdots$,  the $|\mu|\ne 0$ terms are strongly
suppressed with respect to the $|\mu| =0$ or $|\mu|= N$ terms for $x \to y$.
For $\mu$ values with ${N\choose |\mu|} \gg 1$ we have that $T_\mu \ll 1$. Therefore,
we can approximate
the correction to the GUE result by expanding the denominator to first order in $T_\mu$.
In the center of the spectrum this gives the result
  \begin{equation}
  -\frac 1{2D^2}  \sum_{|\mu|}{N \choose |\mu|} T_{\mu}^2 = -2 {N\choose q}^{-1},
  \end{equation}
  so that the correction to the two point correlator is given by
  \begin{equation}
  \delta R_2 = \frac 1{2\pi^2}{N\choose q}^{-1}.
  \end{equation}
In terms of the
unfolded energy difference $ \omega =(x-y)\rho(0)$, we find
\begin{equation}
\frac{\bar \rho_{2c}(\omega)}{\bar\rho^2(0)} = \rho_2^{\rm GUE}(\omega) +
\frac 12 {N\choose q}^{-1},
\end{equation}
which is in agreement with the results in figure~\ref{fig:corr}. Note that we only get the universal GUE result if
we rescale by the saddle point result for the spectral density which differs by  $O(1)$ from the
correct SYK result.

This result agrees with~\cite{Altland-ml-2017eao}, where  it was argued that the corrections to the two-point spectral
correlation function are of the form
\begin{equation}
\rho_2(\omega) = \rho_2^{\rm GUE}(\omega) + \frac {\Delta^2}{2\pi^2}
{\rm Re} \sum_{k=1}^{N/4} {N\choose 2k} \frac 1{(i\omega +\epsilon( 2k))^2},
\end{equation}
where ${N \choose 2k}$ is the degeneracy of massive modes with mass $\epsilon(k)$, and $\Delta=\pi\sigma/D$ is the level spacing in the center of the band.
The $\epsilon(k)$ are given by
\begin{equation}
\epsilon(k) = \sigma(T_k^{-1} -1 ).
\end{equation}
Since $T_k \sim 1/N^q$, the $\epsilon(k)$ are much larger than the span of the spectrum, while $\omega \ll N$.
The correlator of~\cite{Altland-ml-2017eao} is thus well-approximated by
\begin{equation}
\rho_2(\omega) = \rho_2^{\rm GUE}(\omega) + \frac 12 {N\choose q}^{-1},
\end{equation}
which is exactly the result from the above perturbative calculation.

  Although the $\sigma$-model was derived for arbitrary $N$ and even $q$, it cannot be naively applied to
  cases with Dyson index $\beta =1$ or $\beta =4$. As was discussed at end of appendix~\ref{app:A}, the reason
  is that the $|\mu| = 0$ term after the Fierz transformation, gives the GUE result. To extract the GOE result, we
  have exploited the symmetry of the $\gamma$ matrices for $N\pmod 8 =0$ before the Fierz transformation. Then
  the $|\mu| =0 $ terms include the Cooperon contributions characteristic of the GOE result. Similar arguments can be
  made for the $\beta =4$ case.

  \subsubsection{Higher order corrections}
  It becomes increasingly hard to calculate higher order corrections to the loop-expansion of
  the two-point correlation function. Such terms are important for ensemble fluctuations that
  go beyond  scale fluctuations. We only calculate the three-loop diagram
  given by
       \begin{equation}
       \contraction[1ex]{\tr P_{11}} {\alpha} {\,\tr P_{22} \alpha\, \tr\,}{\alpha}
      \contraction[2ex]{\tr  P_{11}\alpha \,\tr P_{22}}{\alpha} {\,\tr \,\alpha \alpha }{\alpha}
       \contraction[3ex]{\tr\, \alpha P_{11} \,\tr \alpha P_{22} \,\tr\, \alpha}{\alpha}{ \alpha\alpha\alpha}{\alpha}
      \contraction[4ex]{\tr\, \alpha P_{11} \,\tr \alpha P_{22} \,\tr\, \alpha \alpha\alpha }{\alpha}
                   {\alpha\alpha}{\alpha}
       \contraction[5ex]{\tr\, \alpha P_{11} \,\tr \alpha P_{22} \,\tr\, \alpha \alpha \alpha\alpha}{\alpha}
        { \alpha\alpha}{\alpha}
       \tr  P_{11}\alpha \, \tr P_{22}\alpha \,\tr\, \alpha \alpha \alpha\alpha\alpha \alpha\alpha\alpha.
\label{dia33}
       \end{equation}
     We find
{\rdmathspace[h]
\begin{align}
&  \frac 1n \frac 1D
  \left \langle
  \frac {\sqrt D}\sigma\tr P_{11} \alpha_0\frac 1n \frac 1D \frac {\sqrt D}\sigma\tr P_{22}\alpha_0 \frac 18
   \Tr \left (\frac {-\bar G\sigma }{\sqrt D} \sum_\mu{}^{'} \alpha_\mu X^U_\mu \right)^8
  \right \rangle \\
 & =   3\frac 1{n^2D^5} \sigma^6
\bar G^4(x)\bar G^4(y)
\scalemath{.97}{\left \langle\!\!
   \tr\, \alpha^{11}_0 \tr \,\alpha^{22}_0
   \sum_\mu{}^{'} \Tr a^{11}_0 \alpha_\mu^{12} X^U_\mu \alpha_0^{22} \alpha_\nu^{21} X^U_\nu \alpha_\rho^{12} X^U_\rho
   \alpha_\mu^{12} X^U_\mu  \alpha_\nu^{21} X^U_\nu \alpha_\rho^{12} X^U_\rho\!\!
  \right \rangle}\nonumber \\
 & =  3 \frac{\sigma^6}{D^5}
  \frac{\bar G^4(x)\bar G^4(y)}{(1-\bar G^2(x))(1-\bar G^2(y))}
\nonumber \\
& \quad \times
  \sum_{\mu\nu\rho}{}^{'}
  \frac{T_{\mu}}{1-T_{\mu} \sigma^2 G(x)G(y)}
  \frac{T_{\nu}}{1-T_{\nu} \sigma^2 G(x)G(y)}
  \frac{T_{\rho}}{1-T_{\rho} \sigma^2 G(x)G(y)}
\Tr X^U_\mu X^U_\nu X^U_\rho X^U_\mu X^U_\nu X^U_\rho .\nonumber
\end{align}}\relax
  There is another remarkable combinatorial identity (note that there is no prime)
  \begin{equation}
\frac 1{D^4}\sum_{\mu\nu\rho} T_\mu T_\nu T_\rho \Tr X^U_\mu X^U_\nu X^U_\rho X^U_\mu X^U_\nu X^U_\rho = 16(-1)^{q/2} {N \choose q}^{-3}\frac{N!}{(N-3q/2)! ((q/2)!)^3},
\end{equation}
which can be proved by applying the Fierz transformation~(\ref{Fierz}) to $\sum_\mu T_\mu X^U_\mu X^U_\nu$ and the same for the sum
over $\nu$ and $\rho$.
The sum is only nonzero when the summation indices are either all even or
all odd. Since the sum over the many-body space is only over the even indices
up to $N/2$, we thus get an overall combinatorial factor of 1/16, so that
the total contribution of the diagram~(\ref{dia33}) to the 33 moment is given by
\begin{equation}
3(-1)^{q/2} {N \choose q}^{-3}\frac{N!}{(N-3q/2)! ((q/2)!)^3},
\end{equation}
which together with the last term of equation~(\ref{cor3a}) gives the correct result for the 33 moment
after using the identity
\begin{equation}
\frac 1{D^2}\frac 14 \sum_{|\mu|=0}^N{N\choose |\mu|}T_{\mu}^3  =  {N \choose q}^{-3}\frac{N!}{(N-3q/2)! ((q/2)!)^3}.
    \end{equation}
    This agrees with the moment calculation in section~\ref{sec:covariance}, and is a very nontrivial
    check of the correctness of the $\sigma$-model calculation.

\subsection[Does the expansion in powers of the resolvent of the semicircle make sense?]{Does the expansion in powers of the resolvent of the semicircle make \texorpdfstring{\newline}{} sense?}\label{sec:resolventCutIssue}

         The $\sigma$-model results in an expansion of the resolvent of the
         SYK model in powers of $g_0(z) = z/2 -\sqrt{z^2-4}/2$:
         \begin{equation}
         G(z) = \sum_k a_k g_0^{2k+1}(z).
         \end{equation}
         Each of the terms has a cut in the complex plane on the interval
         $[-2,2]$, while the resolvent of the SYK model has a cut beyond that.
         In the Q-Hermite approximation, the cut is located on the interval
         \begin{equation}
         \left [ -\frac 2{1-\eta}, \frac 2{1-\eta} \right],
         \end{equation}
         where $\eta$ is given in equation~(\ref{eqn:etaFirstappearance}).
         Since the resolvent of the Q-Hermite spectral density is known analytically,
         we can get the coefficients $a_k$ in this case,
         \begin{equation}
         a_k = \sum_{p=0}^k (-1)^{k+p} {k+p \choose k-p} \tilde M^{\text{QH}}_{2p}(\eta)
         \label{ak-sc}
         \end{equation}
         with
         \begin{equation} \label{eqn:RiordanTouchard}
         \tilde M^{\text{QH}}_{2p}(\eta) = \frac 1{(1-\eta)^p} \sum_{k=-p}^p(-1)^k \eta^{k(k-1)/2}{2p\choose p+k}.
         \end{equation}

 \begin{figure}
\centering
\includegraphics[width=8cm]{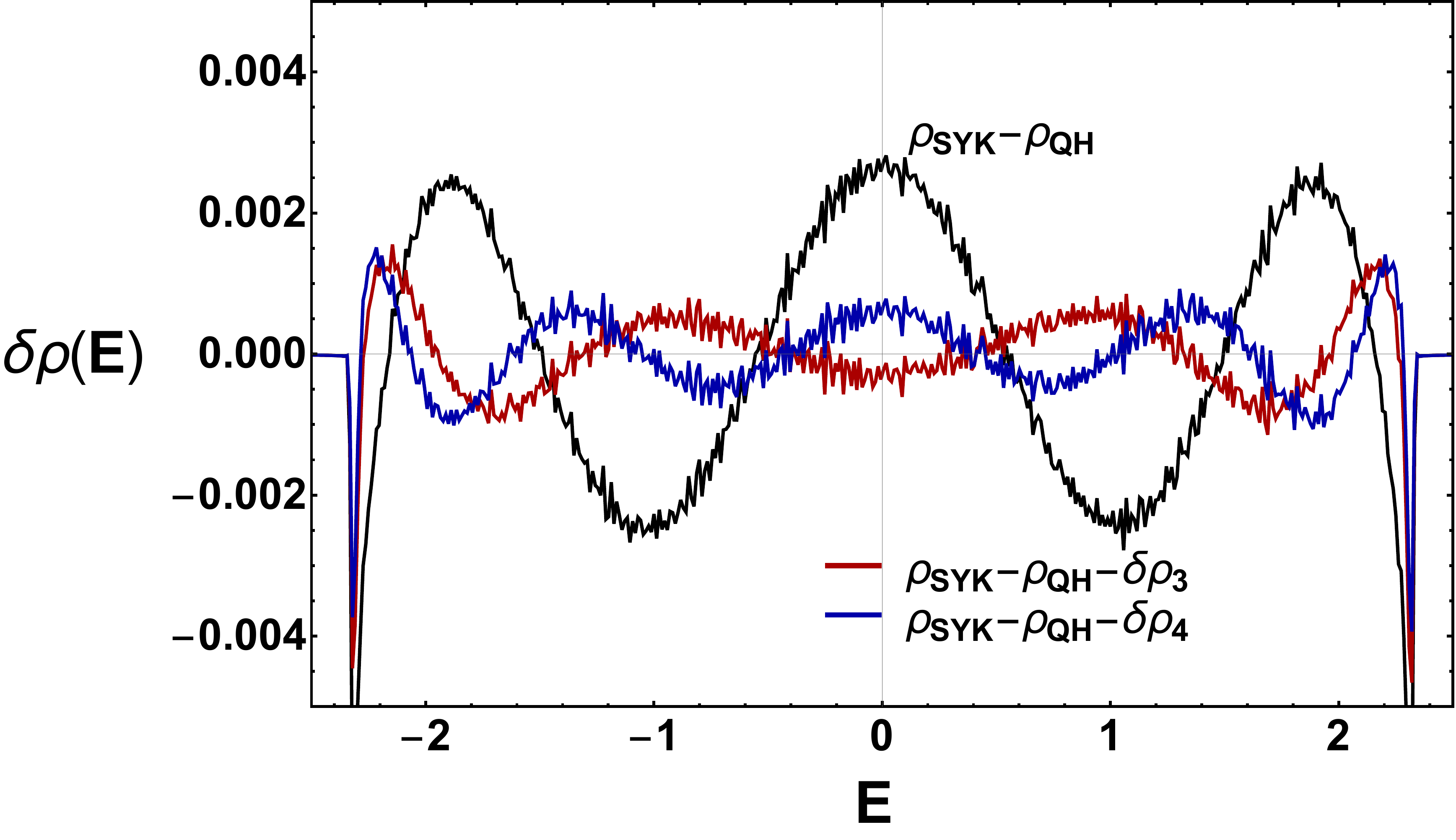}
   \caption{The difference of the spectral density of the SYK model,
     $\rho_{\rm SYK}(E)$, and three different approximations. First,
     the Q-Hermite approximation, $\rho_{\rm QH}(E)$ (black
     curve). Second and third, the expansion in powers of the
     resolvent of the semicircle scaled to the support of the
     Q-Hermite result to 7th order, $\rho_{\rm QH}(E)+\delta \rho_3$
     (blue curve), and 9th order, $\rho_{\rm QH}(E)+\delta \rho_4$
     (red curve).\label{fig:dif}}
  \end{figure}

         The spectral density that can be derived from derived from~(\ref{ak-sc}) is
         strongly oscillating, and to make sense of the result, the asymptotic
         series has to be resummed. Based on the case $\eta = 1$, one might
         think that a Borel resummation might lead to a convergent result~\cite{Verbaarschot1984}, but we were not able to work out the sums for arbitrary
         $\eta$. However, we can also expand the resolvent in powers of
         \begin{equation}
         g_{0\eta}(z) = \sqrt{1-\eta}\left( {\frac z2\sqrt{1-\eta}} -\frac 12 \sqrt{z^2(1-\eta)-4}\right).
         \end{equation}
         Naively, because the spectrum corresponding to this resolvent has
         the same support as the weight function of the Q-Hermite polynomials,
         we expect that this gives a much better expansion. It turns out that
         the expansion is surprisingly  simple~\cite{Cappelli-1998,Berkooz-ml-2018jqr}
         \begin{equation}
       G_{\rm QH}(z) =\sum_{k=0}^\infty \frac{\eta^{k(k+1)/2}}{(\eta-1)^k}
         g_{0\eta}^{2k+1}(z).
         \end{equation}
 Since $|g_{0\eta}| =\sqrt{1-\eta}$ and $|\eta|< 1$, this expansion is convergent for the Q-Hermite spectral density.

         We could also expand the exact resolvent of the SYK model in powers
         of $ g_{0\eta}^{2k+1}(z)$, by requiring that the moments up to a given
         order are the same. To the eighth order in the moments this gives
         \begin{align}
         G_{\rm SYK}(z)  & =  G^{\rm QH}(z) +(M_6^{\rm SYK}-M_6^{\rm QH}) g_{0\eta}^7(z)
         \nonumber \\
& \quad
         +\left(M_8^{\rm SYK}-M_8^{\rm QH} +\frac{7(M_6^{\rm SYK}-M_6^{\rm QH})}{-1+\eta}  \right) g_{0\eta}^9(z)
         \end{align}
with corresponding corrections $\delta \rho_3$ ($k=3$) and $\delta \rho_4$ ($k=4$) to the spectral density.

               For $N=32$ the sixth order result agrees better than $0.1 \%$
               with the exact result, which is much better than the Q-Hermite
               approximation,
               see figure~\ref{fig:dif}. The eighth order correction only gives a slight improvement.

\section{Analytical calculation of the fluctuations of the expansion coefficients}\label{sec:covariance}

In this section, we calculate the covariance matrix of the expansion coefficients
of the spectral density in terms of
the Q-Hermite density and the Q-Hermite polynomials,
\begin{equation}
\rho(x)= \rho_{QH}(x) \sum_{i=0} c_iH_i^\eta (x).
\end{equation}
We will obtain explicit results for the expectation values $\langle c_k c_l \rangle$ for $ k,l \le 6$. Since the numerics in this paper are done in dimensionless units, and spectral densities are normalized to one, it is convenient to study the
rescaled (by the many-body variance  $\sigma$),  and normalized moments
\begin{equation}
\tilde{M}_{m,n}:=\frac{\vev{ \Tr H^m\Tr H^n}}{2^N \sigma^{m+n}},\qquad \tilde{M}_{n} :=\tilde{M}_{0,n},
\end{equation}
which were already introduced in section~\ref{sec:col}.

The covariances of the stochastic coefficients $c_i$ are related to the
double-trace \linebreak moments~by
\begin{align}
\begin{split}
\tilde{M}_{m,n} &= \sum_{i,j=0}\vev{c_ic_j} \int dxdy x^m y^n \rho_{QH}(x)\rho_{QH}(y)H_i^\eta(x)H_j^\eta(y)\\ &=  \sum_{i,j}f_{mi}f_{nj}\vev{c_ic_j} =\sum_{i,j}f_{mi}\vev{c_ic_j}(f^T)_{jn},
\end{split}
\end{align}
where
\begin{equation}
f_{mi}=\int dx x^m  \rho_{QH}(x)H_i^\eta(x).
\end{equation}
Note that because the $H_i^\eta(x)$ are orthogonal with respect to $\rho_{QH}(x)$ we have $f_{mi} =0$ for
$m< i$, which means $f_{mi}$ forms a triangular matrix of infinite size. The coefficients $c_{mi}$ also vanish when $m+i$ is odd.
The covariance matrix $\vev{c_ic_j}$ is given by
\begin{equation}\label{eqn:doubleTrAndCovar}
\vev{c_ic_j} = \sum_{m,n}(f^{-1})_{im}\tilde{M}_{m,n}((f^T)^{-1})_{nj}.
\end{equation}
Since the inverse of the triangular matrix $f_{mi}$ must also be triangular, we can consistently truncate equation~\eqref{eqn:doubleTrAndCovar} up to some finite values of $i, j, m$ and $n$.
An efficient way to calculate the $f_{mi}$, using that  $H_i^\eta(x)=\sum_k a_{ik}^\eta x^k$, is
\begin{equation}
f_{mi}= \sum_{k=0}^i a_{ik}^\eta \tilde M_{m+k}^\text{QH}(\eta),
\end{equation}
where the moments $\tilde M_{m+k}^{\text{QH}}(\eta)$ can be obtained from Riordan-Touchard formula, introduced earlier as equation~\eqref{eqn:RiordanTouchard}:
\begin{equation}
\tilde M^{\text{QH}}_{2p} = \frac 1{(1-\eta)^p} \sum_{k=-p}^p (-1)^k \eta^{k(k-1)/2} {2p \choose p+k}.
\end{equation}

An important caveat is in order: in the numerics we used an irreducible block of the random Hamiltonians to calculate the two-point correlations, which is the appropriate thing to do for probing RMT universalities. This means we really should be looking at
\begin{equation}\label{eqn:blockEffect}
2^{-N+2}\sigma^{-m-n}\left \langle \Tr \left(\frac{1+\gamma_c}{2}H^m \right)\Tr \left(\frac{1+\gamma_c}{2}H^n\right)\right\rangle
\end{equation}
instead of $\tilde{M}_{m,n}$, where $\gamma_c$ is the chirality Dirac matrix in even dimensions. However, since $\gamma_c$ is a product of $N$ different Dirac matrices, we have $\Tr \left(\gamma_c H^m\right) = 0$ realization by realization if $m < N/q$. For $N=32$ and $q=4$, this means equation~\eqref{eqn:blockEffect} coincide with $\tilde{M}_{m,n}$ for the double-trace moments up to $m=7$ and $n=7$, which means we might as well use $\tilde{M}_{m,n}$ for the calculation of $\vev{c_ic_j} $ up to $i=7$ and $j=7$. In this paper we will not go beyond $\tilde{M}_{6,6}$ due to computational complexity, but we do caution that $\Tr \left(\gamma_c H^m\right) $ must be confronted for higher moments.\footnote{Except in the case of $N\text{ mod } 8=2$ or $6$ (GUE universality class), the two blocks have the same eigenvalues realization by realization~\cite{Garcia-Garcia-ml-2016mno}, implying $\Tr \left(\gamma_c H^m\right) = 0$ for any~$m$.}

\pagebreak

\subsection{General expression for double-trace Wick contractions}
\label{sec:dtrAndChord}

In this section we give a general expression for
double-trace contractions contributing to rescaled double-trace moments $\tilde{M}_{m,n}$.

\begin{figure}
 \centering
\begin{tikzpicture}[scale=0.7]
\draw[fill=black] (9,0) circle (1pt);
\draw[fill=black] (10,0) circle (1pt);
\draw[fill=black] (11,0) circle (1pt);
\draw[fill=black] (12,0) circle (1pt);
\draw[fill=black] (13.5,0) circle (1pt);
\draw[fill=black] (14.5,0) circle (1pt);
\draw[fill=black] (15.5,0) circle (1pt);
\draw[fill=black] (16.5,0) circle (1pt);

\draw[] (9,0)--(9,2)--(12,2)--(12,0);
\draw[] (10,0)--(10,1)--(11,1)--(11,0);
\draw[] (13.5,0)--(13.5,2)--(15.5,2)--(15.5,0);
\draw[] (14.5,0)--(14.5,1)--(16.5,1)--(16.5,0);
\draw (8.7,0)--(12.3,0);
\draw (13.2,0)--(16.8,0);
\node at (12.75, -0.5) {($a$)};
\end{tikzpicture}
\begin{tikzpicture}[scale=0.7]

\draw[fill=black] (0,0) circle (1pt);
\draw[fill=black] (1,0) circle (1pt);
\draw[fill=black] (2,0) circle (1pt);
\draw[fill=black] (3,0) circle (1pt);
\draw[fill=black] (4.5,0) circle (1pt);
\draw[fill=black] (5.5,0) circle (1pt);
\draw[fill=black] (6.5,0) circle (1pt);
\draw[fill=black] (7.5,0) circle (1pt);

\draw[] (0,0) -- (0,2.5) -- (4.5,2.5) -- (4.5,0);
\draw[] (1,0) -- (1,2) -- (7.5,2) -- (7.5,0);
\draw[] (2,0) -- (2,1.5) -- (6.5,1.5) -- (6.5,0);
\draw[] (3,0) -- (3,1) -- (5.5,1) -- (5.5,0);
\draw (-.3,0)--(3.3,0);
\draw (4.2,0)--(7.8,0);

\draw[fill=black] (9,0) circle (1pt);
\draw[fill=black] (10,0) circle (1pt);
\draw[fill=black] (11,0) circle (1pt);
\draw[fill=black] (12,0) circle (1pt);
\draw[fill=black] (13.5,0) circle (1pt);
\draw[fill=black] (14.5,0) circle (1pt);
\draw[fill=black] (15.5,0) circle (1pt);
\draw[fill=black] (16.5,0) circle (1pt);

\draw[] (9,0)--(9,2)--(14.5,2)--(14.5,0);
\draw[] (10,0)--(10,1.5)--(12,1.5)--(12,0);
\draw[] (11,0)--(11,1)--(13.5,1)--(13.5,0);
\draw[] (15.5,0)--(15.5,1)--(16.5,1)--(16.5,0);
\draw (8.7,0)--(12.3,0);
\draw (13.2,0)--(16.8,0);
\node at (3.75,-0.5) {($b$)};
\node at (12.75, -0.5) {($c$)};
\end{tikzpicture}
\caption{Three contractions in $\tilde M_{4,4}$: ($a$) is part of the disconnected piece in $\tilde M_{4,4}$; ($b$) has four cross links; ($c$) has two cross links and two single-trace links. \label{fig:fourchords}}
 \end{figure}
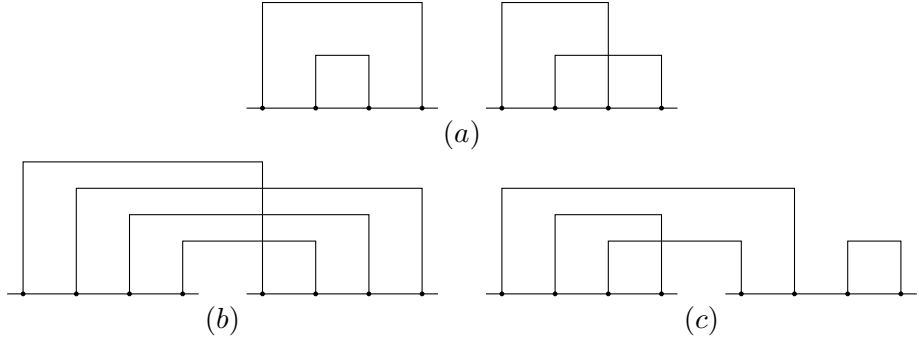

 Since the ensemble average is taken over a Gaussian distribution and thus reduced to a sum over Wick contractions, we can represent the double traces as chord diagrams. Since there are two traces, we will use two different horizontal lines to attach the chords on. Such horizontal lines are called \emph{backbones} in some of the chord diagram literature. For example, in figure~\ref{fig:fourchords} we draw three chord diagrams with two backbones that represent some of the contractions that contribute to $\tilde{M}_{4,4}$. In a self-evident manner they represent (we will not adopt Einstein's summation convention unless otherwise stated)
\begin{equation*}
\begin{split}
 &(a)\quad  2^{-N}\binom{N}{q}^{-4}\sum_{\alpha_1,\alpha_2,\beta_1,\beta_2}\Tr (\Gamma_{\alpha_1}\Gamma_{\alpha_2}\Gamma_{\alpha_2}\Gamma_{\alpha_1})\Tr (\Gamma_{\beta_1}\Gamma_{\beta_2}\Gamma_{\beta_1}\Gamma_{\beta_2})\\
 &(b)\quad  2^{-N}\binom{N}{q}^{-4}\sum_{a_1,a_2,a_3,a_4}\Tr (\Gamma_{a_1}\Gamma_{a_2}\Gamma_{a_3}\Gamma_{a_4})\Tr (\Gamma_{a_1}\Gamma_{a_4}\Gamma_{a_3}\Gamma_{a_2})\\
 &(c)\quad  2^{-N}\binom{N}{q}^{-4}\sum_{a_1,a_2,\beta_1,\beta_2}\Tr (\Gamma_{a_1}\Gamma_{\beta_1}\Gamma_{a_2}\Gamma_{\beta_1})\Tr (\Gamma_{a_2}\Gamma_{a_1}\Gamma_{\beta_2}\Gamma_{\beta_2})
 \end{split}
\end{equation*}
Note ($a$) belongs to the disconnected part of $\tilde{M}_{4,4}$, since all the chords connect only within single traces, and we call such chords \emph{single-trace links}; on the other hand ($b$) and ($c$) belong to the connected part  of $\tilde{M}_{4,4}$ because both contain chords that go from one backbone to the other, and we call such chords \emph{cross links}.  It is important to
keep in mind that notationally we used
\begin{itemize}
\item \emph{Latin-letter subscripts on the cross-linked $\Gamma$'s} {and}
\item \emph{Greek-letter subscripts on the single-trace-linked $\Gamma$'s.}
\end{itemize}
The combinatorics for double-trace moments are much like the single-trace moments
discussed  in~\cite{Garcia-Garcia-ml-2018kvh}, with two additional rules for the $d$ variables which are
the cardinality of the regions in the Venn diagram of overlapping indices. In other words,
$d_{a_{i_1} a_{i_2}\ldots a_{i_k}}$ is the number of elements common and only common to the sets
$a_{i_1}, a_{i_2},\ldots , a_{i_k}$ (naturally, $i_1,\ldots,i_k$ are all different from each other in this definition).  The additional rules are as follows:
\begin{itemize}
\item \emph{The $d$-variables with odd number of Latin-letter subscripts (and an arbitrary number of Greek-letter subscripts) must be set to zero.}
\item \emph{Kronecker deltas are needed to enforce that the total number of indices corresponding to
  one vertex is $q$.}
\end{itemize}
More explicitly, a double-trace chord diagram $G$ has a value of
\begin{equation}\label{eqn:generalDescriptionStarts}
\begin{split}
(-1)^{q E_G}\binom{N}{q}^{-V_G} \sum_{\{d\}}^q&(-1)^{c(G)}\mathcal{M}(\{d\})\Delta(\{d\}),
\end{split}
\end{equation}
where\footnote{The choice of letters $V$ and $E$ comes from ``vertices'' and ``edges'' of the corresponding intersection graphs, see~\cite{Garcia-Garcia-ml-2018kvh} and~\cite{Jia-ml-2018ccl}. In those references the letter $G$ was primarily used to denote intersection graphs, but it should not cause confusion in the present~context.}
\begin{align}
\begin{split}
V_G &= \text{number of chords in } G,\\
E_G &= \text{number of chord intersections in }        G,\\
c(G) &=  \text{sum of $c$-variables whose subscripts denote intersecting chords,} \\
& \quad \text{ see equation~(\ref{c-def}) for a definition},\\
\{d\} &=  \text{set of all $d$-variables with even number of Latin-letter subscripts,}\\
\mathcal{M} &=  \text{multiplicity factor due to partitioning $N$,}\\
\Delta &=  \text{product of Kronecker deltas enforcing the constraints $d_{a_i}=0$.}
\end{split}
\end{align}
More specifically, if there are $L$ cross links and $V_G-L$ single-trace links,
{\rdmathspace[h]
\begin{align}
\{d\} &= \{d_{a_{i_1}\ldots a_{i_r} \beta_{j_1}\ldots\beta_{j_s}}|r \text{ is even}, \{i_1,\ldots,i_r\}\subset \{1,\ldots,L\} ,\{j_1,\ldots,j_s\}\subset \{1,\ldots,V_G-L\} \},\nonumber\\
\mathcal{M}(\{d\})&=\frac{N!}{(N-q V_G+\sum_{k=2}^{V_G} (k-1)d_k)! }\frac{1}{\prod_{\{d\}}d_{\{a\}\{\beta\}}!},\nonumber\\
\Delta(\{d\}) &= \prod_{i=1}^L \delta\left(q- \sum_{\cdots}d_{a_i \ldots}\right),
\end{align}}\relax
in which
\begin{equation}\label{eqn:generalDescriptionEnds}
\begin{split}
d_k &= \text{sum of all $d$-variables with $k$ subscripts,}\\
\prod_{\{d\}}d_{\{a\}\{\beta\}}! &= \text{product of the factorials of all $d$-variables in $\{d\},$}\\
\sum_{\cdots}d_{a_i \ldots} &= \text{sum of all $d$-variables containing $a_i$ as one of the subscripts.}
\end{split}
\end{equation}
As one can see, the most general description of the double trace combinatorics is unfortunately convoluted. We encourage the readers to look into the application of~\eqref{eqn:generalDescriptionStarts}--\eqref{eqn:generalDescriptionEnds} to the examples of $\tilde{M}_{3,3}$ and $\tilde{M}_{3,5}$ illustrated in appendix~\ref{app:mom3}.

\subsection{Some useful properties of double-trace moments}\label{sec:properties}
There are a few other properties of double-trace contractions that will be useful for evaluating double-trace averages:
\begin{enumerate}[label=(\roman{*})]
\item $\tilde{M}_{m,n}= 0 $ if $m+n$ is odd. Hence we only need to concern ourselves with the cases of $m+n$ being even.
\item $\Tr H=0$ even before taking average, so $\tilde{M}_{1,n} = 0 $ for all $n$.
\item $\Tr (\Gamma_\alpha\Gamma_\beta)=\delta_{\alpha\beta}$, this implies $\tilde{M}_{2,2n}-\tilde{M}_{2}\tilde{M}_{2n}= \binom{N}{q}^{-1} 2n \tilde{M}_{2n} $, or more generally
\begin{equation}\label{eqn:2crossFactorization}
  \tilde{M}_{m,n}^{(2)}
  = \binom{N}{q}^{-1} 2mn \tilde M_{2m}\tilde M_{2n},
\end{equation}
where   $\tilde{M}_{m,n}^{(2)}$ denotes the sum over all contractions that contribute to  $\tilde{M}_{m,n}$ with exactly two cross links. This gives a precise meaning to equation~\eqref{eqn:mommn}.
\item For any $N$, there exists a charge conjugation matrix, either $C_+$ or $C_-$ (or both), such that $C^{-1}_\pm\gamma_i C_\pm= \pm \gamma_i^T$, this implies the following reflection identity\footnote{Choosing $C_+$ or $C_-$ won't make a difference: a potential difference only arises when $q$ is odd, in which case we have an extra factor $(-1)^{kq}$, but when $q$ is odd, $k$ must be even for the trace to be~nonzero.}
\begin{equation}\label{eqn:reflectionIden}
\Tr \left( \Gamma_{\alpha_1}\Gamma_{\alpha_2}\ldots\Gamma_{\alpha_k}\right)=(-1)^{kq(q-1)/2}\Tr \left( \Gamma_{\alpha_k}\Gamma_{\alpha_{k-1}}\ldots\Gamma_{\alpha_1}\right),
\end{equation}
where we also used
\begin{equation}
\gamma_{i_q}\gamma_{i_{q-1}}\cdots \gamma_{i_2}\gamma_{i_1} = (-1)^{q(q-1)/2}\gamma_{i_1}\gamma_{i_2} \cdots\gamma_{i_{q-1}}\gamma_{i_q}.
\end{equation}
This reflection, together with the cyclic permutations of the $\Gamma$'s, gives rise to a natural dihedral group action on the traces.\label{item:dihedral}
\item If all but one subscript in the traces are summed, the result is a constant independent of the remaining unsummed subscript, see appendix D of reference~\cite{Garcia-Garcia-ml-2018kvh}. This implies every contraction's value must contain a factor of $\binom{N}{q}$.  This means each term contributing to $2^{-N}\vev{ \Tr H^m\Tr H^n}$ will factorize into $\binom{N}{q}$ times another integer (and times $\sigma^{m+n}$).
  Analogous to the single trace situation, this implies a further factorization property of a special class of double-trace contractions: if a single-trace link intersects with exactly one cross link and nothing else, then this chord diagram factorizes into a double-trace diagram with this single-trace link removed and a single-trace diagram of two chords with one intersection (whose value is $\eta$ of~\eqref{eqn:etaFirstappearance}). Figure~\ref{fig:331factor} illustrates such an example.

\pagebreak

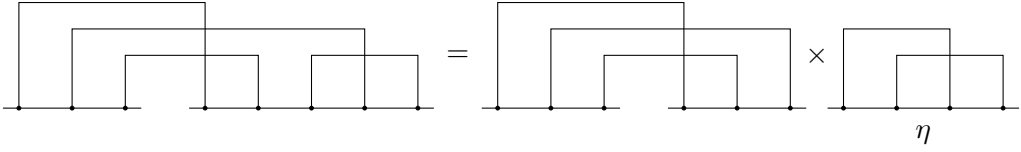
\begin{figure}
\centering
\begin{tikzpicture}[scale=0.7]
\draw[fill=black] (0,0) circle (1pt);
\draw[fill=black] (1,0) circle (1pt);
\draw[fill=black] (2,0) circle (1pt);
\draw[fill=black] (3.5,0) circle (1pt);
\draw[fill=black] (4.5,0) circle (1pt);
\draw[fill=black] (5.5,0) circle (1pt);
\draw[fill=black] (6.5,0) circle (1pt);
\draw[fill=black] (7.5,0) circle (1pt);

\draw[] (0,0)--(0,2)--(3.5,2)--(3.5,0);
\draw[] (1,0)--(1,1.5)--(6.5,1.5)--(6.5,0);
\draw[] (2,0)--(2,1)--(4.5,1)--(4.5,0);
\draw[] (5.5,0)--(5.5,1)--(7.5,1)--(7.5,0);
\draw (-0.3,0)--(2.3,0);
\draw (3.2,0)--(7.8,0);
\node at (8.25,1) {$=$};

\draw[fill=black] (9,0) circle (1pt);
\draw[fill=black] (10,0) circle (1pt);
\draw[fill=black] (11,0) circle (1pt);
\draw[fill=black] (12.5,0) circle (1pt);
\draw[fill=black] (13.5,0) circle (1pt);
\draw[fill=black] (14.5,0) circle (1pt);
\draw[] (9,0)--(9,2)--(12.5,2)--(12.5,0);
\draw[] (10,0)--(10,1.5)--(14.5,1.5)--(14.5,0);
\draw[] (11,0)--(11,1)--(13.5,1)--(13.5,0);
\draw (8.7,0) --(11.3,0);
\draw (12.2,0)--(14.8,0);
\node at (15,1) {$\times$};

\draw[fill=black] (15.5,0) circle (1pt);
\draw[fill=black] (16.5,0) circle (1pt);
\draw[fill=black] (17.5,0) circle (1pt);
\draw[fill=black] (18.5,0) circle (1pt);

\draw[] (15.5,0)--(15.5,1.5)--(17.5,1.5)--(17.5,0);
\draw[] (16.5,0)--(16.5,1)--(18.5,1)--(18.5,0);
\draw (15.2,0)--(18.8,0);
\node at (17,-0.5) {$ \eta$};
\end{tikzpicture}
\caption{The factorization of a chord diagram.\label{fig:331factor}}
\end{figure}

\end{enumerate}

\subsection{Low-order covariances of Q-Hermite expansion coefficients}
In this section we give explicit results for the covariances  up to sixth order
for $N=32$ and $q=4$ which is the case we study numerically later. Apart from the
moments $\tilde{M}_{2,2m}$ given by equation~\eqref{eqn:2crossFactorization}, we also need the following results:
\begin{alignat*}{2}
\label{eqn:doubleTraceValues}
\tilde M_{3,3}&=\frac{1701}{161640200}&\qquad
\tilde M_{3,5}&= \frac{4835943}{72657269900}\\
\tilde M_{4,4}&=\frac{14912736088383}{2906290796000}&\qquad
\tilde M_{4,6}&= \frac{201028717157105439}{13063777128020000}\\
\tilde M_{5,5}&=\frac{8812289619}{20902043404832}&\qquad
\tilde M_{6,6}&= \frac{2710088957667107403387}{58721678190449900000}.
\end{alignat*}
Then using equation~\eqref{eqn:doubleTrAndCovar} up to $i, j, m, n=6$, we obtain for $N=32, q=4$ the following nonzero covariances up to $i,j\le6$:
\begin{alignat*}{3}
  \vev{c_0^2} &= 1, &\quad  \vev{c_0 c_6} &=  -6.4397\times10^{-3},
& \vev{c_0 c_2}&=0 ,\quad \vev{c_0 c_7} =0
  ,\\
\vev{c_2^2} &=  3.47581\times10^{-5},&\quad
\vev{c_2 c_4} &=  1.78799\times10^{-5},&\quad
\vev{c_2 c_6} &=  1.11629\times10^{-5}, \\
\vev{c_3^2} &=  3.68918\times10^{-6},&
\quad
\vev{c_3 c_5} &=  2.74962\times10^{-6}, && \\
\vev{c_4^2} &=  9.59628\times10^{-6},&\quad
\vev{c_4 c_6} &=  6.154203\times10^{-6}, &&
\\
\vev{c_5^2} &=  2.11476\times10^{-6},&\quad
\vev{c_6^2} &=  4.55032\times10^{-5}. &&
\end{alignat*}
Note that $c_0=1$ and $c_1=0$ even before averaging.\footnote{That $c_0=1$ is simply because $\rho_\text{QH}$ is normalized to unity and spectral density is normalized to unity for every realization. That $c_1=0$ is because $\Tr \left((1+\gamma_c) H \right)=0$ for every~realization.}
These numbers agree reasonably well with the numerical results presented in figure~\ref{fig:coeffs}.
General expressions for low-order moments are given in appendix~\ref{sec:lowOrderMoments}.

\section{Numerical analysis of spectral correlations}\label{sec:numberVarAndSpecForm}

In this section we analyze the spectral correlations of the eigenvalues obtained by diagonalization
of the SYK Hamiltonian for $N=32$. In section~\ref{sec:num} we discuss the number variance. The
fluctuations due to ensemble averaging are analyzed in section~\ref{sec:ensemble} and spectral
form factor is evaluated in~\ref{sec:spectralFormFactor}.

\subsection{Number variance}
\label{sec:num}

If $\langle \rho(x)\rangle$ is the average spectral density, then the
average number of levels in an interval of width $\Delta$ is given by
\begin{equation}
\bar n=\int_{x-\Delta/2}^{x+\Delta/2} \langle \rho(y)\rangle dy.
\end{equation}
The actual number of levels in the interval is equal to
\begin{equation}
n=\int_{x-\Delta/2}^{x+\Delta/2}  \rho(y) dy,
\end{equation}
so that the deviation from the average number is given by
\begin{equation}
 \delta n =\int_{x-\Delta/2}^{x+\Delta/2}(  \rho(y)-\langle \rho(y)\rangle) dy.
\end{equation}
The variance of $\delta n$ as a function of $\bar n$ is known as the number variance,
\begin{equation}
\Sigma^2(\bar n) \equiv {\rm var}\{ \delta n \}.
\end{equation}
The average eigenvalue density can be obtained in two ways. First as the
ensemble average of the spectral density for a very large ensemble. Second,
for large matrices, as a fit of a smooth function, $\rho_\text{smo}(x)$,
to the spectral density of a single disorder realization. If the two are the same for $N\to \infty$, we speak of spectral ergodicity~\cite{Pandey-1979}. Even if
$\rho_\text{smo}(x) - \langle \rho(x) \rangle \sim 1/N$, it can give a large contribution to the number variance for intervals where $n$ is no longer much smaller
than $N$.
In particular, this may happen in many-body systems where the number of levels
increases exponentially with the number of particles.

  To calculate the number variance as a function of the average number of levels
  in an interval we need to know $x$ as a function of the number of
  levels with energy less than $x$.  In other words, we have to invert the \textit{mode number function}
  defined as
  \begin{equation}
  N(x) = \int_{-\infty}^x \langle \rho(y)\rangle dy.
\label{cumul}
  \end{equation}
  Since the number of eigenvalues in an interval does not change under
  a coordinate transformation, it is simplest to invert~(\ref{cumul}) in
  coordinates where the level density is constant. If the constant is
  equal to one, the transformation is particularly simple:
  \begin{equation}
  dx' = \langle \rho(x)\rangle  d x.
  \end{equation}
  The spacing of the levels in the new coordinates is thus given by
  \begin{equation}
  x_{k+1}' - x_k' = \langle \rho(x_k)\rangle(x_{k+1} - x_k),
  \end{equation}
  and the new level sequence is obtained by adding the differences,
  \begin{equation}
  x_{k}' = \sum_{p=1}^{k-1}\langle \rho(x_{p-1})\rangle( x_p -x_{p-1}) \approx
  \int_{-\infty}^{x_k} dy\langle  \rho(y)\rangle .
  \end{equation}
  This procedure is usually referred to as unfolding. We emphasize that this procedure is not necessary for the calculation of the number variance, but it makes it
  much more convenient to invert the mode number function $N(x)$.

  \begin{figure}
\centering
  \includegraphics[width=7.3cm]{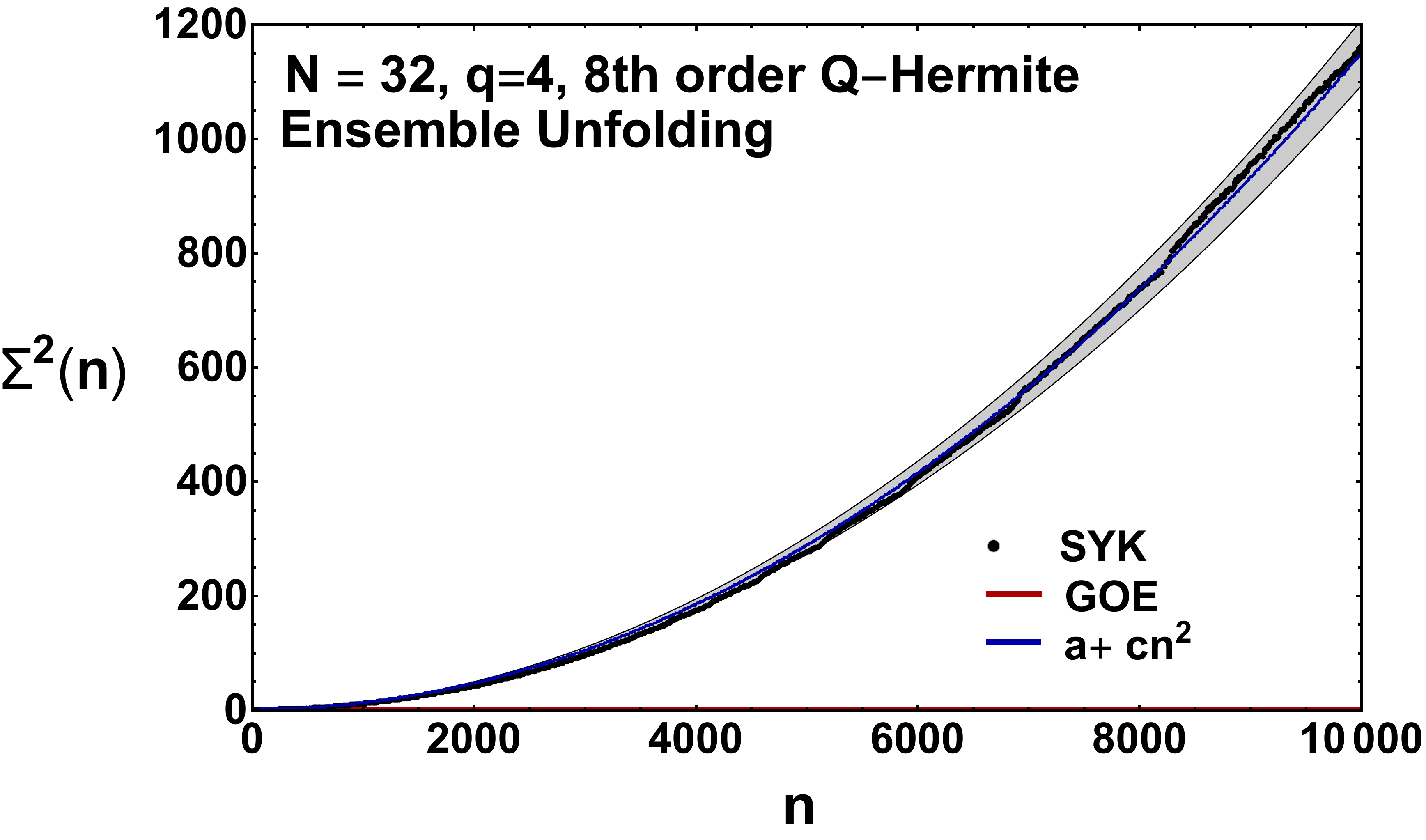}
  \includegraphics[width=7cm]{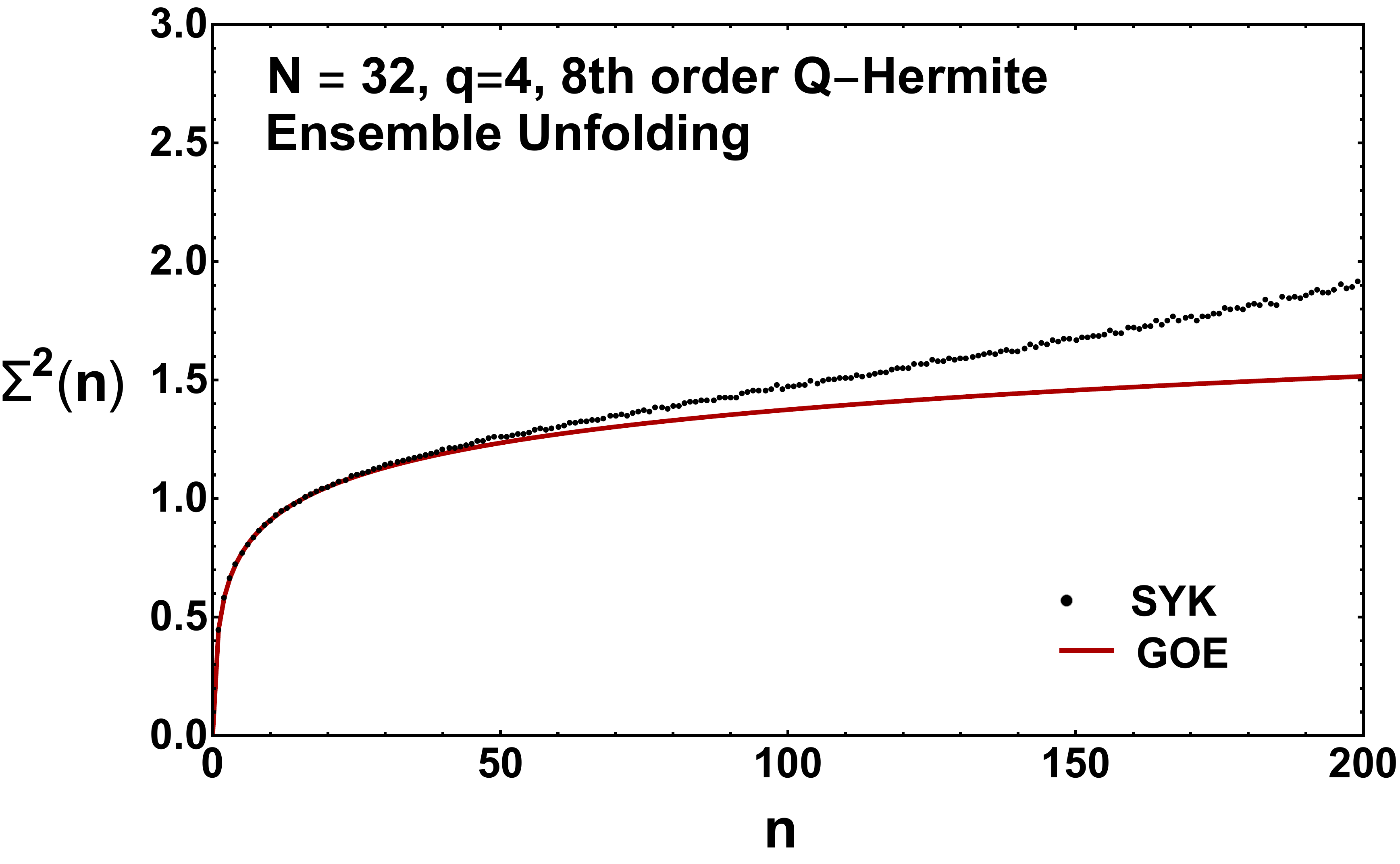}
  \caption{The number variance of the eigenvalues of an ensemble
    of 400 SYK Hamiltonians versus the length of the interval $n$. In the left figure we also show a
    quadratic fit (red curve) and the grey band depicts the error bars. The
    right figure which is a blow-up of the left figure also shows the number
  variance of the GOE (red curve).\label{fig:ensemble}}
  \end{figure}

  As was argued in section~\ref{sec:col}, due to the relative error in the spectral density of ${N\choose q}^{-1/2}$,
  for large $\bar n$ the number variance
  behaves as
  \begin{equation}
  \Sigma^2(\bar n) \sim\frac12  {N \choose q}^{-1}\bar n^2,
  \end{equation}
  which agrees with results obtained previously~\cite{Altland-ml-2017eao,Garcia-Garcia-ml-2018ruf} for $q=4$ and $q=3$. The Thouless energy scale
  can thus be estimated  as
\begin{equation}\label{eqn:thoulessEnergy}
n_{\rm Th} \sim   {N\choose q}^{1/2}.
  \end{equation}

  To calculate the number variance, we unfold the spectrum by
  means of a fit to the
  ensemble average of the spectral density including up to eighth order
  Q-Hermitian polynomials.
  The results are shown
  in figure~\ref{fig:ensemble} where we plot
  $\Sigma^2(\bar n)$ (black points) versus the average number of levels
  $\bar n$ in an interval for an ensemble of 400 SYK Hamiltonians with $N=32$.
  For small $\bar n < 40$ we see excellent agreement with
  the GOE (red curve) (see right figure), but for large distances, the number variance grows
  quadratically, $\Sigma^2(\bar n) \sim (\bar n/293)^2$. The latter result
  is in good agreement with analytical result of $\bar n^2/2{N\choose q}$  (which gives $(\bar n/268)^2$ for $N=32$ and $q=4$).

  \subsection{The effect of ensemble averaging}
\label{sec:ensemble}

  As mentioned before, the ensemble fluctuations contribute to the number variance as
  \begin{equation}
  \Sigma^2(\bar n) = \frac 12 {N \choose q}^{-1}\bar n^2.
  \label{quad}
  \end{equation}
  This contribution can be subtracted by rescaling the eigenvalues of each realization of the ensemble
  according to its width~\cite{French1973,Flores_2001,Altland-ml-2017eao,Garcia-Garcia-ml-2018ruf,Gharibyan-ml-2018jrp}.
  However, this is only the first term in a ``multi-pole'' expansion, and in this section we systematically
  study the long-wavelength fluctuations that give rise to the
  discrepancy  between spectral correlations in
  the SYK model and random matrix theory.

  \begin{figure}
\centering
\includegraphics[width=6cm]{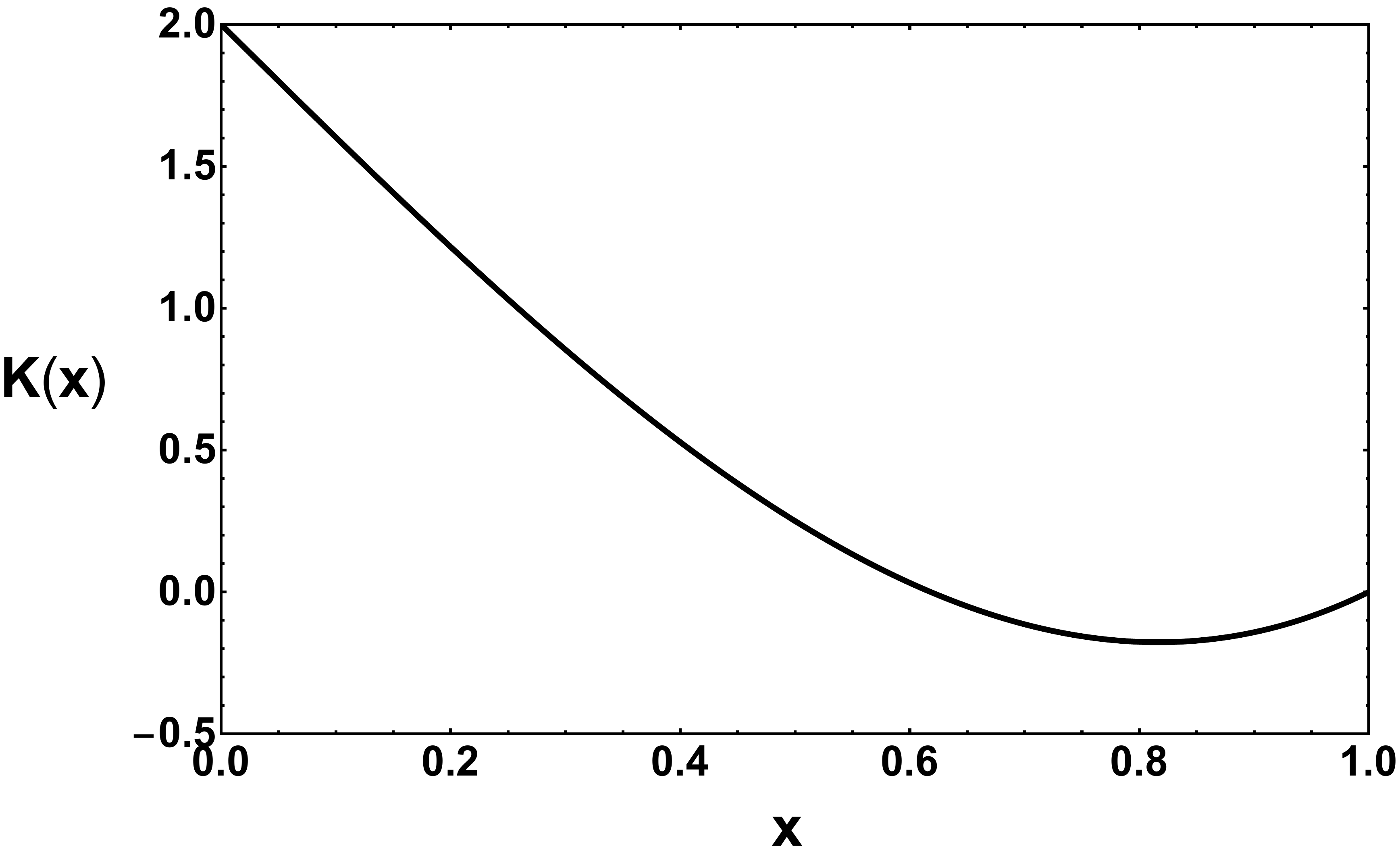}\quad
  \includegraphics[width=6cm]{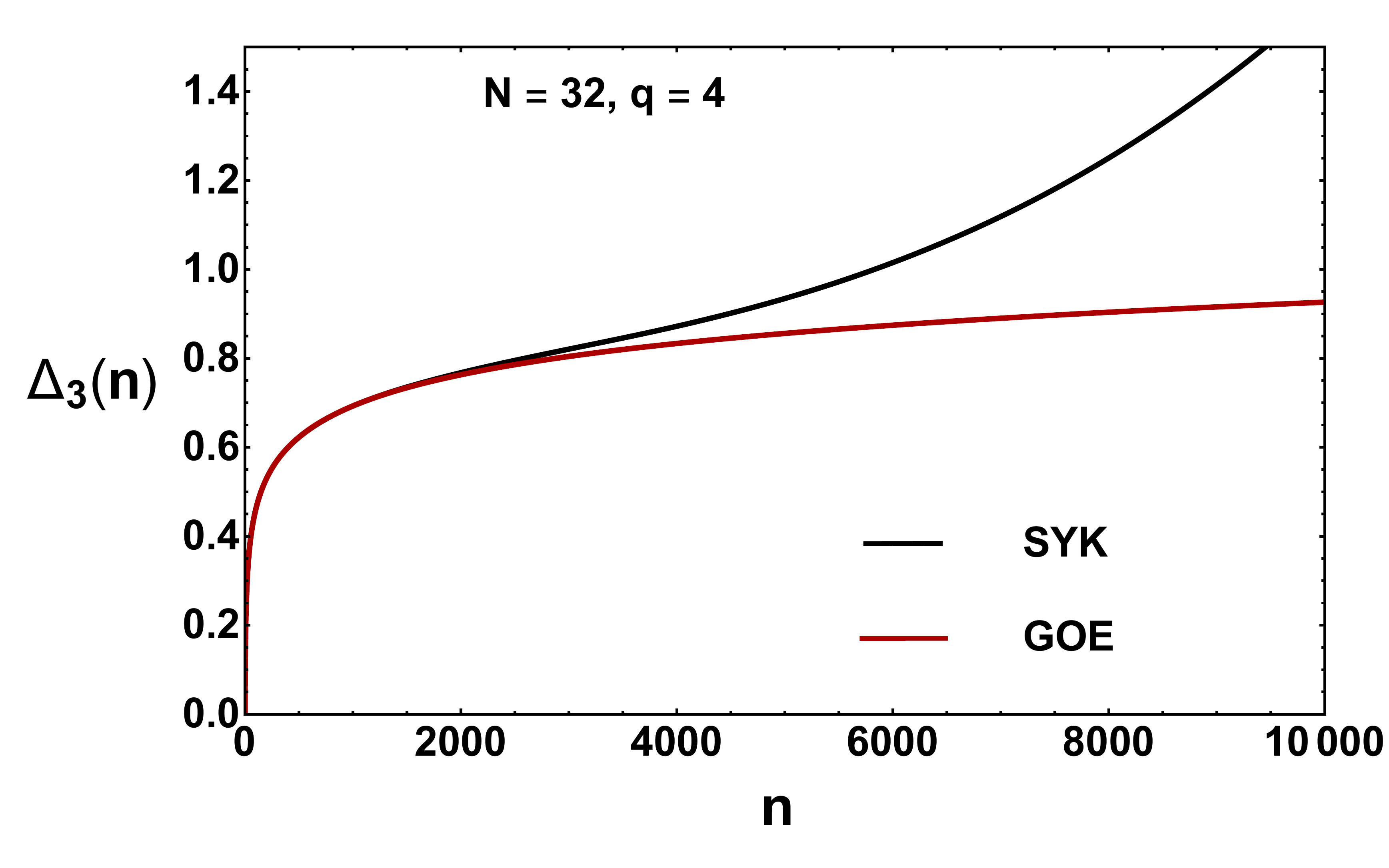}
 \caption{The kernel of the $\Delta_3$ statistic (left) projects out the quadratic part
  of the number variance resulting in agreement with random matrix theory to much
  larger distances (right figure).\label{fig:delta3}}
 \end{figure}

  This quadratic contribution~(\ref{quad})
  due to the scale fluctuations   is projected out
  in the $\Delta_3$ statistic defined by
  \begin{equation}
  \Delta_3(\bar n) = \int_0^1du K(u) \Sigma^2(u\bar n),
  \end{equation}
  with smoothening kernel $K(u) $ given by (see figure~\ref{fig:delta3})
  \begin{equation}
  K(u)=2(1-2u +u^3).
\end{equation}
Since $ \int_{\sqrt 2 -1}^1 K(u)du =0$, in case where the number variance depends
only weakly on $\bar n$, the $\Delta_3(\bar n)$ statistic in essence only measures level
fluctuations up to $(\sqrt 2 -1) \bar n$ and roughly corresponds to the number
variance at $\bar n/4$.
Indeed, as is shown in figure~\ref{fig:delta3}, the $\Delta_3$ statistic agrees with the GOE to  much larger values of $\bar n$ with a Thouless energy that is about
2000 level spacings.

Next we explore the origin of other contributions to the discrepancy between the SYK model and
  the Wigner-Dyson ensembles. We do this by eliminating the ``collective''
  fluctuations of the eigenvalues of each realization of the ensemble
  in which a macroscopic
  number of eigenvalues moves together relative  to the ensemble
  average. This is achieved by calculating the average mode number
  by fitting a smooth function
  to the spectral density of a single configuration. The smoothened spectral density
  can be obtained in a systematic way by expanding it in Q-Hermite polynomials, truncated at $l$-th order:
  \begin{equation}
       \bar \rho_{QH,l}(x) = \rho_{QH}(x)\left[  1 + \sum_{k=1}^l c_k
       H^\eta_{k}(x)\right ].
       \label{rhobar}
    \end{equation}
    The coefficients $c_k$ can be calculated by minimizing the $L_2$ norm excluding
    a small fraction of the eigenvalues in both tails of the spectrum.
    Generically, all even and odd coefficients are nonvanishing with an
    ensemble average that agrees with the fit to the average spectral density obtained in
    equation~(\ref{rhoavexp}).

      \begin{figure}
\centering
      \includegraphics[width=7cm]{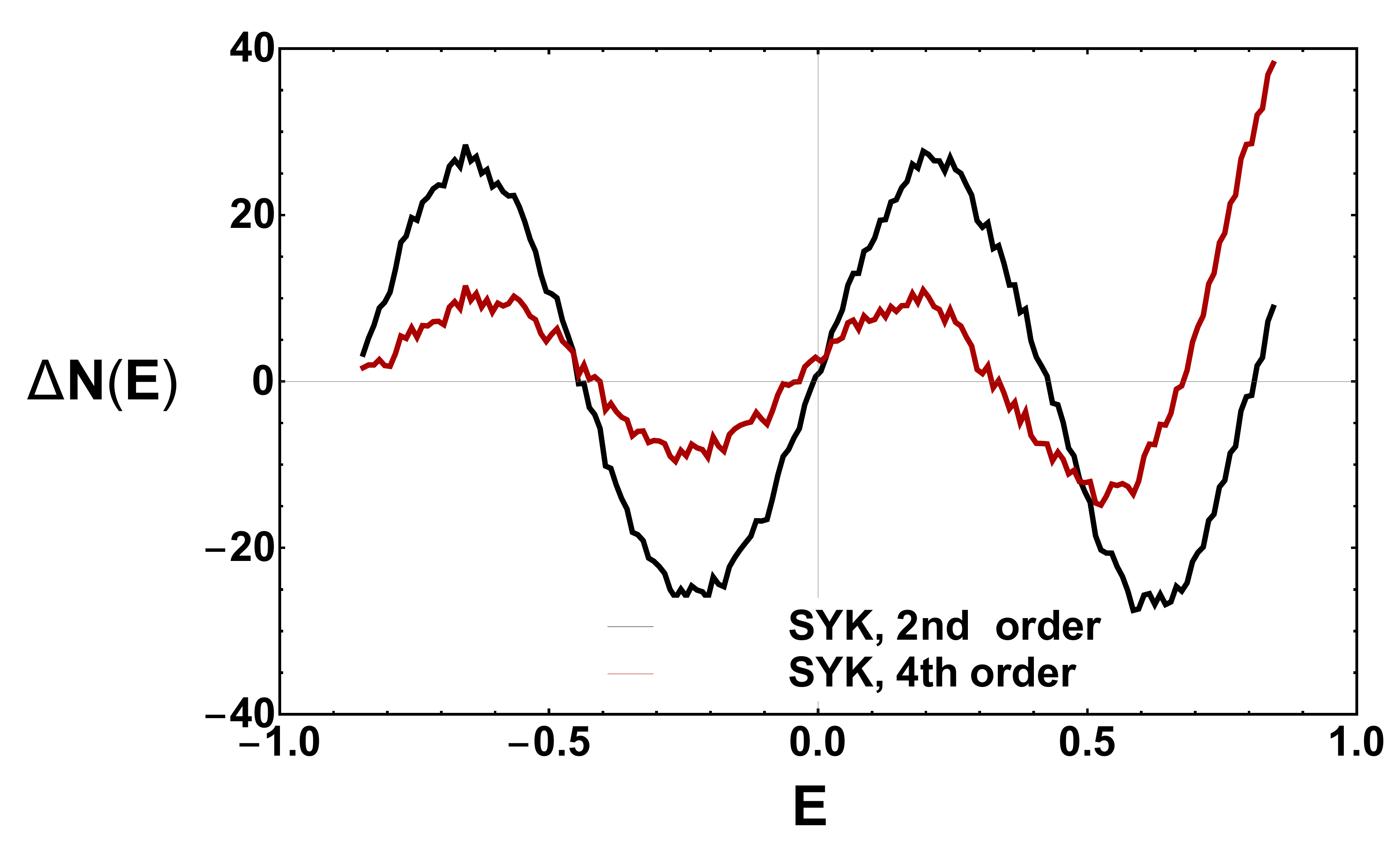}\quad
\includegraphics[width=7cm]{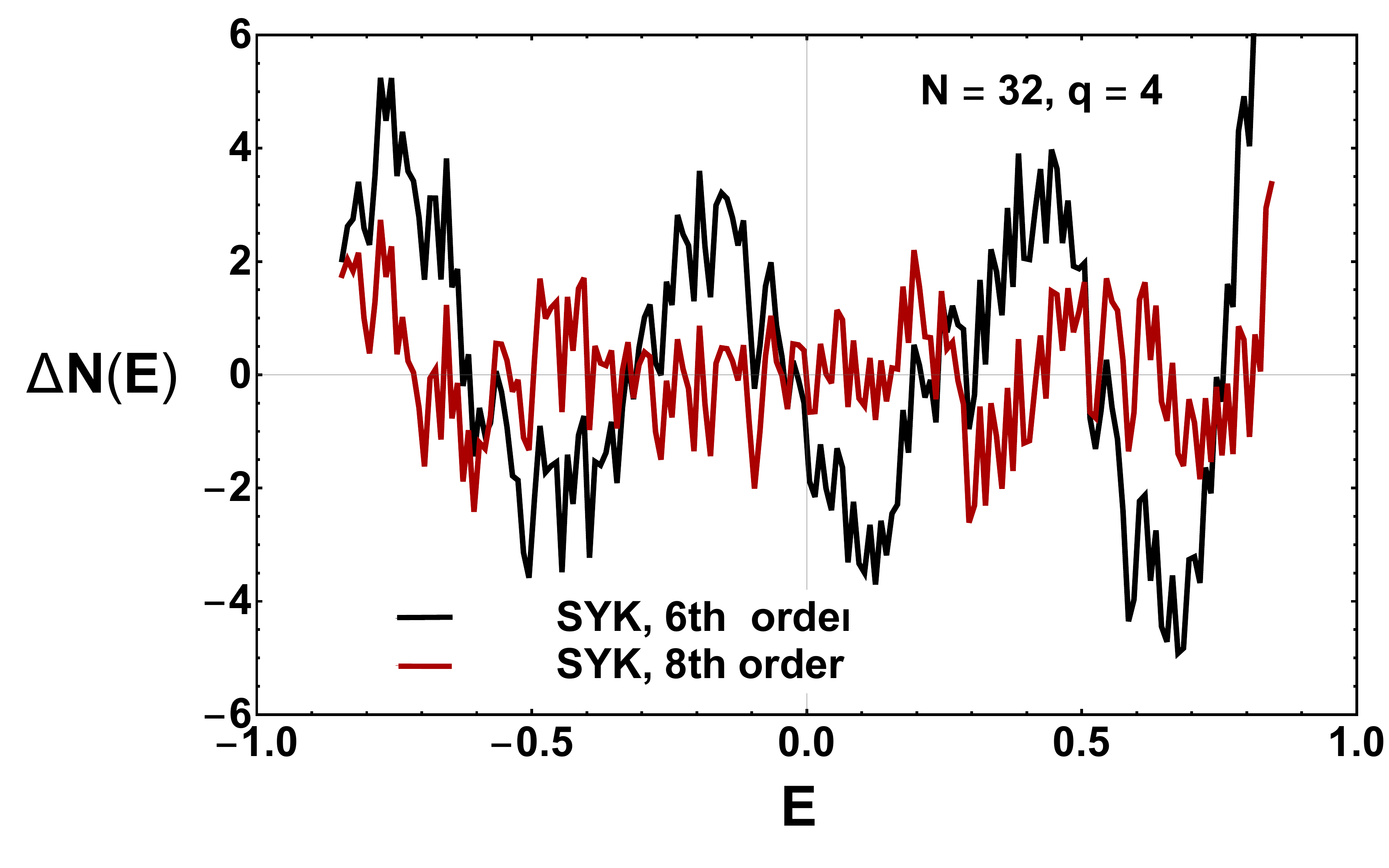}
\caption{The difference between the mode number function of the SYK model and the
  Q-Hermite approximation to order 2 (left, black), 4 (left, red), 6 (right, black) and 8 (right, red). At order 8, the systematic fluctuations are of the same order as the statistical fluctuations.\label{fig:cums}}
    \end{figure}

  \begin{figure}
  \centering

  \includegraphics[width=6.5cm]{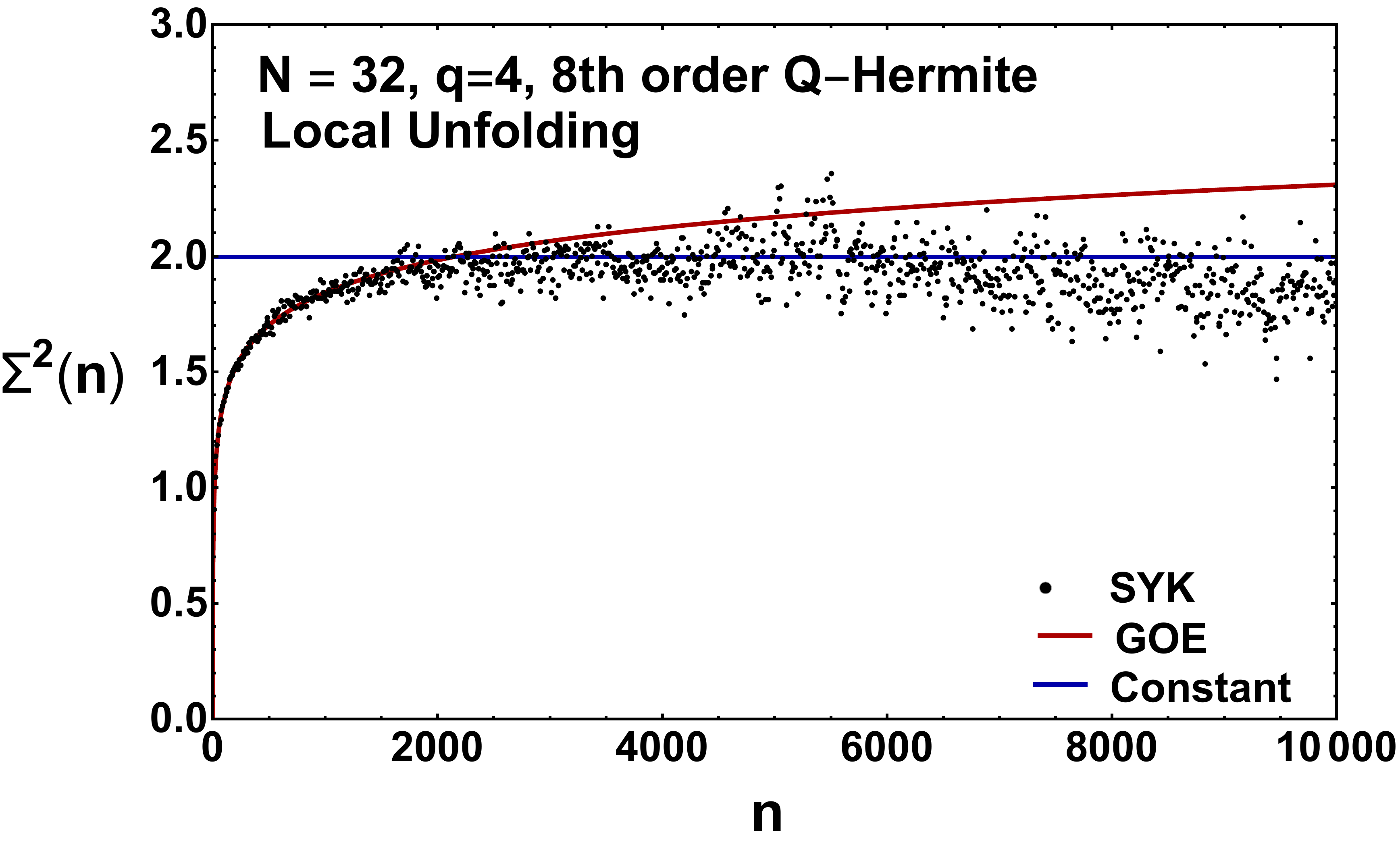}\quad
  \includegraphics[width=6.5cm]{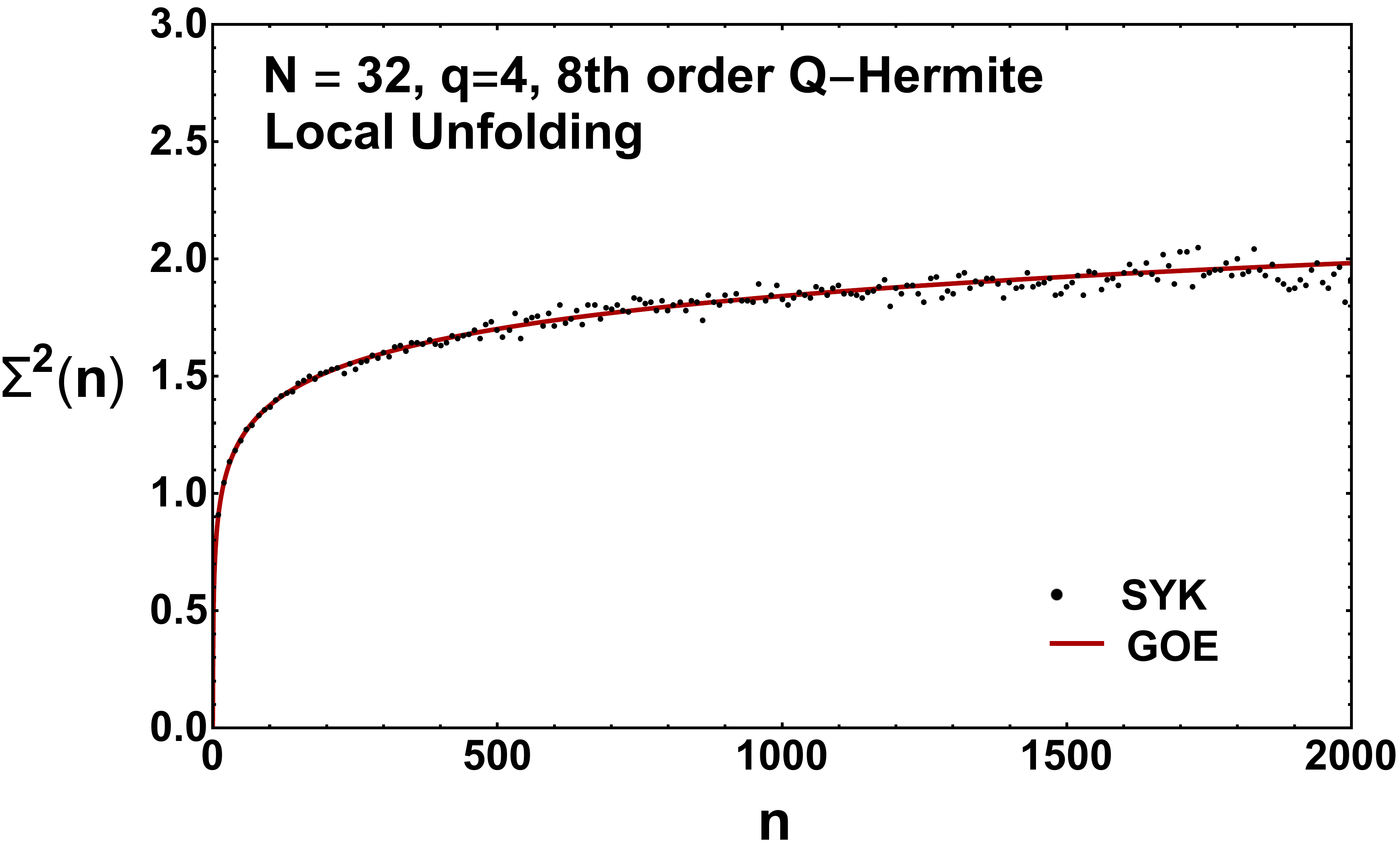}

  \caption{Number variance (black points) versus the average number of
    levels $n$ in the interval. The average number of eigenvalues in
    the interval is calculated from an eighth order Q-Hermite fit to
    the mode number density of each configuration.  In the left figure
    we compare the data to a constant determined by the data for
    $2500<\bar n<4000$. The right figure is a blow-up of the left
    figure for smaller $\bar n$.  In both figures, the GOE result is
    depicted by the red curve.
\label{fig:order8}}
\end{figure}

    In figure~\ref{fig:cums} we show the difference of the cumulative spectral density of the
    SYK model and $\bar \rho_{\rm QH, 2}$, $\bar \rho_{\rm QH, 4}$, $\bar \rho_{\rm QH, 6}$
    and $\bar \rho_{\rm QH, 8}$. It is clear that for $l=8$ the systematic dependence
    is of the same order as the statistical fluctuations, and no further gains can be made
    by including higher values of $l$.

    We now calculate the number variance using $\bar \rho_{QH,l}$ to determine
    the number of eigenvalues in the interval for each configuration. We use
    both ensemble and spectral averaging to reduce the statistical fluctuations. For the
    latter we choose half-overlapping intervals.
    In figure~\ref{fig:order8} we show
    results for $l=8$. The Thouless energy is about
    2000 level spacings,  but for large
    $\bar n$ the number variance saturates to a constant and decreases beyond
    $\bar n\approx 6000$. This has two
reasons. First, by unfolding configuration by configuration, we have eliminated fluctuations with wavelength
larger than about $2^{N/2}/16 =2048$. Second, because the total number of eigenvalues is fixed, the variance
is suppressed when $\bar n$ becomes of the same order as $2^{N/2}/2$.

    \begin{figure}
\centering
  \includegraphics[width=6cm]{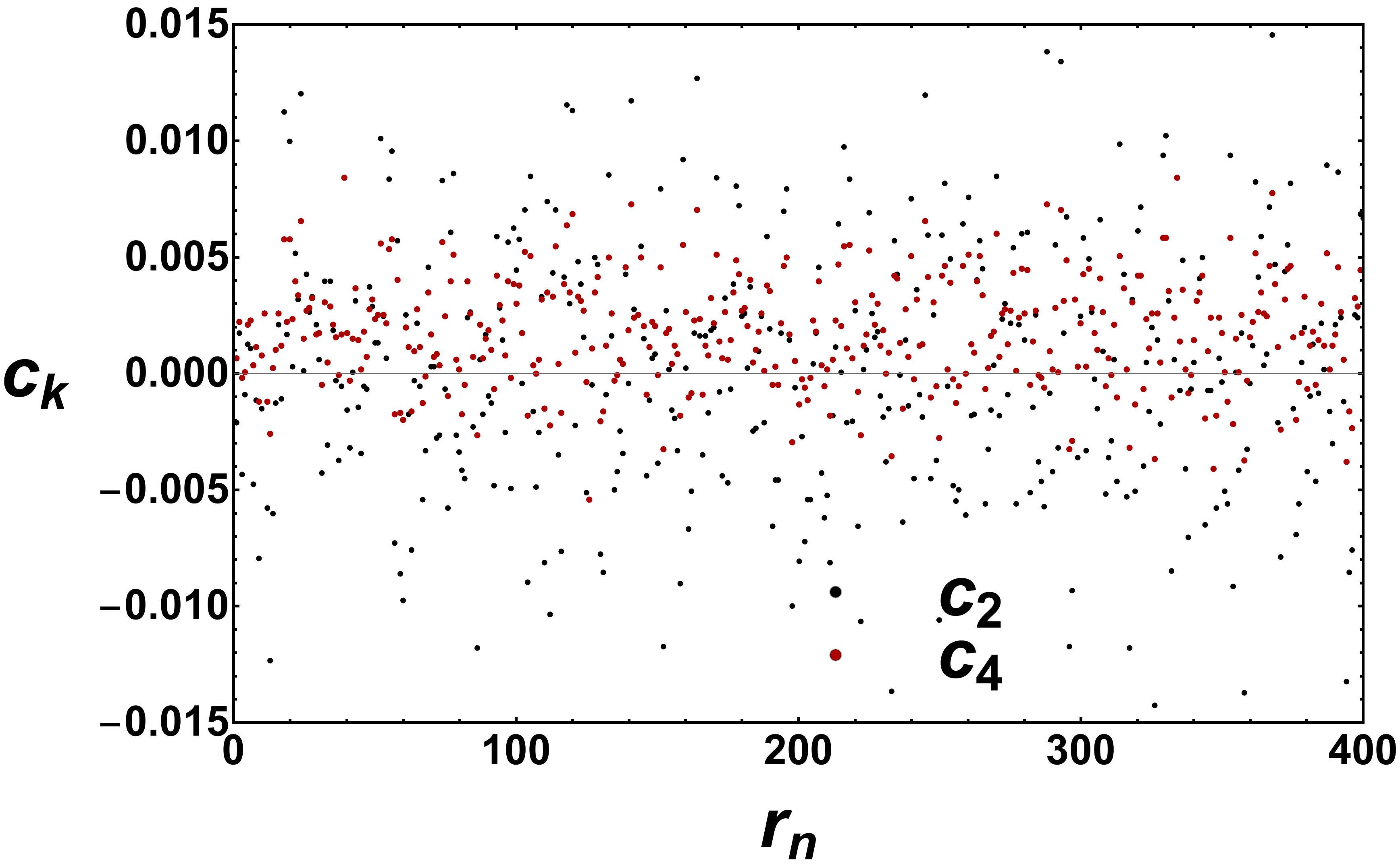}\quad
  \includegraphics[width=6cm]{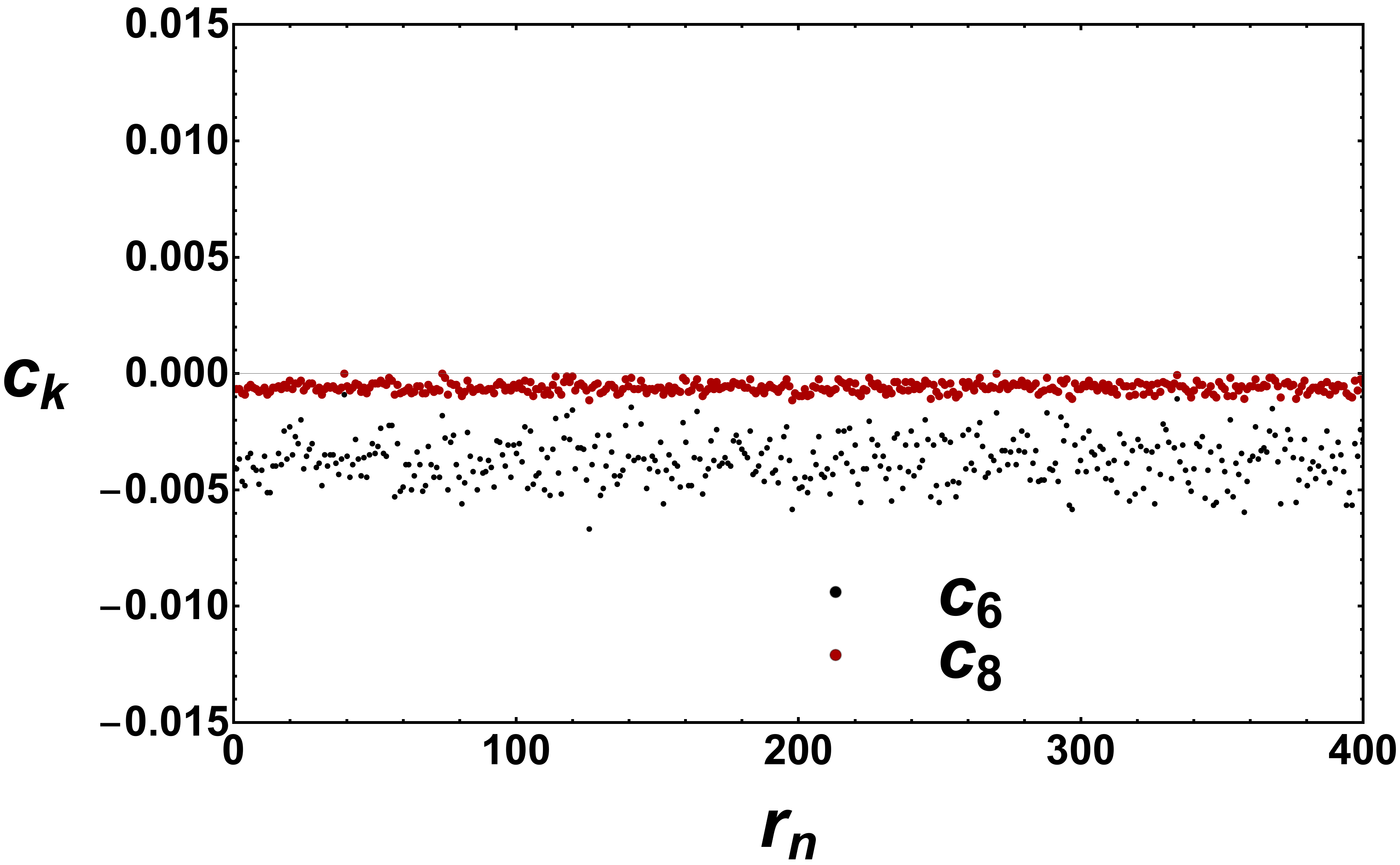}\\[5pt]
\includegraphics[width=6cm]{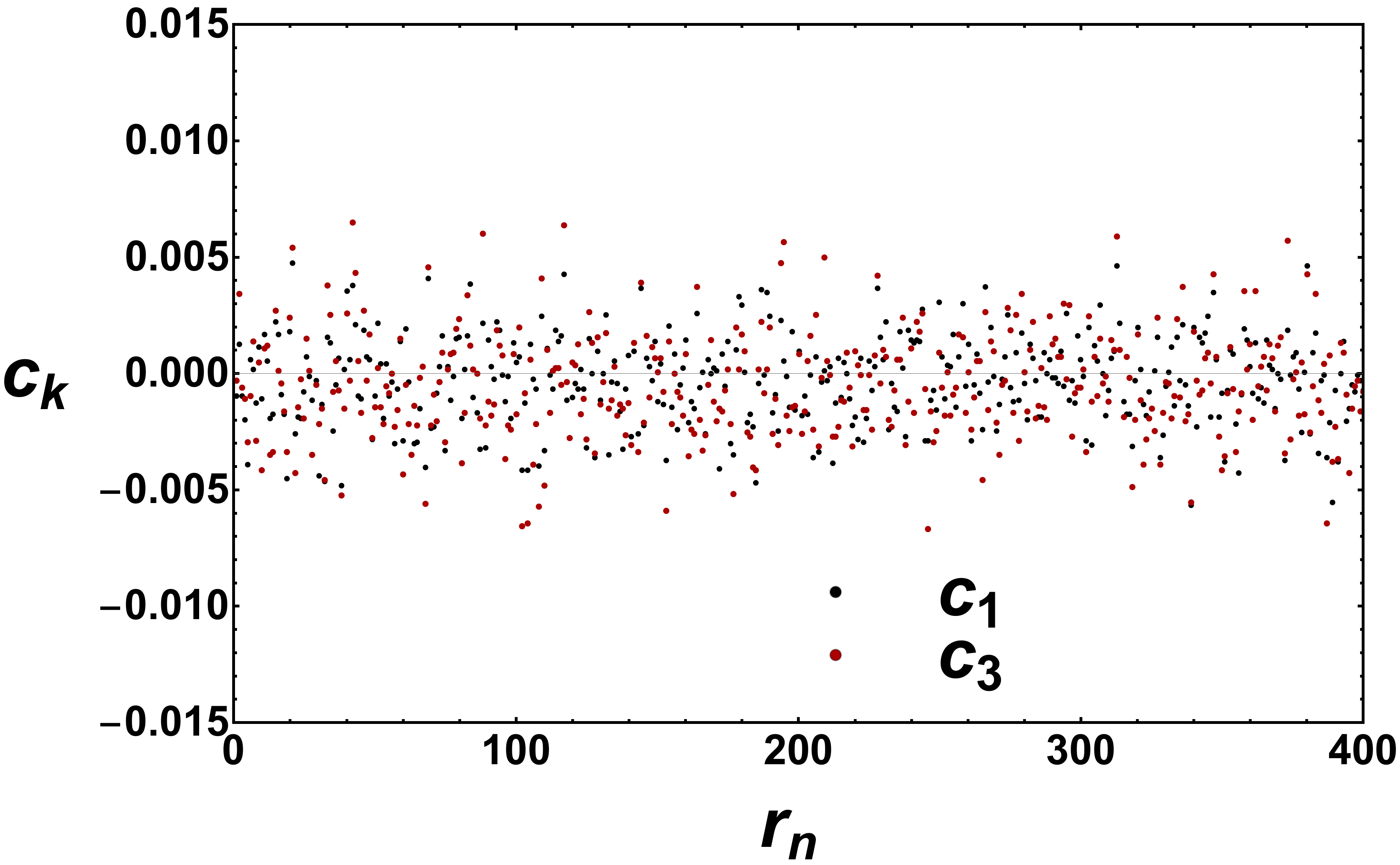}\quad
  \includegraphics[width=6cm]{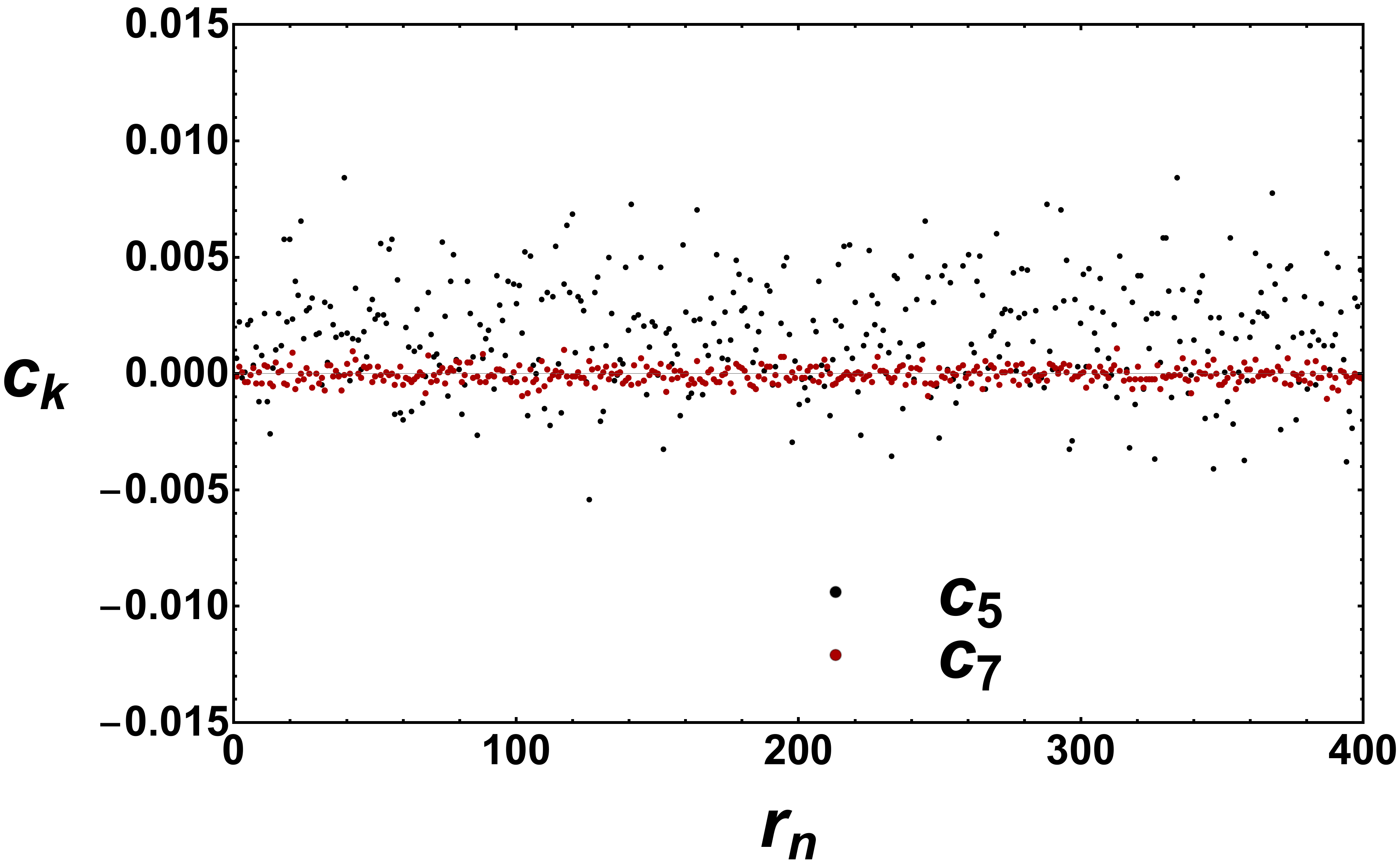}
  \caption{The expansion coefficients $c_k$ of the mode number $N(x)$ in the first eight
    Q-Hermite polynomials versus the realization number $r_n$ for an ensemble
    of 400 realizations.\label{fig:coeffs}}
    \end{figure}

  The coefficients of the expansion of the mode number function $N(x)$ in Q-Hermite polynomials
  as a function of the ensemble realization number are shown in figure~\ref{fig:coeffs}.
In agreement with our analytical results,
  the first four coefficients and $c_5$ and $c_7$ fluctuate about zero, while
  $c_6$ and $c_8$ fluctuate about a nonzero value.

\begin{figure}[t]
  \centering
  \includegraphics[width=6cm]{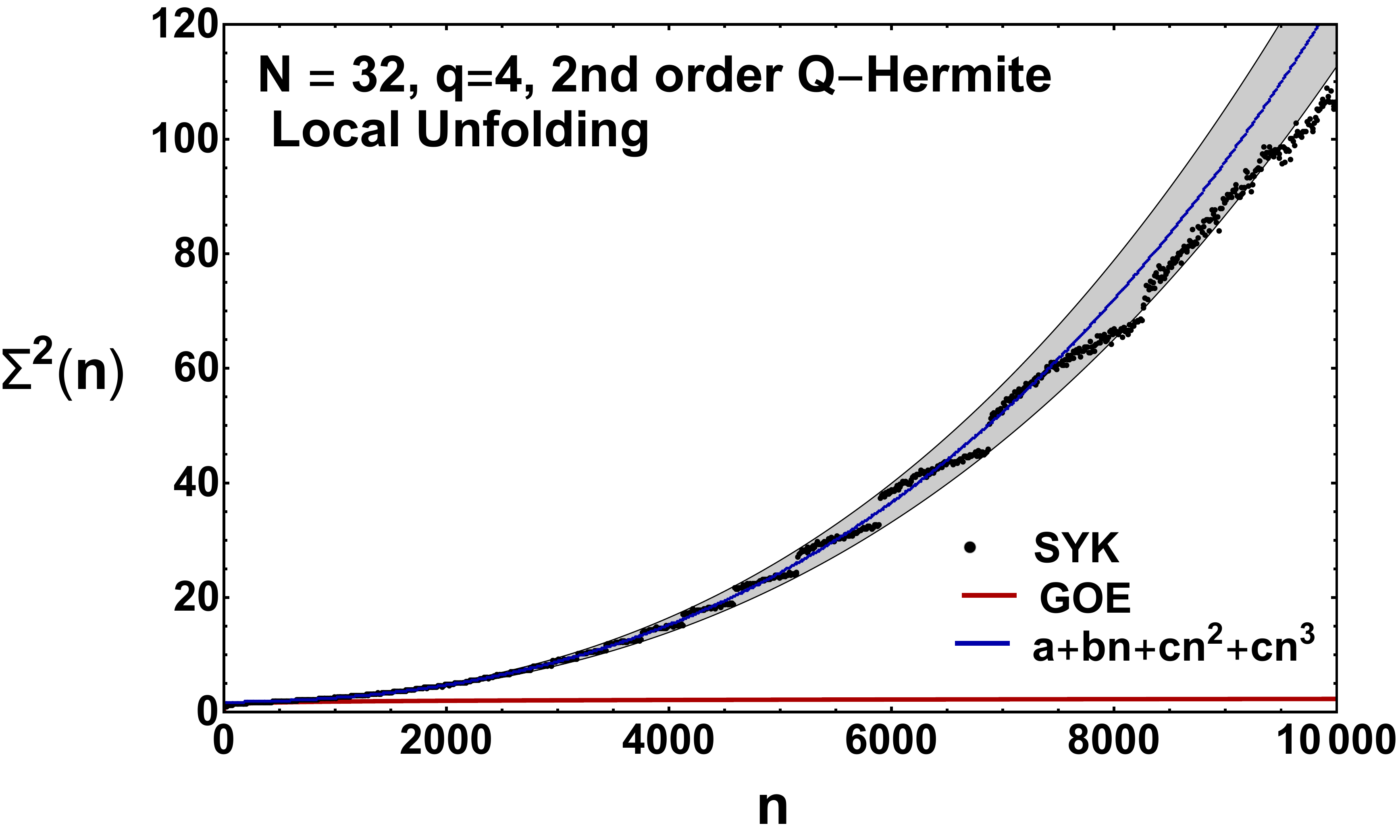}\quad
  \includegraphics[width=6cm]{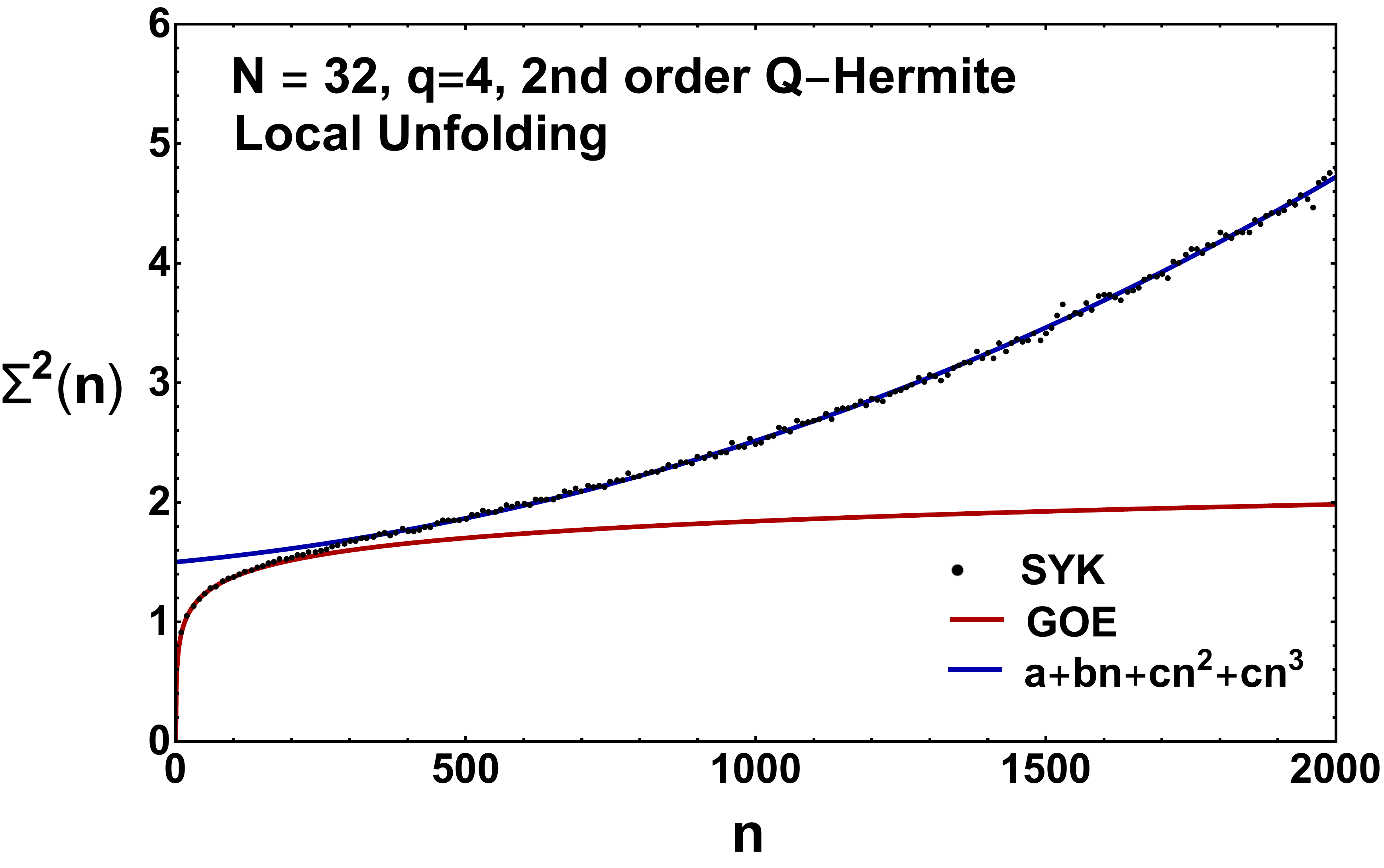}\\[5pt]
  \includegraphics[width=6cm]{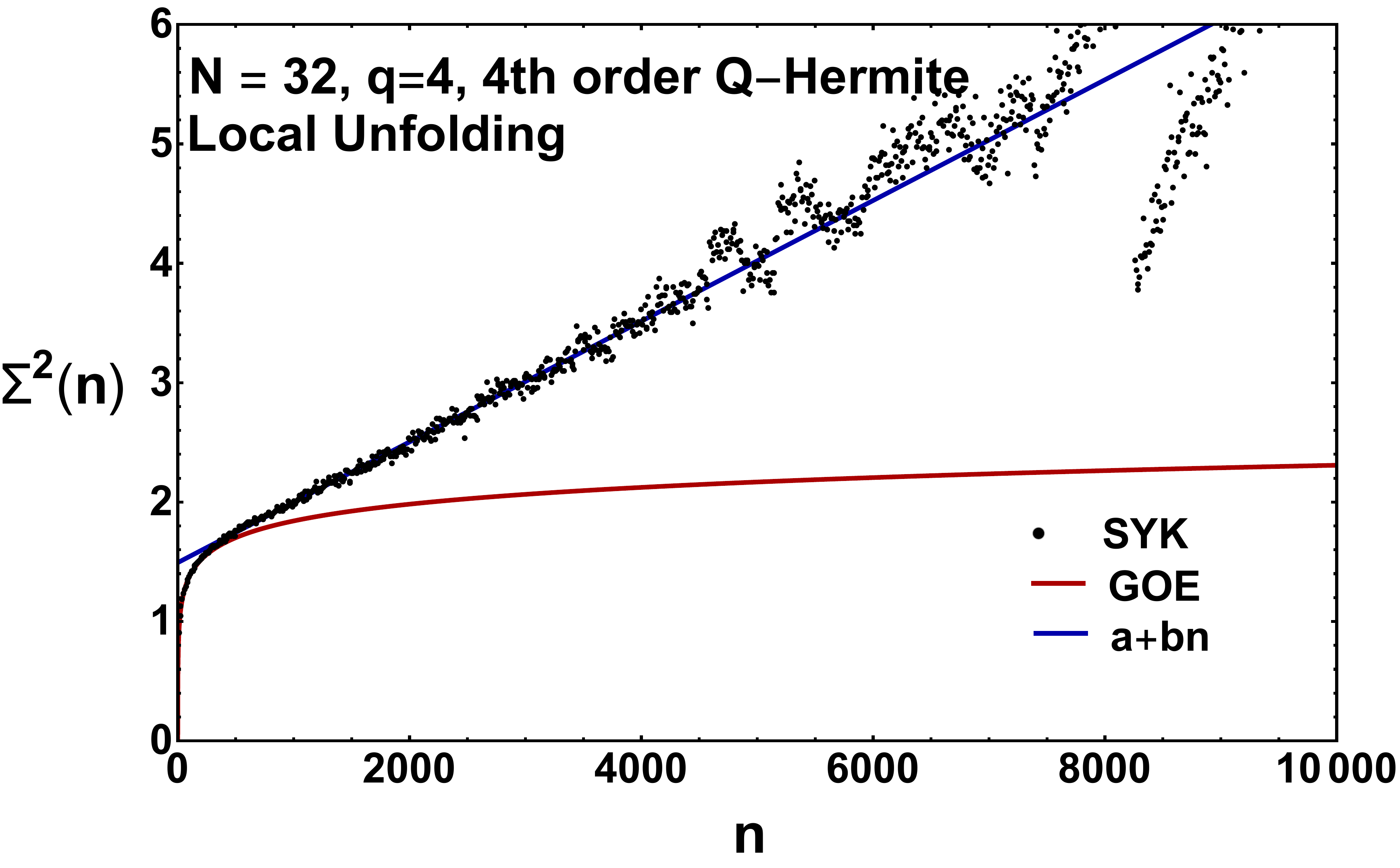}\quad
  \includegraphics[width=6cm]{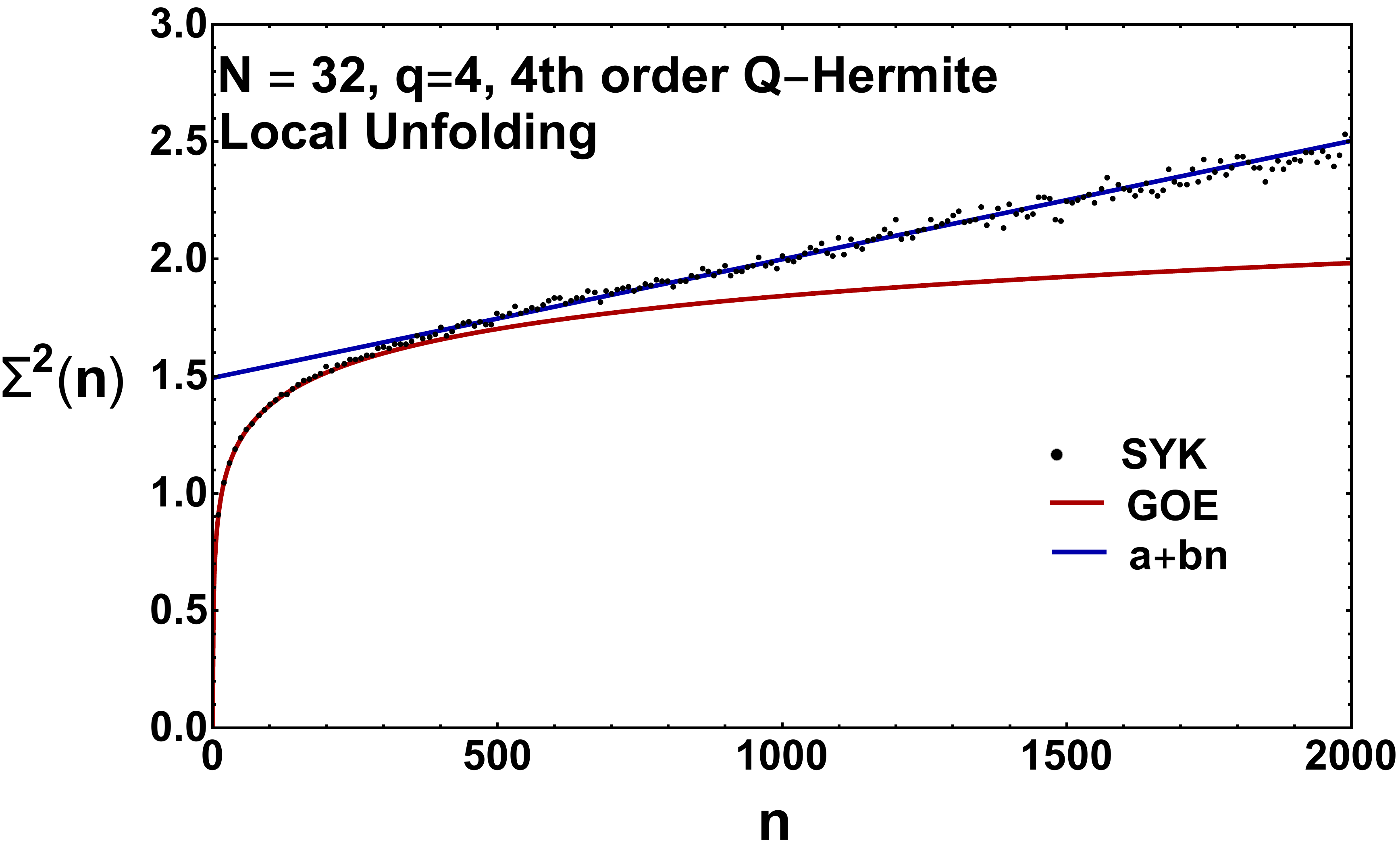}\\[5pt]
  \includegraphics[width=6cm]{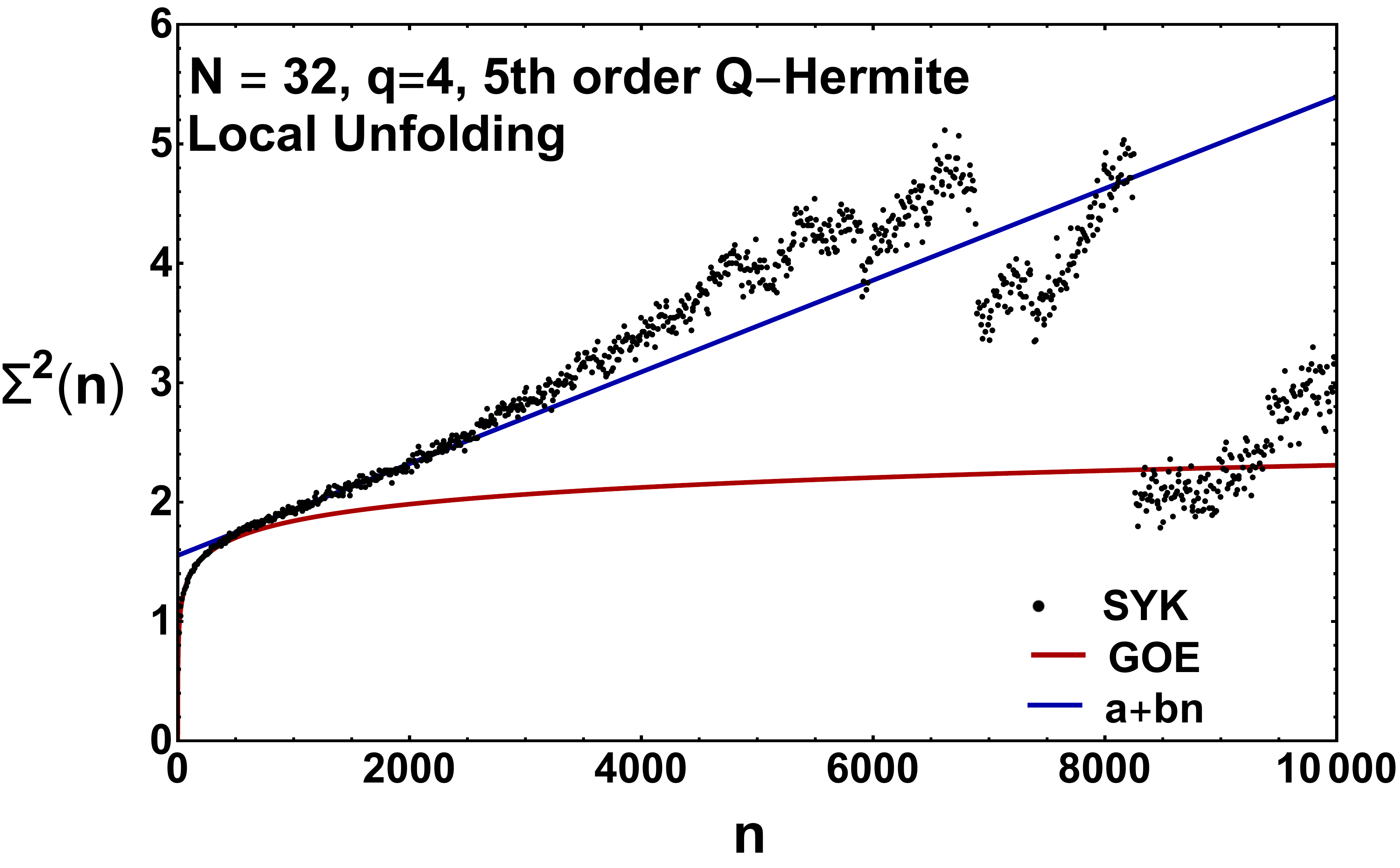}\quad
  \includegraphics[width=6cm]{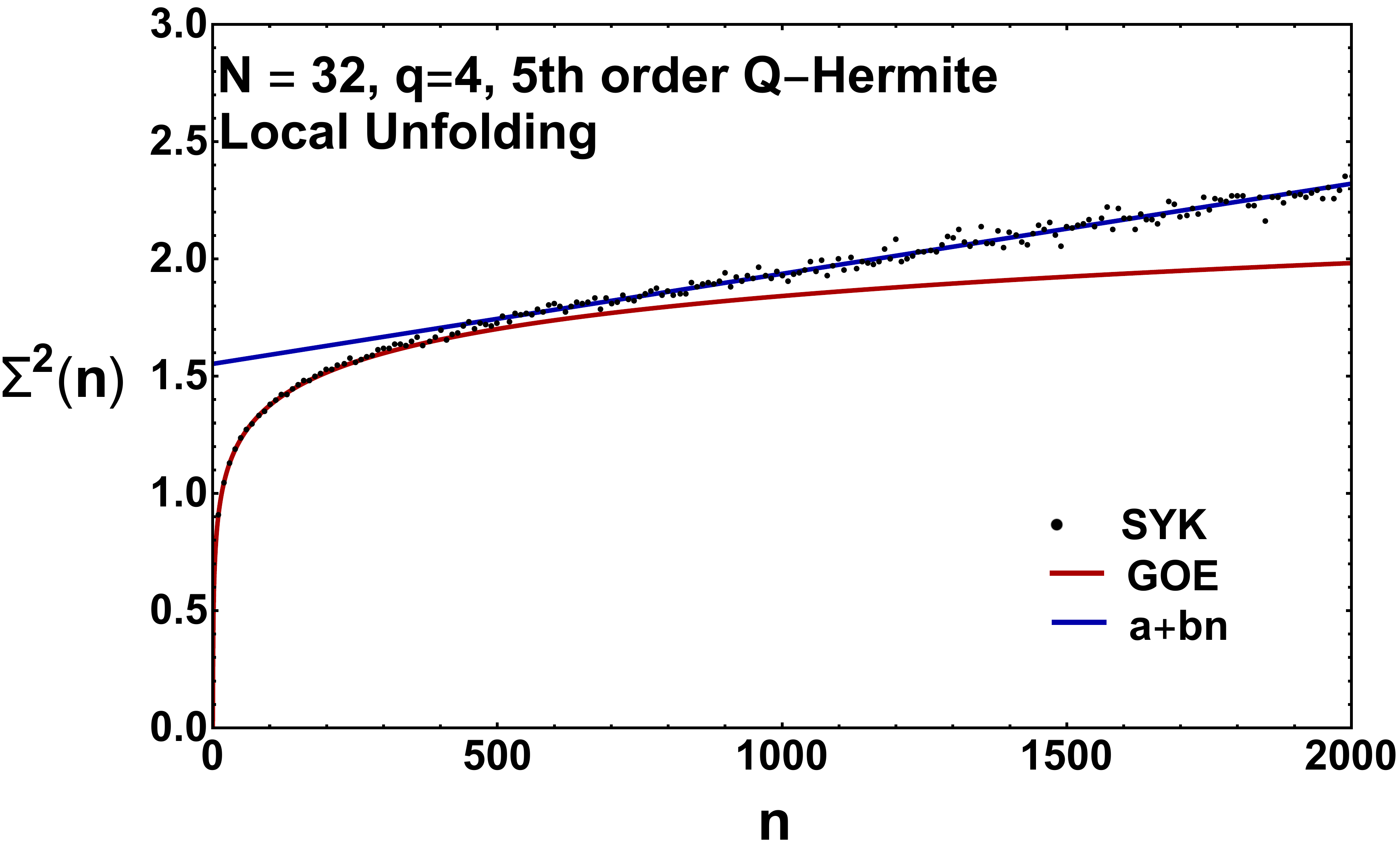}\\[5pt]
 \includegraphics[width=6cm]{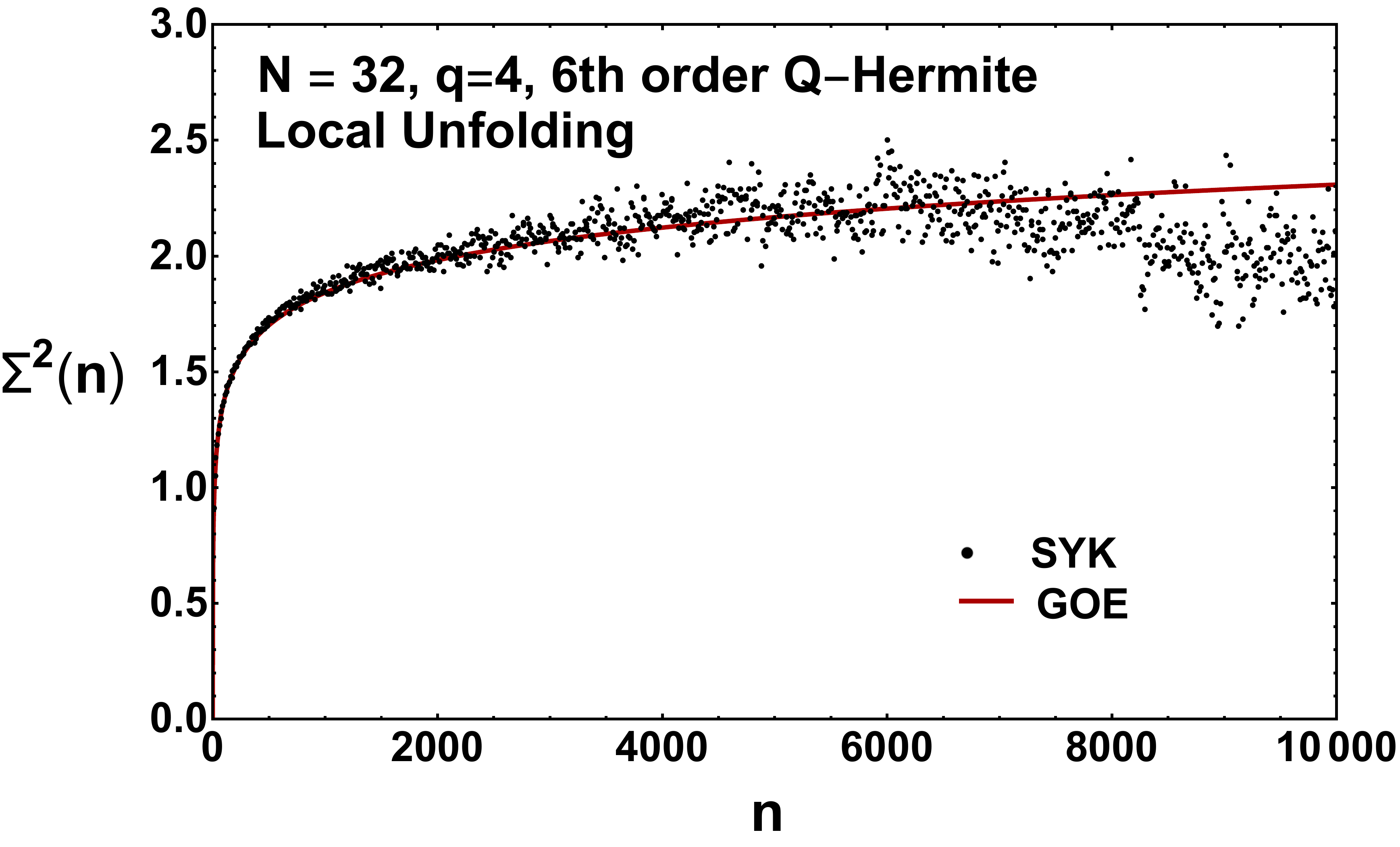}\quad
  \includegraphics[width=6cm]{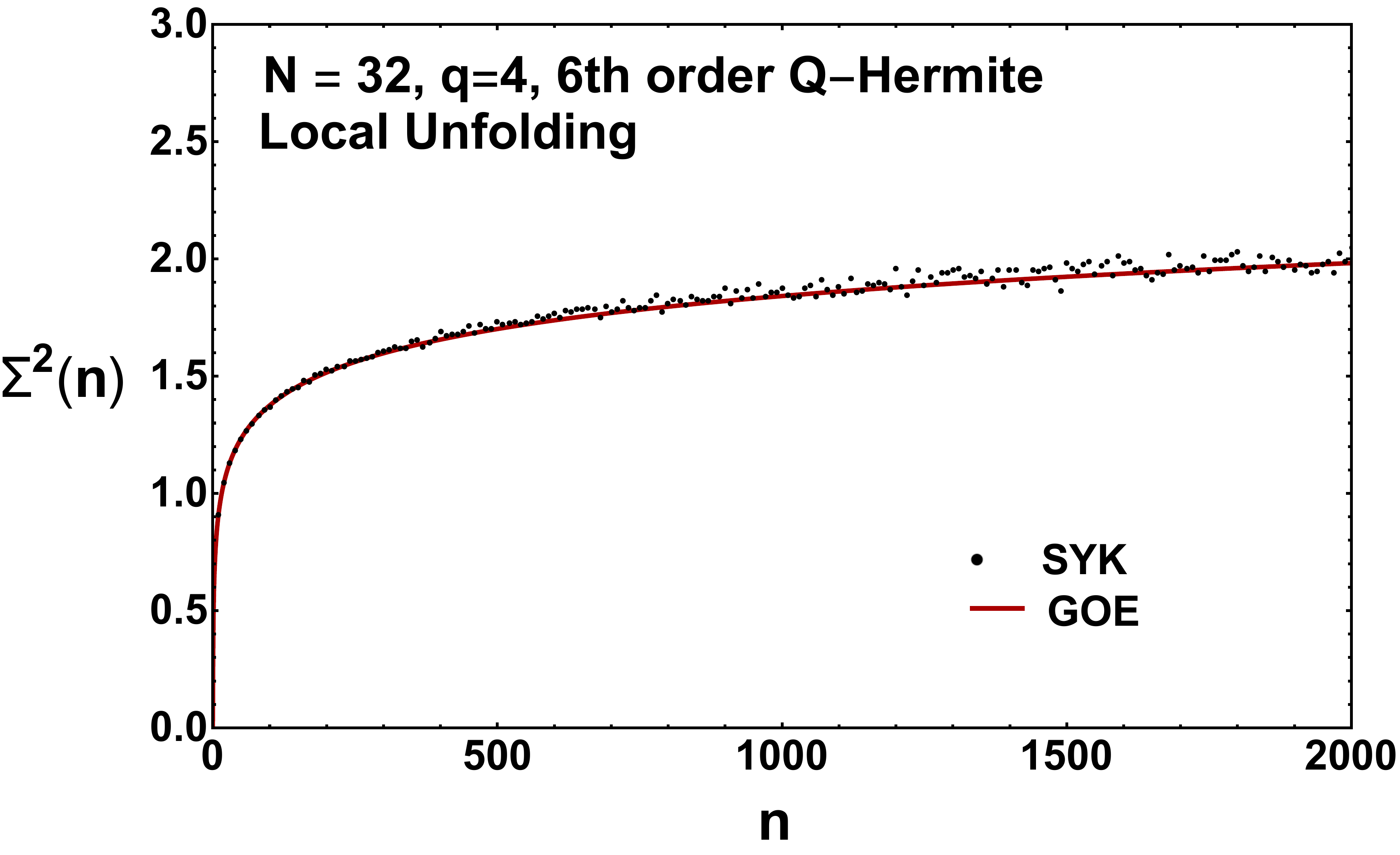}
  \caption{Number variance (black points) versus the average number of levels $n$ in the interval. In the upper row $c_1$ and $c_2$ are fitted to the mode number
    for each configuration, in the second row the same for $c_1, c_2, c_3, c_4$,
    and in the third (fourth) row the same for $c_1, c_2, c_3, c_4, c_5, (c_6)$. In the first
    row the SYK data are fitted to a quadratic $\bar n$-dependence, and in the next
    two rows to a linear one (blue curve). The GOE result is depicted by the
    red curve. A blow-up of the left figures for smaller $n$ values is shown in
    the right column.\label{fig:nv2-6}}
   \end{figure}

Next we study the dependence of the deviation
of the number variance from the GOE result on the number of
Q-Hermite polynomials that
 have been taken into account.
  In figure~\ref{fig:nv2-6}  we show the number variance when an increasing
  number of coefficients has been fitted to the spectral density of
  each configuration. In the upper row only $c_2$ has been fitted while $c_3$ until $c_8$ have been put to the their ensemble-averaged values.
  \relax
     In the second row, $c_2$, $c_3$ and $c_4$ have been fitted and $c_5$, $c_6$, $c_7$  and $c_8$ have been put equal to their ensemble-averaged values.  In the third (fourth) row, the first five
  (six)  coefficients have been obtained by fitting while the remaining ones up
  to order eight have been put equal to their ensemble-averaged values.
   Because the bulk of the deviation of the
    spectral density from the Q-Hermite approximation
    is given by a sixth order polynomial, the number variance
    is sensitive to whether the length of the interval is commensurate
    with the distance of the zeros of the polynomial. That is why we observe
    large jumps (see figure~\ref{fig:nv2-6} lower left) when the number of intervals
    used to calculate the number variance changes. When the length of the
    interval becomes smaller than about half the distance between the zeros of the sixth
    order Q-Hermite polynomial, this
    effect is no longer visible. The right column which shows a blow-up of the
    figures in the left column for small $\bar n$ illustrates that the Thouless
    energy increases gradually with the number of Q-Hermite polynomials that
    have been taken into account in fitting the local spectral density, until
    it saturates at a value of $\bar n$ corresponding to the shortest wavelength
    used for unfolding.

    For second-order local unfolding, the number variance for large $\bar n$ is
    still quite accurately given by a quadratic $\bar n$-dependence, but with a much
    smaller coefficient than in the case of ensemble averaging, For fourth
    order and beyond, the quadratic contribution is negligible and the
    number variance beyond the Thouless energy is very well fitted by a linear $\bar n$-dependence.

    \subsection{Spectral form factor}\label{sec:spectralFormFactor}

    Alternatively spectral correlations can be studied by means of the
    connected spectral form factor defined as
    \begin{equation}
    K_c(t) =\int dxdy e^{it(x-y)}\rho_{2c}(x,y)= \int dxdy e^{it(x-y)} (\langle \rho(x) \rho(y)\rangle
    -  \langle \rho(x)\rangle\langle \rho(y)\rangle).
\label{kc}
    \end{equation}
    The second term is the disconnected part.
    One could also study the spectral
    form factor without this subtraction~\cite{Cotler2016}, but in this paper we only consider
    the connected spectral form factor. Since spectral correlations are
    universal on the scale of the average level spacing, we have
    \begin{equation}
    \rho_{2c}(x,y)= \langle \rho(x)\rangle \langle \rho(y)\rangle \rho_{\rm 2, unv}\left((x-y)\langle \rho(x)\rangle\right),
    \end{equation}
    where $\rho_{\rm 2,unv}$ is the universal random matrix result. Therefore,
    the spectral form factor becomes universal in terms of the variable
    $t/\langle \rho(x)\rangle$. Its large $N$ limit is thus given by a double scaling limit
    which receives contributions  from all orders in $1/N$ of moments.

    We calculate the spectral form factor for the eigenvalues of the
    Hamiltonian of the SYK model. In order to have a well-behaved sum in
    equation~(\ref{kc}) we need to perform some degree of smoothening which we
    will do by introducing a Gaussian cutoff of width  $w$ with centroid
    $\bar E =0$~\cite{Saad-ml-2018bqo},
    \begin{equation}
    K_c(t) = 2 \sqrt \pi w \int dxdy e^{it(x-y)}
    \frac{e^{-(x-\bar E)^2/2w^2}}{\sqrt{2\pi} w} \frac{e^{-(y-\bar E)^2/2w^2}}{{\sqrt{2\pi} w}}
    \rho_{2c}(x,y).
\label{kcsm}
\end{equation}
We will evaluate the spectral form factor for the unfolded spectrum
with $\langle \rho_{\rm unf}(x) \rangle =1$ and have  chosen the
prefactor  such that for
large $t$, when we can use the diagonal approximation, it is equal to unity.

\begin{figure}
\centering
  \includegraphics[width=7.0cm]{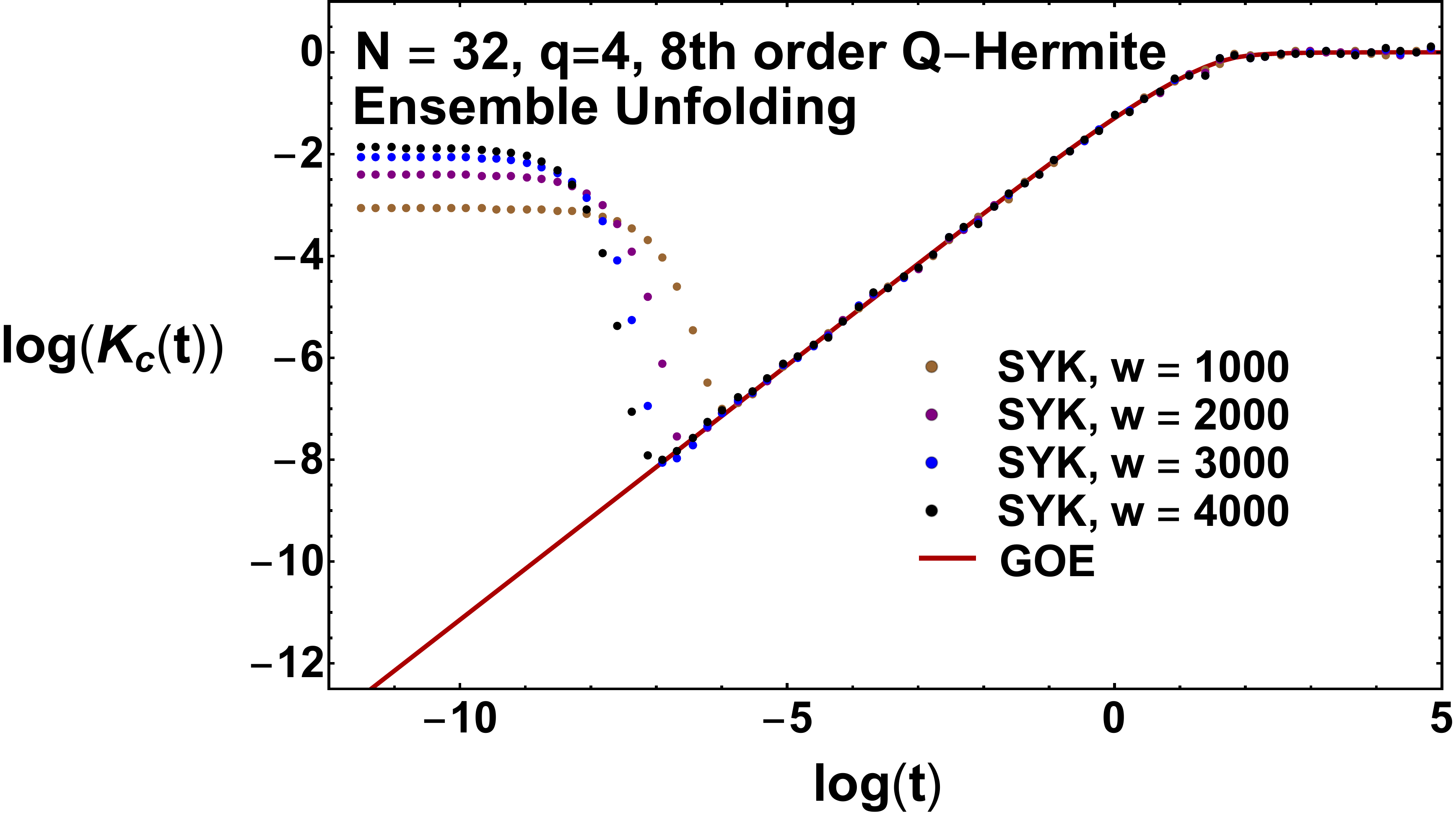}\quad
  \includegraphics[width=7.0cm]{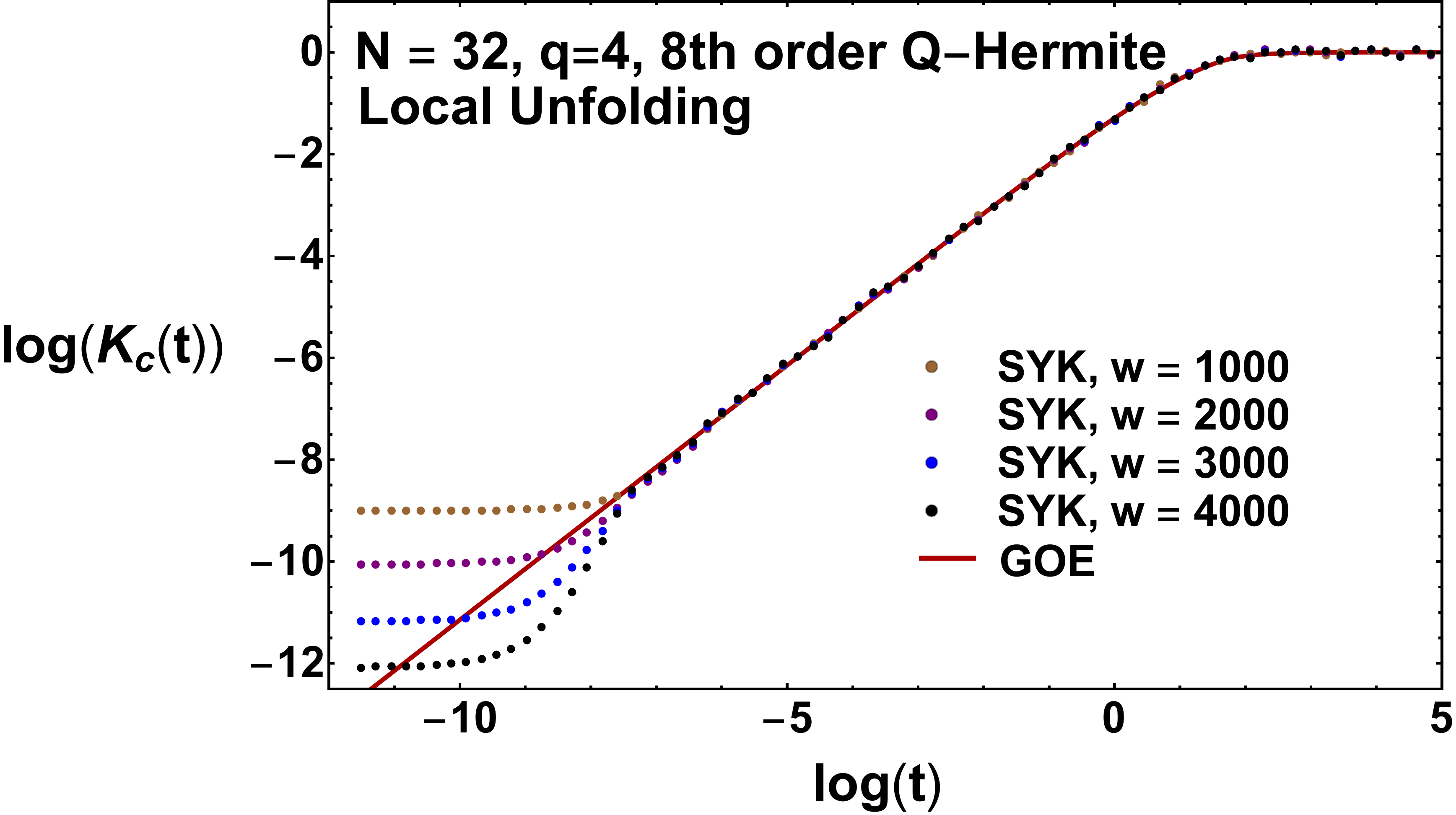}
  \caption{The connected spectral form factor $K_c(t)$ versus $t$ using the ensemble
    averaging for unfolding (left) and unfolding configuration by configuration including
  Q-Hermite polynomials up to eighth order (right).\label{fig:specform0}}
  \end{figure}

\begin{figure}
\centering
  \includegraphics[width=7.0cm]{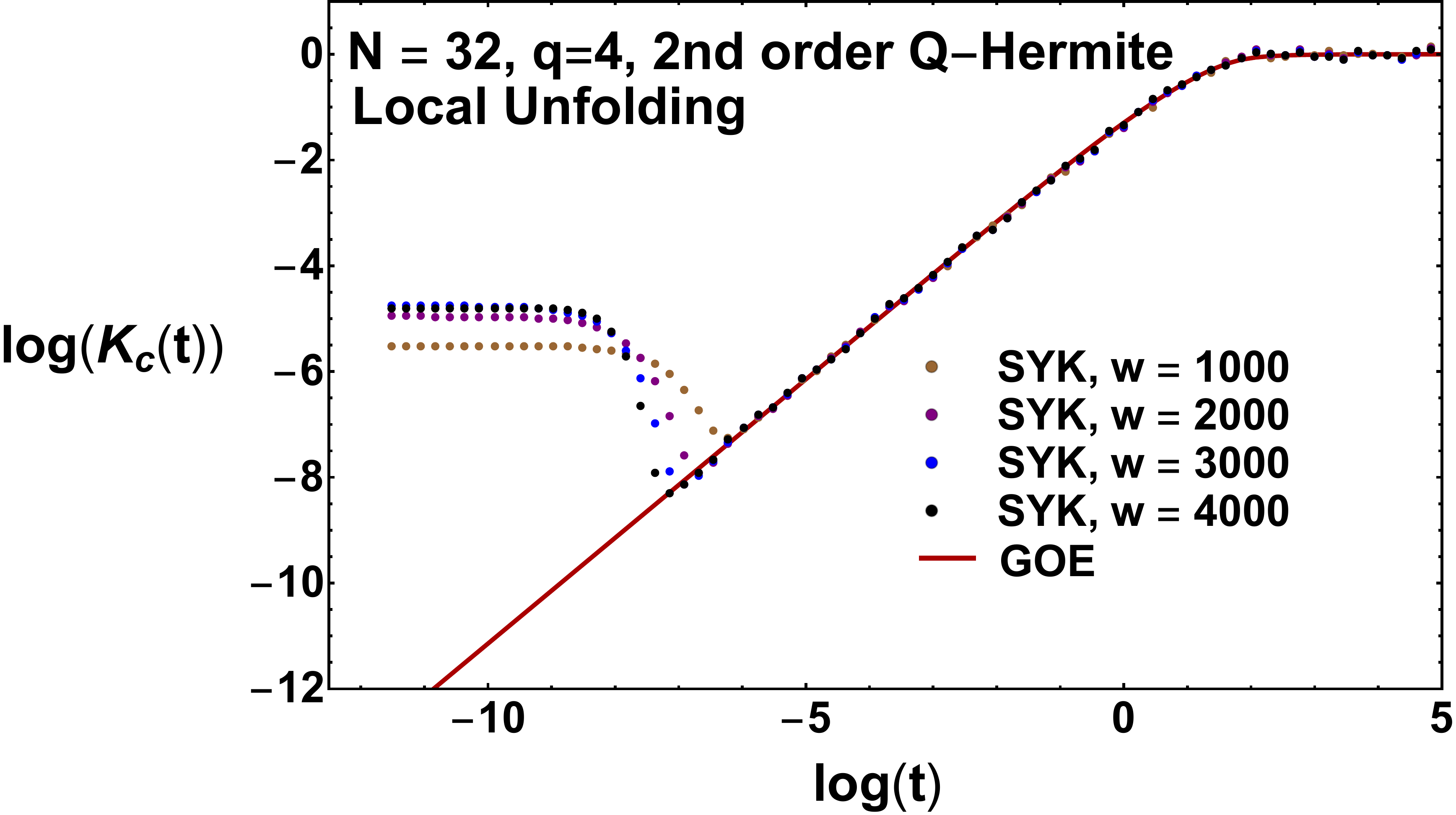}\quad
  \includegraphics[width=7.0cm]{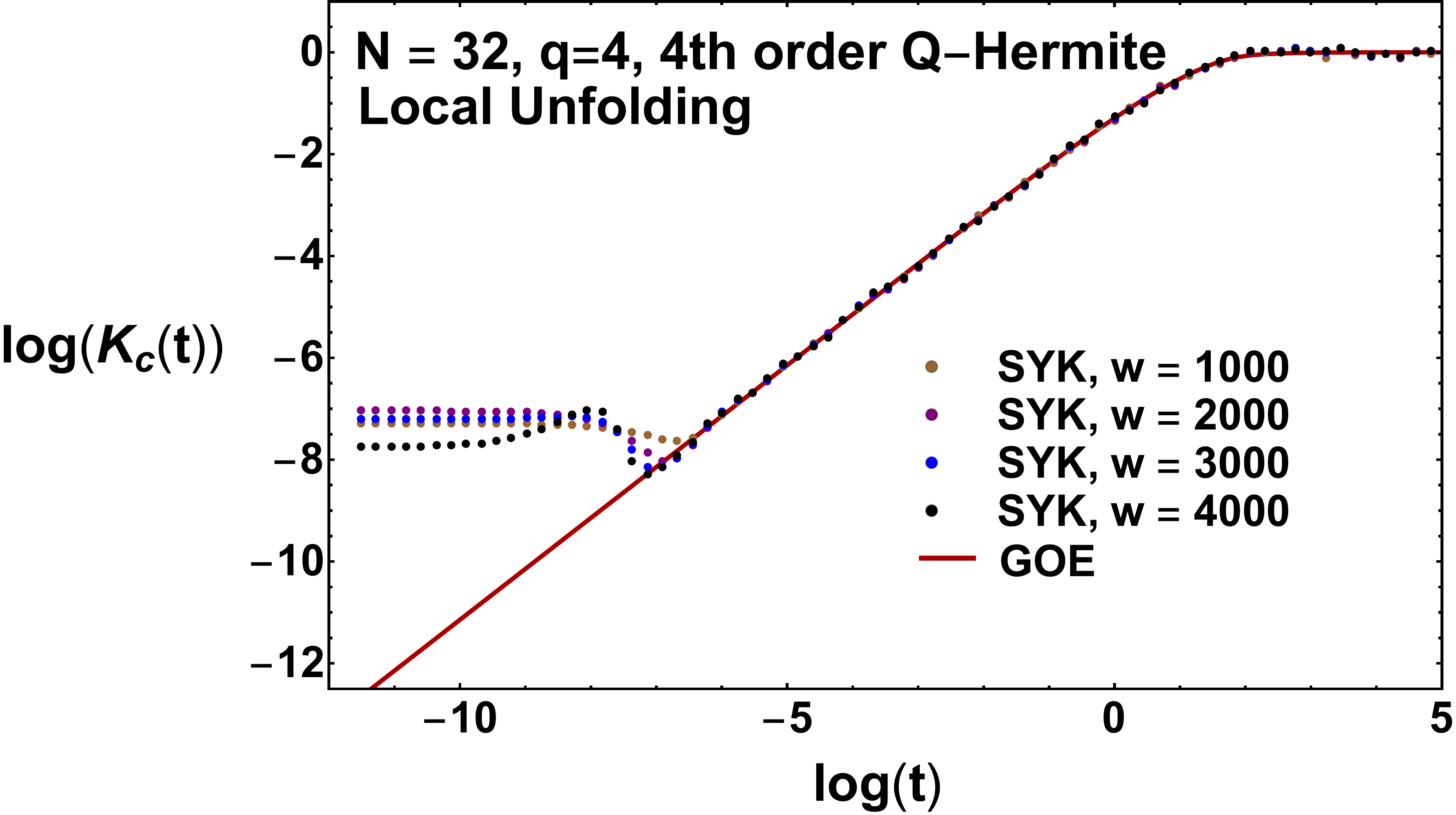}\\[5pt]
  \includegraphics[width=7.0cm]{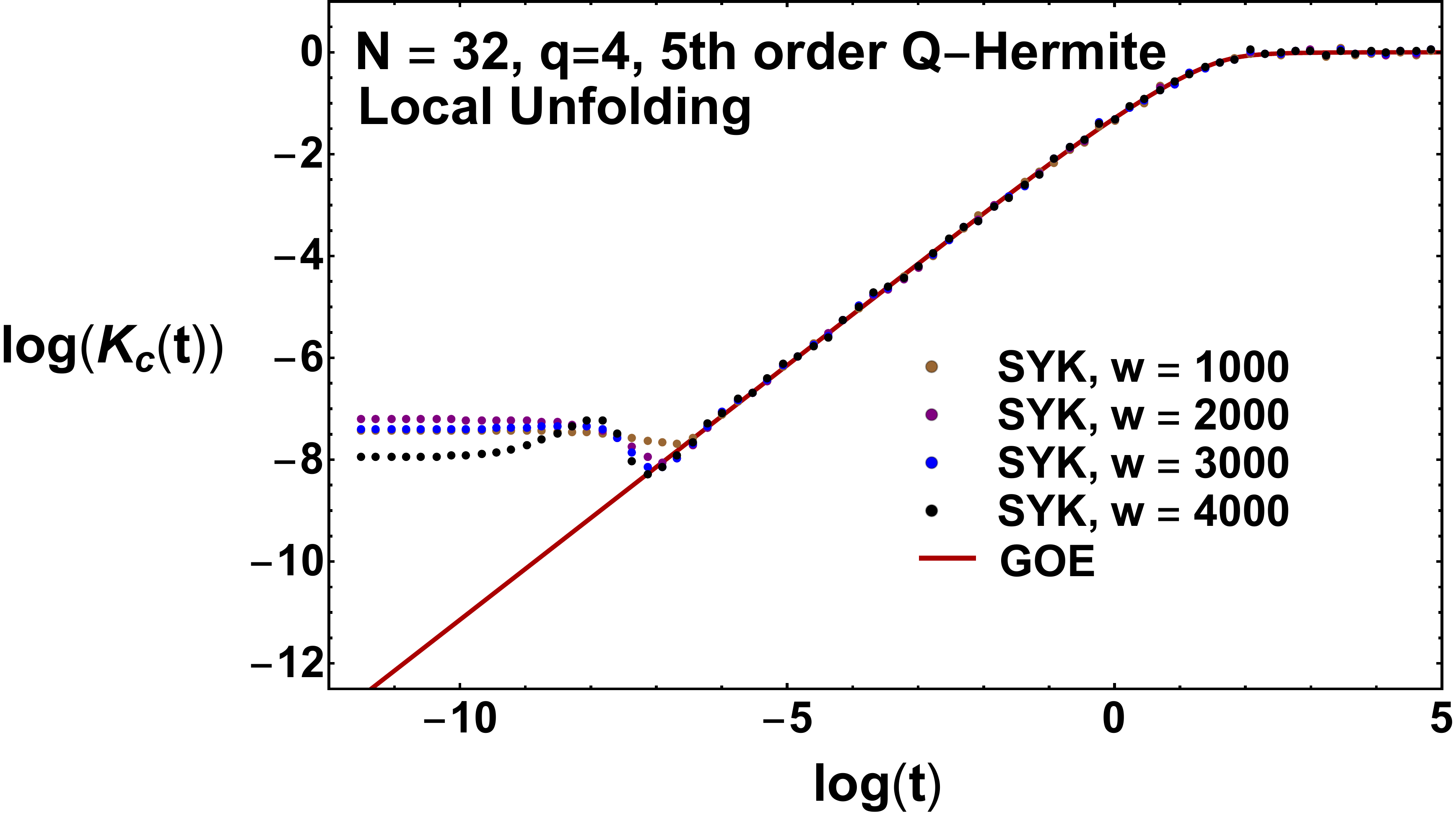}\quad
  \includegraphics[width=7.0cm]{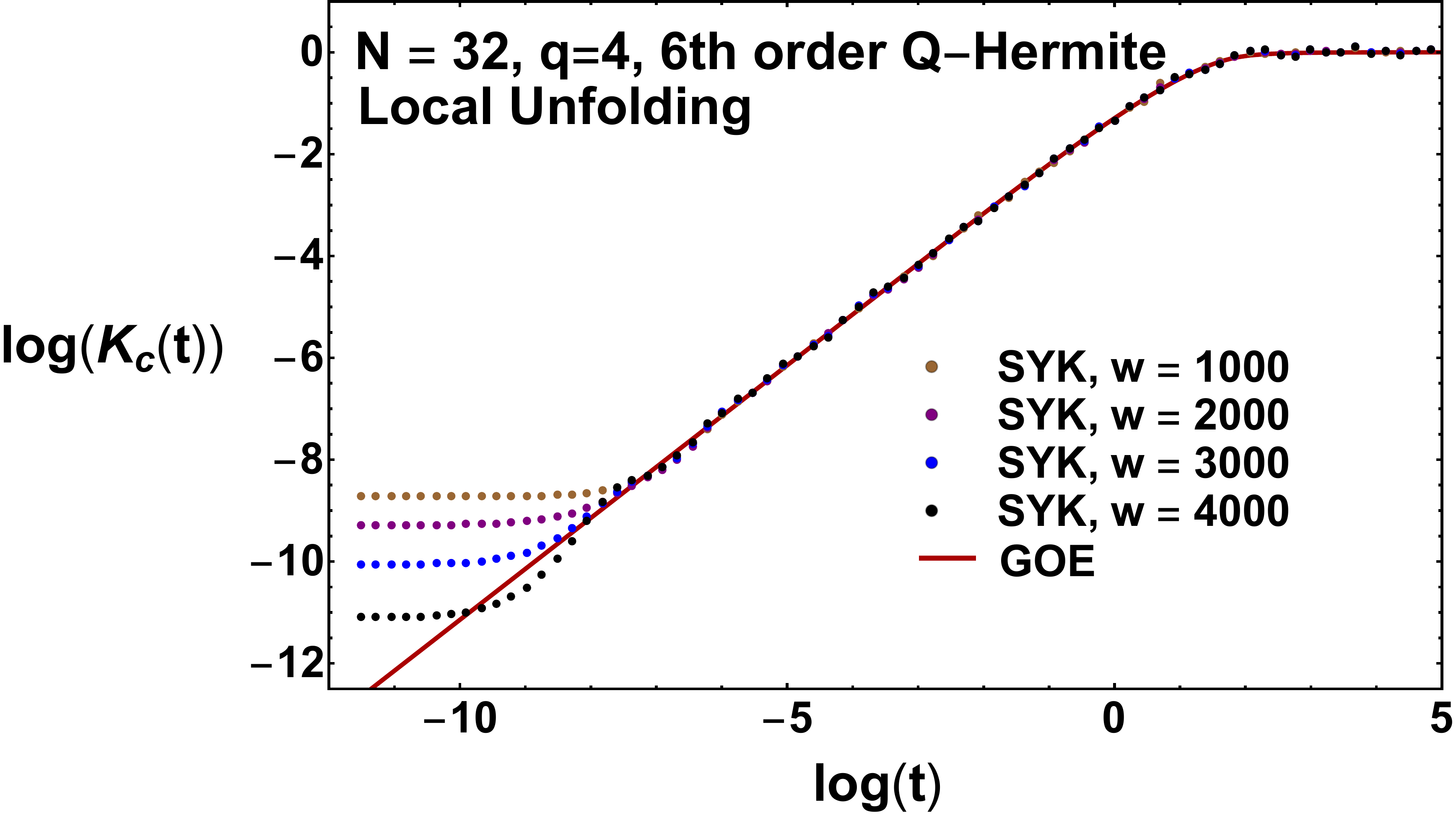}
  \caption{Double logarithmic plot of the spectral form factor for
    increasing order of the Q-Hermite polynomial used to fit the
    smoothened spectral density for each
    configuration.\label{fig:specform1}}
  \end{figure}

\looseness=-1 In figures~\ref{fig:specform0} and~\ref{fig:specform1} we show the spectral form factor for the
unfolded spectra that were used to calculate the number variance
in the previous section. In figure~\ref{fig:specform0},
the results for unfolding with the ensemble
average are given in the left figure, and for unfolding configuration
by configuration with eighth order Q-Hermite polynomials, in the right figure.
We used four different values for the width of the Gaussian cutoff. For
still larger values of the cutoff, the results are affected significantly by finite
size effects. Already for $w = 4000$ we observe considerable finite size effects.
The GOE result given by
{\rdmathspace[h]
\begin{equation}
K_{\rm GOE}(t/2\pi)=\left . \theta(1-u)\left(2u -u\log(1+2u)\right) + \theta(u-1)\left(2-u\log
  \frac{ 2u+1}{2u -1}\right)\right |_{u \to t/2\pi}
\label{korm}
  \end{equation}}\relax
is represented by the red curve in both figures. It agrees with
the SYK spectra up to very short times.  Comparing the two figures,
we observe that the spectral fluctuations due to the ensemble average result
in a peak at small times. The peak should not be confused
with the contribution from the disconnected part of the two-point correlator
which is many orders of magnitude larger. For local unfolding there is no peak, and
spectral form factor approaches zero for $t \to 0$. Note that
without the Gaussian cut-off the
spectral form factor $K_c(t=0) =0$ because of the normalization of the spectral density.
With the cut-off, the spectral form factor at $t=0$ measures the fluctuations of the number
of levels inside the Gaussian window. It becomes only zero when the window is larger than
the spectral support.

        \begin{figure}
       \centering
         \includegraphics[width=7.cm]{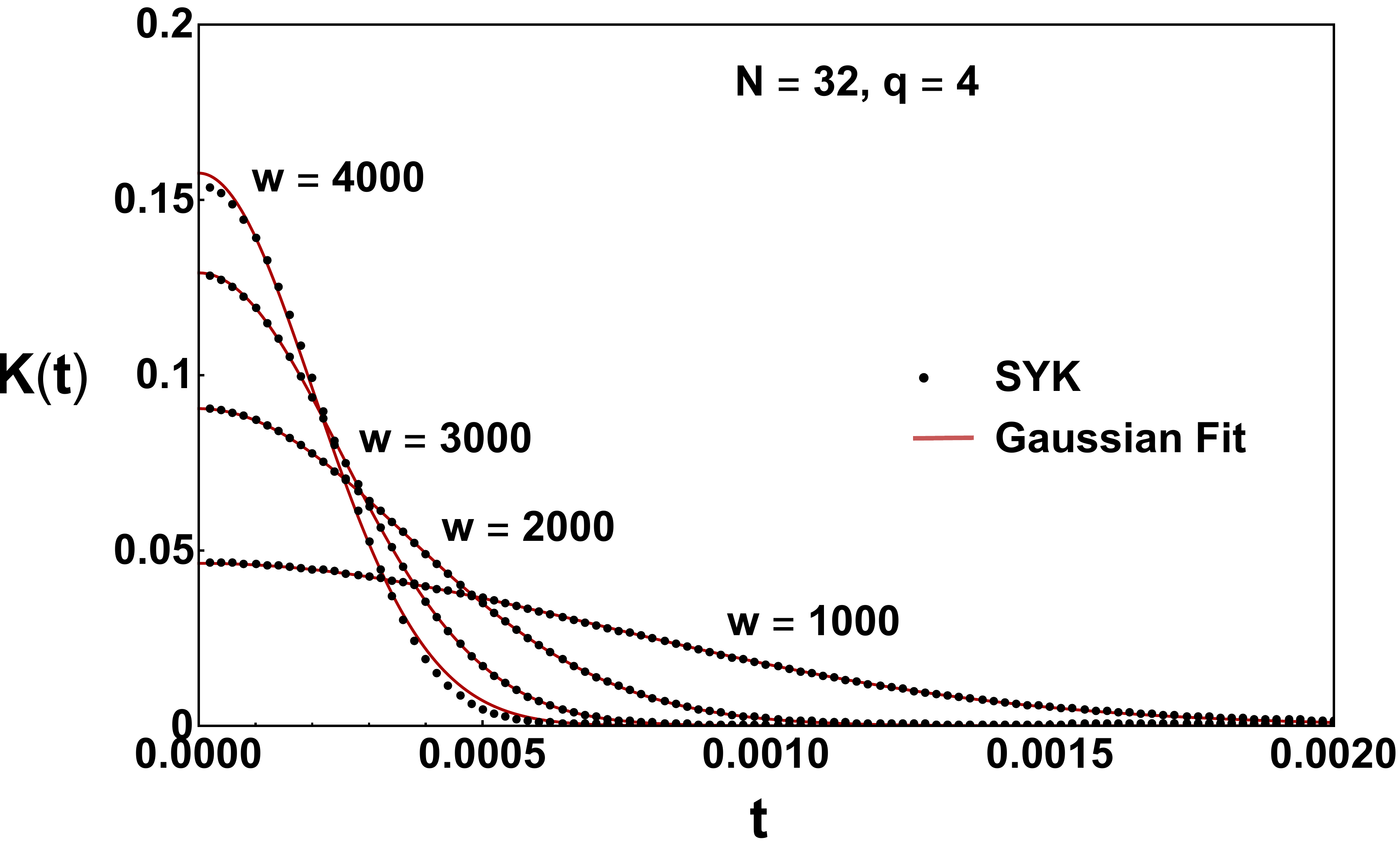}
  \caption{The spectral form factor for small $t$ obtained by using the ensemble average of the
  spectral density for unfolding (black points). The results are fitted to a Gaussian (red curve).\label{fig:peak}}

        \end{figure}

        In figure~\ref{fig:peak} we show the form factor
for small times on a linear scale. Then
the constant part of the two-point correlator gives the contribution
\begin{equation}
2\sqrt \pi w b_N e^{-t^2w^2}.
\label{gauss}
\end{equation}
For $w \to \infty$ this converges to the $\delta$-function
\begin{equation}
2 \pi b_N \delta(t).
\end{equation}
The constant  $b_N$ can be approximated by (see equation~(\ref{eqn:scaleFluc}))
\begin{equation}
b_N = \frac {1}{2} {N \choose q}^{-1}= 1.39\times 10^{-5}.
\label{cn}
\end{equation}
In figure~\ref{fig:peak}, we fit the parameters $b_N$ and $w$ to the SYK data.
For $w\le 3000$, the fitted value of $b_N \approx 1.26\times 10^{-5}\approx 1/282^2$
is close to the theoretical value given in equation~(\ref{cn}),
while for $w=4000$ it is  15 \% smaller.

        The number variance is related to the spectral form factor by a
        an integral over a smoothening kernel~\cite{Delon-1991}:
        \begin{equation}
        \Sigma^2(\bar n) = \frac {\bar n^2}{2\pi} \int_{-\infty}^\infty dt K(t) \left( \frac{ \sin ( \bar nt/2)} {\bar  nt/2}\right )^2.
              \label{kform}
         \end{equation}
         Therefore, the $\delta(t)$  contribution to the spectral form factor leads to a quadratic dependence
         of the number variance,
\begin{equation}
         \Sigma^2(\bar n) = b_N \bar n^2.
\end{equation}
If $K(t)$ remains finite for $t\to 0$ the large $\bar n$ limit of the number variance
is given by
\begin{equation}
\Sigma^2(\bar n) \approx \bar n\lim_{\bar n\to\infty}K(1/\bar n).
\end{equation}

In figure~\ref{fig:specform1} we show the spectral form factor for an
increasing number of fitted coefficients while the remaining ones up
to eighth order are kept equal to the ensemble average. For a fourth
or fifth order fit, the spectral form factor seems to approach a
constant for $ t \to 0$. From equation~\eqref{kform} it is clear that
this will result in a linear dependence of the number variance. The
coefficient of the linear term for fifth order unfolding is equal to
$3.9\times 10^{-4}$ which is in good agreement with the spectral form
factor for $t\to 0$ (which is $5.7\times 10^{-4}$, $7.1\times
10^{-4}$, $5.7\times10^{-4}$ and $3.0\times 10^{-4}$ for $w =1000$,
$w= 2000$, $w=3000$ and $w=4000$, respectively.)

\begin{figure}
\centering
\includegraphics[width=7cm]{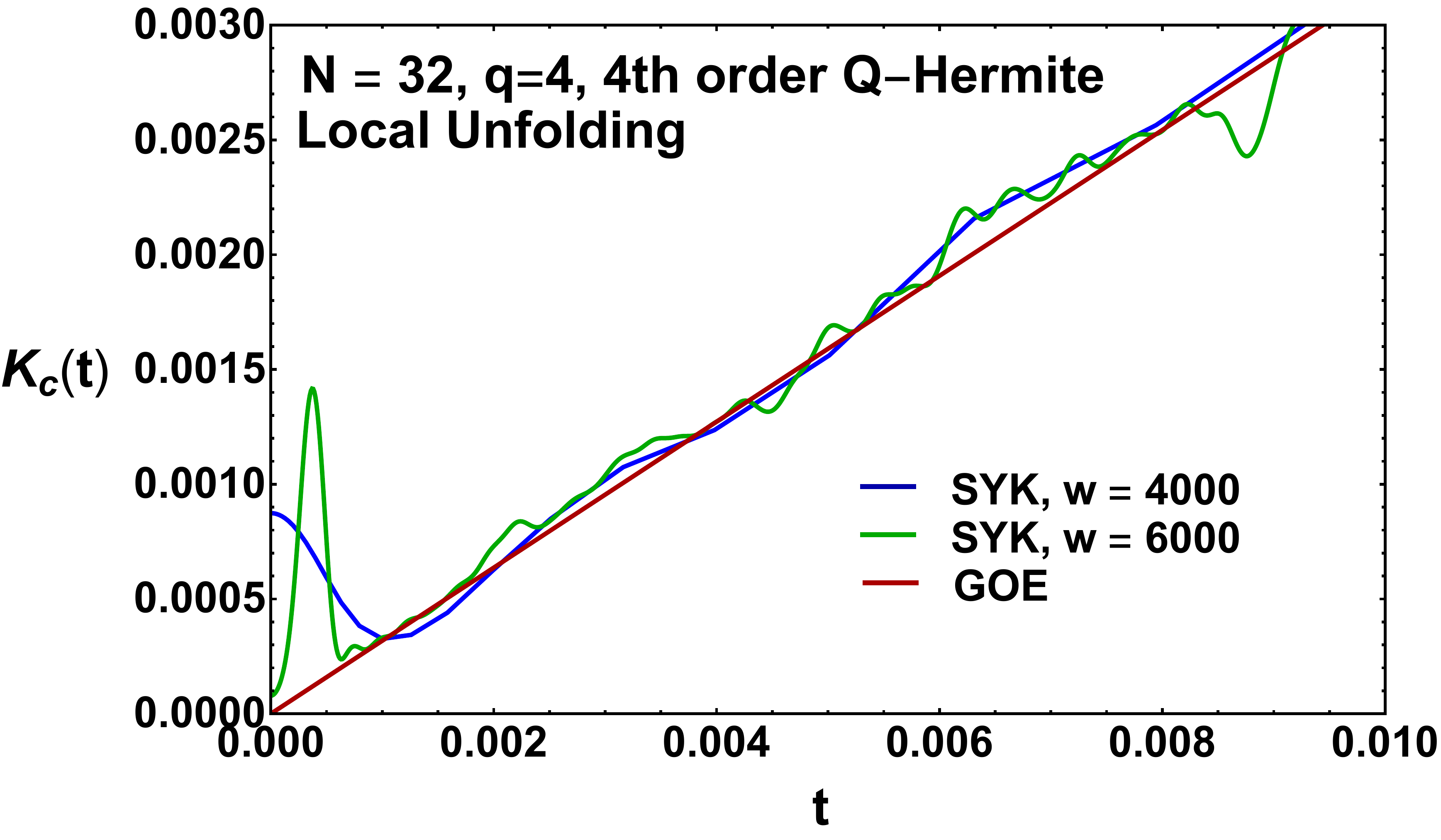}\quad
\includegraphics[width=7cm]{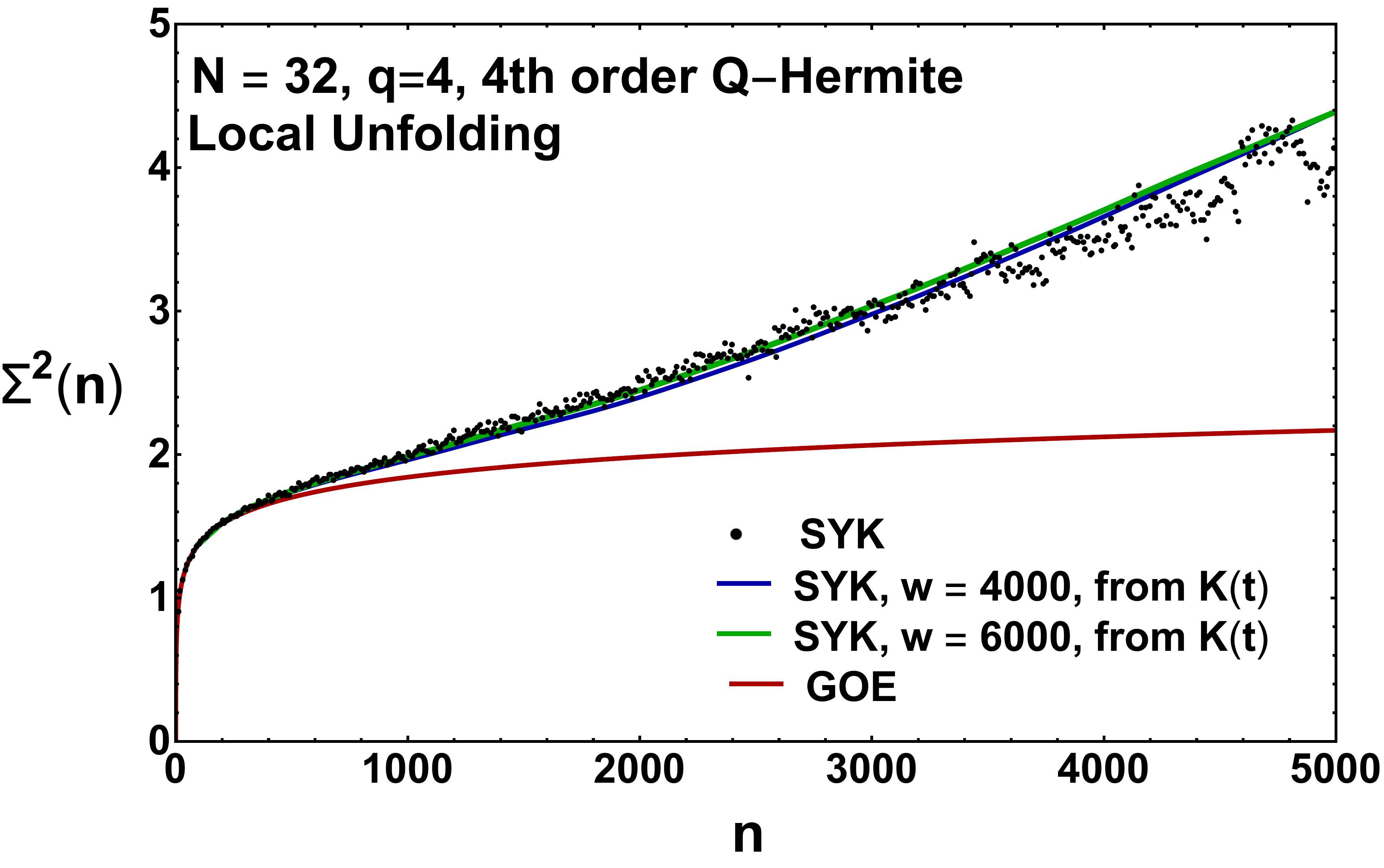}\\[5pt]
\includegraphics[width=7cm]{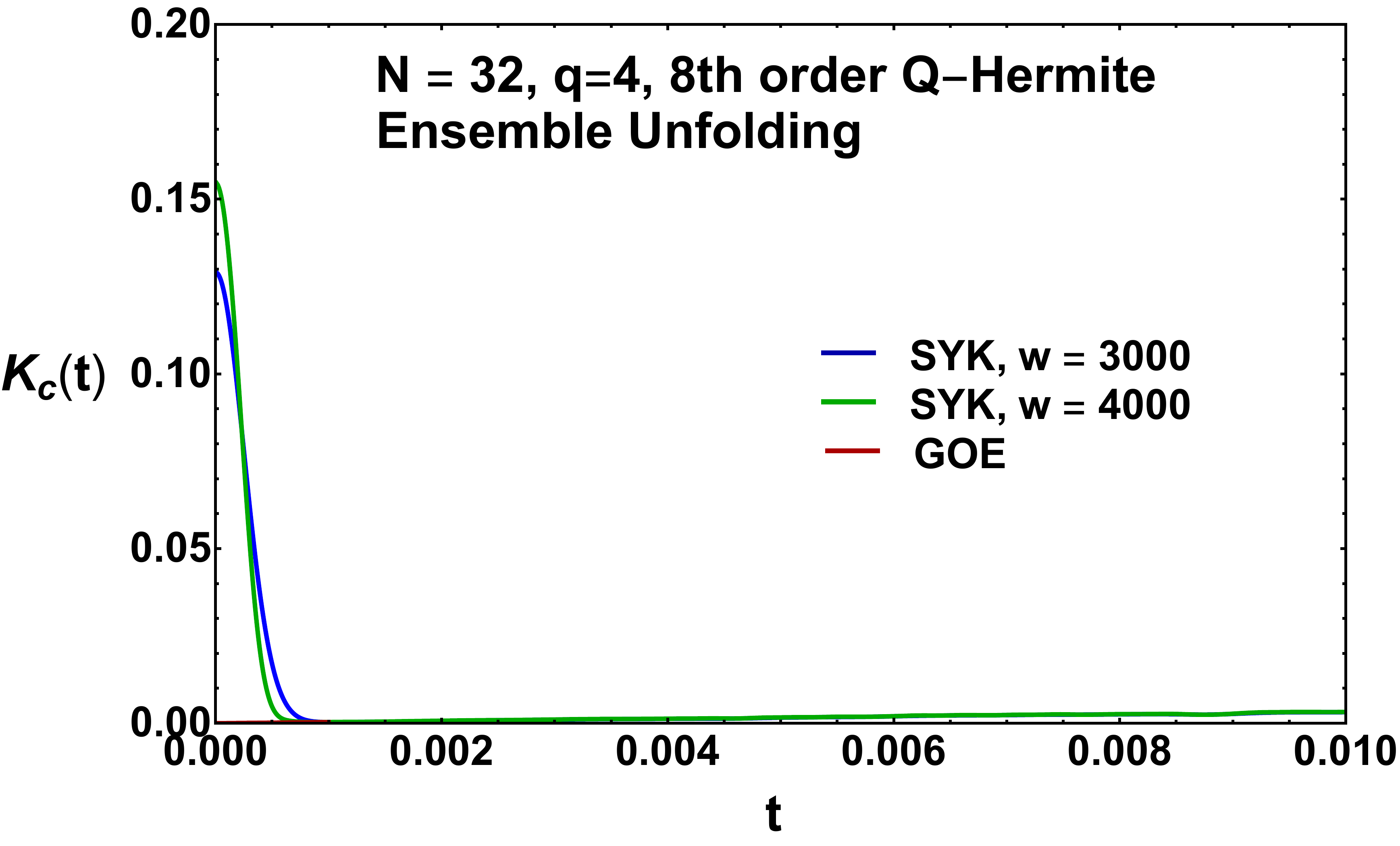}\quad
\includegraphics[width=7cm]{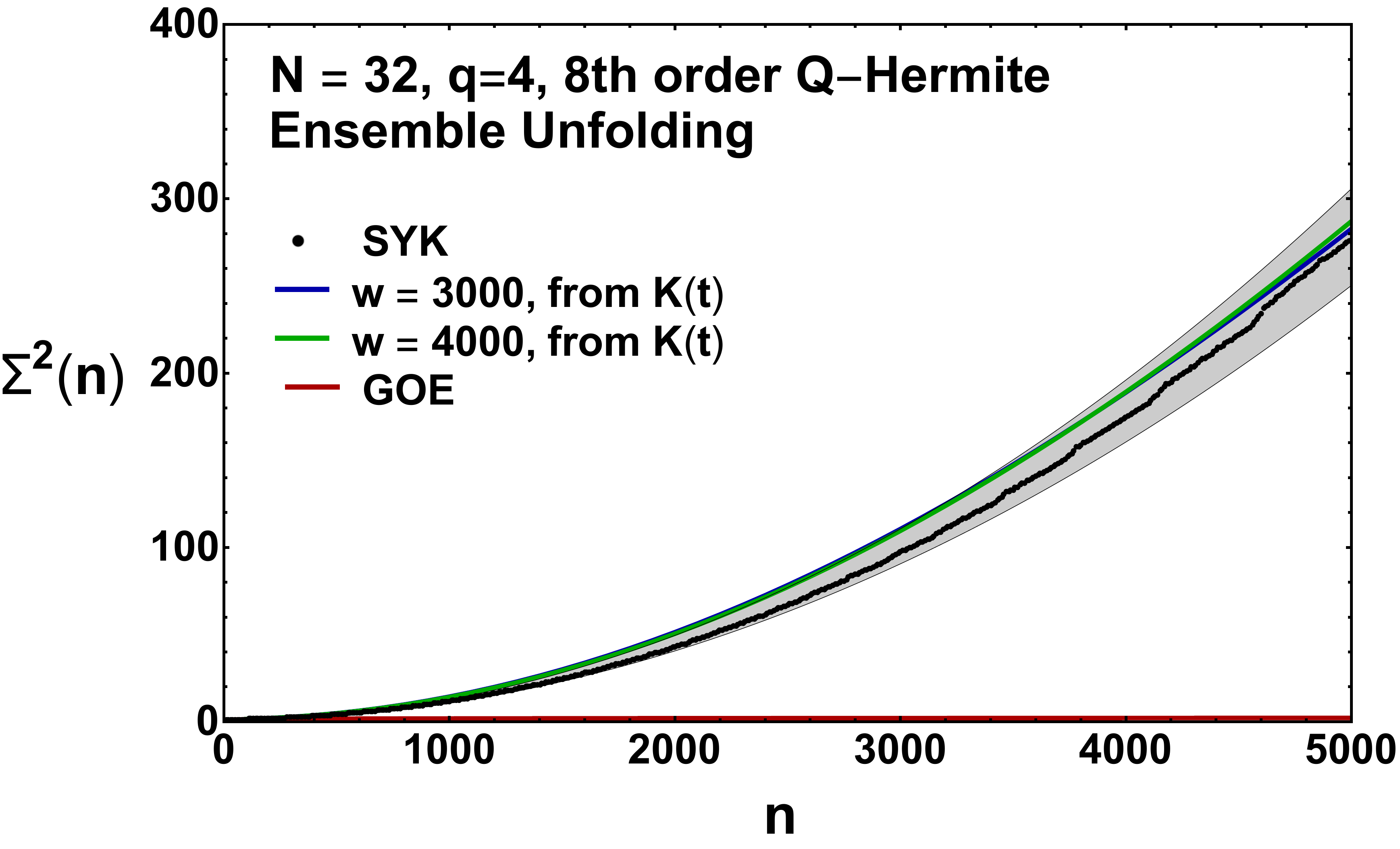}
\caption{Number variances in the right figures are obtained from the spectral form
  factor in the left figures by evaluating the integral~(\ref{kform}). The deviations
  in the number variance from the GOE result are due to the small $t$ peak in the spectral form
  factor.\label{fig:concis}}
\end{figure}

 In order to see the consistency of the calculation  we compare two different ways to calculate the number variance: directly from eigenvalues and from using equation \eqref{kform}. The results are shown in figure \ref{fig:concis}.
We conclude that a seemingly
insignificant deviation of the form factor from the GOE result
gives rise to a large deviation from the random matrix result for
the $\bar n$-dependence of the number variance.

\section{Double-trace moments and the validity of random matrix correlations}\label{sec:doubleTrRMTvalidity}
In this section we explore the question whether double-trace moments can
shed light on the convergence to random matrix correlations. In subsection~\ref{sec:double1}
we discuss the calculation of high-order double-trace moments, and
in section~\ref{sec:momFastConvergence} we evaluate the subleading corrections to estimate the
convergence to the RMT result.

\subsection{The calculation of high-order moments}
\label{sec:double1}

The connected two-point
correlator is given by
\begin{equation}
\langle \rho(x) \rho(y) \rangle_c =\frac 1{4\pi^2}
\int_{-\infty}^{\infty} \int_{-\infty}^{\infty} ds dt e^{-ixs} e^{-iyt}
\sum_{m,n} \frac {(it)^m(is)^n}{m!n!} \left( \tilde{M}_{m,n}-\tilde{M}_m\tilde{M}_n \right)
  \end{equation}
  Since the universal random matrix result scales as
  \begin{equation}
  \frac 1{D^2 (x-y)^2},
  \end{equation}
  high-order moments have to  be suppressed by $2^N$, yet a naive subscripts counting when Wick-contracting $\Gamma$'s would suggest double traces are suppressed by a power law $N^{-kq}$ where $k$ is the number of cross links.

 An important observation was first made in the appendix F of~\cite{Cotler2016} on the nested cross-linked-only chord diagrams. The authors of~\cite{Cotler2016} proved that
 \begin{equation}\label{eqn:Cotler}
 \begin{split}
   t_{1234\ldots k|k(k-1)\ldots1}:\!&=
  2^{-N}\binom{N}{q}^{-k}\sum_{a_1\ldots a_k} \Tr \left(\Gamma_{a_1}\Gamma_{a_2}\ldots\Gamma_{a_k}\right)\Tr \left(\Gamma_{a_k}\Gamma_{a_{k-1}}\ldots\Gamma_{a_1}\right)\\
&=   2^{-N}\sum_{m=0}^N\binom{N}{m}T_m(N,q)^k,
\end{split}
 \end{equation}
 where $T_m$ was first defined in equation~\eqref{eqn:TmuDef}:
 \begin{equation*}
 T_m:= \binom{N}{q}^{-1}\sum_{p=0}^q \binom{m}{p}\binom{N-m}{q-p} (-1)^p.
 \end{equation*}
 The observation was that
\begin{equation}\label{eqn:CotlerLimit}
\lim_{k\to\infty}t_{1234\ldots k|k(k-1)\ldots1} = 2^{1-N}
\end{equation} for the simple reason that this limit is dominated by only two terms $m=0, N$, both of which give $T_m=1$ (other terms have $|T_m|<1$). This shows the naive expectation for moments is
wrong and may serve as a starting point for understanding the RMT ramp from a moment method perspective.

In this section we offer a proof of a generalized version of equation~\eqref{eqn:Cotler}, and in turn offer an interesting generalization of the limiting scenario~\eqref{eqn:CotlerLimit}.

The starting point of our proof is the completeness relation
\begin{equation}
  \label{eqn:matrixCompleteness}
2^{-N/2}\sum_{\mu} \left(X_{\mu}\right)_{ij}\left(X_{\mu}\right)_{kl} = \delta_{il}\delta_{jk}.
\end{equation}
The $X_\mu$'s are as defined at the beginning of section~\ref{sec:sigmaModel}: $\{X_{\mu}\}$ is the set of all linearly independent matrices formed by products of Dirac gamma matrices, with appropriate prefactors such that $X_{\mu}^2=\mathbb{1}$, and $\mu$ is a multi-index containing all the subscripts of the Dirac matrices in the product. The set $\{X_{\mu}\}$ has $2^N$ elements because they form a basis for the vector space of $2^{N/2}\times 2^{N/2}$ matrices. It will be useful to organize $X_\mu$'s by the number of Dirac matrices in the product, that is, the length $|\mu|$ of the multi-index $\mu$, then the completeness relation can be rewritten as
\begin{equation}\label{eqn:matrixCompleteness2}
2^{-N/2}\sum_{m=0}^N\sum_{\mu_m} \left(X_{\mu_m}\right)_{ij}\left(X_{\mu_m}\right)_{kl} = \delta_{il}\delta_{jk},
\end{equation}
where $\mu_m$ is a multi-index of length $m$, namely $|\mu_m|=m$.

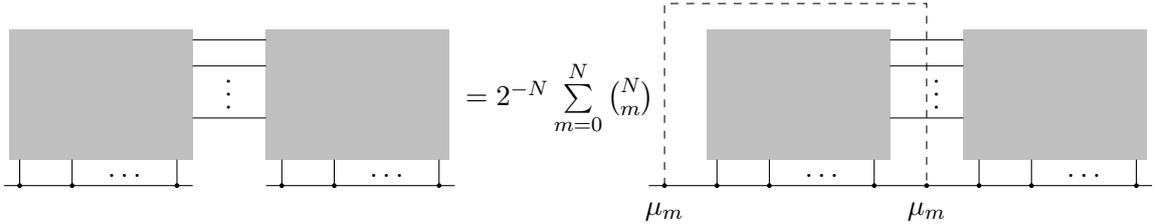
\begin{figure}
\centering
\resizebox*{1\textwidth}{!}{%
\begin{tikzpicture}[scale=0.7]
\draw[fill=black] (1,0) circle (1pt);
\draw[fill=black] (2,0) circle (1pt);
\node at (3,0.2) {$\ldots$};
\draw[fill=black] (4,0) circle (1pt);
\draw[fill=black] (6,0) circle (1pt);
\draw[fill=black] (7,0) circle (1pt);
\node at (8,0.2) {$\ldots$};
\draw[fill=black] (9,0) circle (1pt);
\draw (0.7,0)--(4.3,0);
\draw (5.7,0)--(9.3,0);
\fill [lightgray] (0.8,0.5) rectangle (4.3,3);
\fill [lightgray] (5.7,0.5) rectangle (9.2,3);
\draw (1,0)--(1,0.5);
\draw (2,0)--(2,0.5);
\draw (4,0)--(4,0.5);
\draw (6,0)--(6,0.5);
\draw (7,0)--(7,0.5);
\draw (9,0)--(9,0.5);
\draw (4.3,2.8)--(5.7,2.8);
\draw (4.3,2.3)--(5.7,2.3);
\node[rotate=90] at (5, 1.8) {$\cdots$};
\draw (4.3,1.3)--(5.7,1.3);
\node at (11.3,1.7) {$=2^{-N}\sum\limits_{m=0}^N\binom{N}{m}$};
\draw[fill=black] (13.3,0) circle (1pt);
\draw[fill=black] (14.3,0) circle (1pt);
\draw[fill=black] (15.3,0) circle (1pt);
\node at (16.3,0.2) {$\ldots$};
\draw[fill=black] (17.3,0) circle (1pt);
\draw[fill=black] (18.3,0) circle (1pt);
\draw[fill=black] (19.3,0) circle (1pt);
\draw[fill=black] (20.3,0) circle (1pt);
\node at (21.3,0.2) {$\ldots$};
\draw[fill=black] (22.3,0) circle (1pt);
\node at (13.3,-0.5) {$\mu_m$};
\node at (18.3,-0.5) {$\mu_m$};
\fill [lightgray] (14.1,0.5) rectangle (17.6,3);
\fill [lightgray] (19,0.5) rectangle (22.5,3);
\draw (14.3,0)--(14.3,0.5);
\draw (15.3,0)--(15.3,0.5);
\draw (17.3,0)--(17.3,0.5);
\draw (19.3,0)--(19.3,0.5);
\draw (20.3,0)--(20.3,0.5);
\draw (22.3,0)--(22.3,0.5);
\draw (17.6,2.8)--(19,2.8);
\draw (17.6,2.3)--(19,2.3);
\node[rotate=90] at (18.5, 1.8) {$\cdots$};
\draw (17.6,1.3)--(19,1.3);
\draw[dashed] (13.3,0)--(13.3,3.5)--(18.3,3.5)--(18.3,0);
\draw (13,0)-- (22.6,0);
\end{tikzpicture}}\relax
\caption{Diagrammatic representation of equation~\eqref{eqn:doubleToSingle}. The dashed chord represents the inserted $X_{\mu_m}$. Chords can intersect in arbitrary ways in the shaded regions.\label{fig:insertion}}
\end{figure}\relax
Inserting the completeness relation~\eqref{eqn:matrixCompleteness} in to double-trace moments, we have
\begin{equation}\label{eqn:doubleToSingle}
\tilde{M}_{l,n} = 2^{-N}\sum_{m=0}^N\sum_{\mu_m}\binom{N}{m}\frac{\vev{\Tr\left( X_{\mu_m} H^l X_{\mu_m} H^n\right)}}{2^{N/2}\binom{N}{m}\sigma^{l+n}},
\end{equation}
thus reducing every double trace to a sum over single traces. At the level of chord diagrams, this means every double-trace chord diagram can be written as a sum of single trace diagrams with one extra chord inserted, see figure~\ref{fig:insertion} for illustration. From equation~\eqref{eqn:doubleToSingle} we can further calculate each single trace by the combinatorics developed in~\cite{Garcia-Garcia-ml-2018kvh}, however, this does not warrant a straightforward application of the Q-Hermite approximation because this approximation has errors in polynomials of $1/N$, whereas the $\binom{N}{m}$ term is exponentially large in $N$ when $m\sim N/2$.

Let us calculate the contraction $t_{1234\ldots k|k(k-1)\ldots1}$ which in terms of a single
trace is given by
\begin{align}\label{eqn:parallelContractionWithInsertion}
  t_{1234\ldots k|k(k-1)\ldots1}=2^{-N} \sum_{m=0}^N\binom{N}{m}\sum_{\mu_m}\sum_{a_1,\ldots,a_k}\frac{\Tr \left(X_{\mu_m}\Gamma_{a_1}\ldots\Gamma_{a_k}X_{\mu_m}\Gamma_{a_k}\ldots\Gamma_{a_1}\right)}{2^{N/2}\binom{N}{m} \binom{N}{q}^k}.
\end{align}
Since (with $c_{a\mu_m }$ the number of indices that $a$ and $\mu_m$ have in common)
\begin{equation}
X_{\mu_m}\Gamma_{a}=(-1)^{mq+c_{a\mu_m }}\Gamma_{a}X_{\mu_m},
\end{equation}
we have for fixed $m$ that
\begin{align}\label{eqn:etaGeneralized}
\begin{split}
\sum_{\mu_m}\sum_{a}\frac{\Tr \left(X_{\mu_m}\Gamma_{a}X_{\mu_m}\Gamma_{a} \right)}{2^{N/2}\binom{N}{m}\binom{N}{q}} &= (-1)^{mq}\sum_{c_{a\mu_m }=0}^{q}\binom{m}{c_{a\mu_m}}\binom{N-m}{q-c_{a\mu_m}}
  (-1)^{c_{a\mu_m}} \\ &= (-1)^{mq}T_m.
\end{split}
\end{align}
Substituting equation~\eqref{eqn:etaGeneralized}  into equation~\eqref{eqn:parallelContractionWithInsertion}, we arrive at
\begin{equation}
  t_{1234\ldots k|k(k-1)\ldots1}=
2^{-N}\sum_{m=0}^N\binom{N}{m} (-1)^{kmq} T_m^k.
\end{equation}
This result is an application of
the so-called ``cut-vertex factorization'' property for single traces~\cite{Garcia-Garcia-ml-2018kvh}.\footnote{The adjective ``cut-vertex'' is understood at the level of intersection graphs~\cite{Garcia-Garcia-ml-2018kvh}.}
This  property states that the chord intersections that do not form a closed loop factorize into products of individual intersections. In this instance we have $k$ independent intersections, each giving a factor of $(-1)^{mq} T_m$.

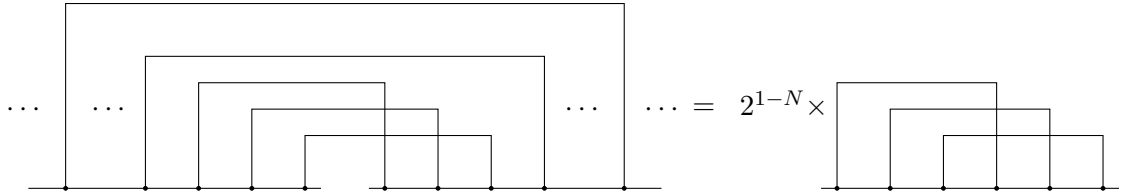
\begin{figure}
\centering
\begin{tikzpicture}[scale=0.7]
\draw[fill=black] (1,0) circle (1pt);
\draw[fill=black] (2,0) circle (1pt);
\draw[fill=black] (3,0) circle (1pt);
\draw[fill=black] (4,0) circle (1pt);
\draw[fill=black] (5.5,0) circle (1pt);
\draw[fill=black] (6.5,0) circle (1pt);
\draw[fill=black] (7.5,0) circle (1pt);
\draw[fill=black] (8.5,0) circle (1pt);
\draw[fill=black] (-0.5,0) circle (1pt);
\draw[fill=black] (10,0) circle (1pt);

\draw (2,0) -- (2,2)--(5.5,2)--(5.5,0);
\draw (3,0) -- (3,1.5)--(6.5,1.5)--(6.5,0);
\draw (4,0) -- (4,1)--(7.5,1)--(7.5,0);
\draw (1,0) -- (1,2.5)--(8.5,2.5)--(8.5,0);
\draw (-0.5,0) -- (-0.5, 3.5) -- (10,3.5)--(10,0);
\draw (-1.2,0)--(4.3,0);
\draw (5.2,0)--(10.7,0);
\node at (0.35,1.5) {$\cdots$};
\node at (-1.25,1.5) {$\cdots$};
\node at (9.25,1.5) {$\cdots$};
\node at (10.75,1.5) {$\cdots$};
\node at (11.5,1.5) {$=$};
\node at (13,1.6) {$2^{1-N}\times$};

\draw[fill=black] (14,0) circle (1pt);
\draw[fill=black] (15,0) circle (1pt);
\draw[fill=black] (16,0) circle (1pt);
\draw[fill=black] (17,0) circle (1pt);
\draw[fill=black] (18,0) circle (1pt);
\draw[fill=black] (19,0) circle (1pt);
\draw (14,0) -- (14,2)--(17,2)--(17,0);
\draw (15,0) -- (15,1.5)--(18,1.5)--(18,0);
\draw (16,0) -- (16,1)--(19,1)--(19,0);
\draw (13.7,0)--(19.3,0);
\end{tikzpicture}
\caption{An example of equation~\eqref{eqn:parallelLinksRemovalLimit}.
  The ellipses on the left represent infinite number of nested cross links. On the right we have same contraction with all the nested links removed, as a single trace diagram.\label{fig:parallelLinksRemovalLimit}}
\end{figure}

We thus have demonstrated that equation~\eqref{eqn:doubleToSingle} reproduces equation~\eqref{eqn:Cotler} for even $q$ and even $k$, and trivially for when both $q$ and $k$ are odd because both formulas give zero.
We will now use~\eqref{eqn:doubleToSingle} to derive a modest generalization to  equations~\eqref{eqn:Cotler} and~\eqref{eqn:CotlerLimit}. The diagrams we consider are the ones with a  number of intersecting cross links and a number of nested links. Repeating the previous analysis, we conclude
that these contributions are of the form
\begin{equation}
   2^{-N}\sum_{m=0}^N\binom{N}{m} (-1)^{kmq}
  T_m^k \times \text{rest}(N,m,q),
\end{equation}
\looseness=-1 where $k$ is the number of nested cross links  and $\text{rest}(N,m,q)$ is whatever  remains of the single-trace chord diagram after taking out the intersections between the dashed chord and nested links. We remind the reader that the dashed chord represents the insertion of $X_\mu$'s. Again in this limit the sum is dominated by two terms $m=0$ and $m=N$, so the limit is simply
\begin{equation}\label{eqn:parallelLinksRemovalLimit}
2^{-N}\left[\text{rest}(N,0,q)+\text{rest}(N,N,q)\right].
\end{equation}
Note that
\begin{equation}
\text{rest}(N,0,q)=\text{rest}(N,N,q)= \text{single-trace diagram with dashed chord removed},
\end{equation}
because the $m=0$ and $m=N$ terms correspond to the insertions of identity matrix and Dirac chirality matrix in equation~\eqref{eqn:doubleToSingle}, and the chirality matrix insertion is equivalent to the identity matrix insertion when $q$ is even,\footnote{For odd $q$ it is possible that $\text{rest}(N,0,q)=-\text{rest}(N,N,q)$ for certain diagrams.}
we conclude that a diagram with a large number of nested cross links
is equal to $2^{1-N}$ times the corresponding single-trace diagram with all nested links removed ---
see figure~\ref{fig:parallelLinksRemovalLimit} for a concrete example.

The approach developed in this section can be viewed as complementary to that of section~\ref{sec:dtrAndChord}: the method of section~\ref{sec:dtrAndChord} gives a completely combinatorial formula~\eqref{eqn:generalDescriptionStarts}--\eqref{eqn:generalDescriptionEnds} in which the number of summations scales as a power in $q$ and is independent of $N$, hence is  useful for calculating exact numerics for low-order moments with large $N$, as we have seen. However, for exactly the same reason it is not very useful  for discussing the large $q$ behaviors, neither is it helpful for discussing the asymptotic behavior with large number of cross links due to its complexity. On the other hand, the approach of this section gives a symbolically simple formula in terms of single traces, hence makes various asymptotic behaviors amenable to discussion for a large number of nested cross links
as we have seen and will see more soon.

\subsection{Long-wavelength fluctuations}\label{sec:momFastConvergence}
Previously we saw a somewhat mysterious numerical phenomenon: we do not need to remove exponentially many long-range modes to get very close to the random matrix result. With the techniques discussed in section~\ref{sec:doubleTrRMTvalidity}, we give a partial explanation to this phenomenon and estimate how the number of subtractions scales with $N$. We first summarize the gist of this section:
\begin{itemize}[leftmargin=*]
\item \emph{The mode numbers correspond to the numbers of cross links in double-trace contractions.}
\item \emph{The double-trace contractions converge rapidly to the RMT result as the number of cross links increases.}
\end{itemize}
Here by ``RMT result'' we simply mean the weak statement that after a certain point the contractions stop being suppressed by the number of cross links but by $2^{-N}$ instead, as would be the case for an RMT theory. As we have seen in equations~\eqref{eqn:doubleTrFromScaleFluc} and~\eqref{eqn:2crossFactorization}, the double traces from diagrams with precisely two cross links can be entirely attributed to scale fluctuations --- the lowest long-wavelength mode. For higher modes and larger number of cross links, we cannot make the correspondence as precise, but it is intuitively clear that both probe finer structures of the correlations. In the case of actual RMT ensembles, such a correspondence can be made precise~\cite{brody1981}. Since the goal of this section is humble, this level of understanding should suffice as a starting point.

We will estimate speed of convergence of equation~\eqref{eqn:CotlerLimit}. In fact $t_{1234\ldots k|k(k-1)\ldots1}$ is not quite the right quantity to study. Near equation~\eqref{eqn:blockEffect}, it was pointed out that for higher moments (large number of cross links) the contribution $\vev{\Tr \left(\gamma_c H^m\right)\left(\Tr \gamma_c H^n\right)}$ must be included. The completely nested diagram contribution in this double trace is
 \begin{equation}\label{eqn:CotlerWithGamma5}
 \begin{split}
   u_{1234\ldots k|k(k-1)\ldots1}:\!&=
  2^{-N}\binom{N}{q}^{-k}\sum_{a_1\ldots a_k} \Tr \left(\gamma_c\Gamma_{a_1}\Gamma_{a_2}\ldots\Gamma_{a_k}\right)\Tr \left(\gamma_c\Gamma_{a_k}\Gamma_{a_{k-1}}\ldots\Gamma_{a_1}\right)\\
&=   2^{-N}\sum_{m=0}^N\binom{N}{m}(-1)^mT_m^k,
\end{split}
 \end{equation}
where the second equality can again be proved by the insertion of completeness relation~\eqref{eqn:matrixCompleteness}. Hence we have
\begin{equation}
t_{1234\ldots k|k(k-1)\ldots1}+u_{1234\ldots k|k(k-1)\ldots1}=2^{1-N}\sum_{m=0}^{N}\frac{1}{2}[1+(-1)^{m}]\binom{N}{m} T_m^k,
\end{equation}
and so we see only the even $m$ terms contribute. It is clear
\begin{equation}
\lim_{k\to \infty} [t_{1234\ldots k|k(k-1)\ldots1}+u_{1234\ldots k|k(k-1)\ldots1}]=2^{2-N},
\end{equation}
and this is the quantity whose speed of convergence we wish to study. Its finite-$k$ relative deviation from the limiting result is
\begin{equation}\label{eqn:relativeError}
\frac{1}{2^{2-N}}\left(t_{1234\ldots k|k(k-1)\ldots1}+u_{1234\ldots k|k(k-1)\ldots1}\right) -1  = \frac{1}{2} \sum_{m=2}^{N-2}\frac{1}{2}[1+(-1)^{m}]\binom{N}{m}T_m^k.
\end{equation}
For $m\neq 0, N$, we have $|T_m|<1$, and the remaining largest terms are
\begin{equation}
T_2=T_{N-2}= 1-\frac{4q}{N-1}+\frac{4q^2}{N(N-1)}.
\end{equation}
So for large enough $k$, we would expect the summand $\binom{N}{m}T_m^k$ to be sharply peaked near $m=2, N-2$. Hence
\begin{equation}\label{eqn:errorEstimate}
\frac{1}{2} \sum_{m=2}^{N-2}\frac{1}{2}[1+(-1)^{m}]\binom{N}{m}T_m^k \approx \binom{N}{2} \left(1-\frac{4q}{N-1}+\frac{4q^2}{N(N-1)}\right)^k
\end{equation}
So we see that contractions with $k > N/(2q)\log N$ are suppressed, and to obtain
the RMT result we only have to subtract long wavelength fluctuations with
$k < N/(2q)\log N$.\footnote{This is a large-$N$ estimate. Using equation~\eqref{eqn:errorEstimate} gives $k>-\log \binom{N}{2} /\log\left(1-\frac{4q}{N-1}+\frac{4q^2}{N(N-1)}\right)$.}
For $N=32$ and $q=4$, we subtracted the first eight long-range modes
to get a decent agreement to the RMT result, which is in the right ball park of our estimate  $N/(2q)\log N\approx 14$.  This analysis also applies to the slightly broader class of diagrams that include the ones described near equation~\eqref{eqn:parallelLinksRemovalLimit}.

For a more complete analysis, we would need to get a handle on the cases with linked intersections and the unfolding, which is outside the scope of this paper.

\section{Discussion and conclusions}
\label{sec:conclusion}
How can we analytically understand the observations of this paper?
To get some perspective, we first discuss spectral correlations of
the eigenvalues of the QCD Dirac operator. The first results were
obtained in~\cite{Halasz-ml-1995vd} were the number variance was calculated from
the quenched Dirac eigenvalues of a single lattice QCD configuration.
Agreement with random matrix theory was found up to as many level spacings
as allowed by the statistics (which is about 100 level spacings for a
$12^4$ lattice). However, it was soon realized that when we compare to
the ensemble average, which is what should be done, deviations are
already visible on the scale of a couple of level spacings. The deviations
are now well understood and can be obtained analytically from chiral
perturbation theory~\cite{Osborn-ml-1998qb,BerbenniBitsch-ml-1999ti}.  The scale where deviations from random matrix theory
occur is the quark mass scale for which the corresponding pion
Compton wave length is equal to the size of the box.

The situation with the SYK model is similar. Comparing the estimate
of the scale where spectral correlations start deviating from RMT with
statistical fluctuations of the spectral density from one realization
of the ensemble to the next,
we see that the deviations from the RMT predictions can mostly be accounted for by such long-wavelength fluctuations.
This is further confirmed by  our numerical study, where after the first
few long-wavelength modes are subtracted realization by realization of the ensemble,
the spectral form factor becomes RMT-behaved until very short times
and the number variance has RMT spectral rigidity until the energy scale
given by the wavelength of the subtracted modes. This can be partly explained by the fact that the Hilbert space is $2^{N/2}$-dimensional whereas there are only $\binom{N}q$ model parameters $J_\alpha$, implying that very few matrix entries
of the Hamiltonian can  fluctuate independently. This scenario is to be contrasted with actual random matrix ensembles,
where all matrix entries (barring Hermiticity and symmetry constraints) can fluctuate independently.
However, this does not quite explain the separation of scales between the
long-wavelength modes and scale of universal RMT fluctuations.
We have gained some insight
into this scale separation by looking at the convergence of a class of double-trace chord diagrams toward those of RMT,
and analytically demonstrated that their early convergence behavior is consistent with our numerical results.
It is desirable to have analytically controlled estimates of
all chord diagrams to make more quantitative comparisons. We did develop a combinatorial formula for all chord diagrams in section~\ref{sec:dtrAndChord}, and it has been used to calculate the first few moments, but it is complicated and hard to apply to
general high-order moments.  We hope to investigate this in future works.

In the same vein of understanding QCD Dirac operator's spectral statistics through chiral perturbation theory,
we  studied the nonlinear sigma model formulation of the spectral determinant of the SYK model.
Although it is reassuring that the large energy expansion of the sigma model correctly reproduces the moments
in a nontrivial manner, and that the wide correlator is given by the sum
of the GUE result and corrections that are in agreement with numerical results,
only few low-order correction terms could be calculated and it is not clear
how the contribution of all other corrections to the two-point function cancels.
 In our view the following two issues have to be addressed
to complete our understanding of the spectral density and spectral correlations in terms of
the $\sigma$-model:
\begin{enumerate}
\item The most straightforward calculation gives a one-point resolvent that has the wrong branch cut.
  It is not known how to perform a resummation at the level of the $\sigma$-model action. In particular,
  we have not been able to obtain a Gaussian spectral density when $N \gg q$.
\item Although the correction terms to the RMT result explain the numerical results for $q=4$, it has not
  been shown that all other corrections in the loop expansion are small --- in fact
  we do not know the range of validity of the loop expansion. Since naive application of
  the result to $q=2$, which
  has Poisson statistics, also gives RMT spectral  correlations
  albeit with a much smaller Thouless energy, and we conclude that the loop
  expansion has to break down in this case.

\end{enumerate}
Neither issue implies that the $\sigma$-model approach is inherently flawed, but they do suggest that
our understanding of the $\sigma$-model is not complete. Similar issues arise in  the calculation of nested cross-linked diagrams
where the  $q=2$ case is not qualitatively different from the $q=4$ case.
We hope to address some of these questions in future work.

\acknowledgments{
We  acknowledge partial support from U.S.\ DOE Grant
No. DE-FAG-88FR40388. Antonio Garc\'ia-Garc\'ia is thanked for collaboration in
the early stages of this project. Alexander Altland, Dmitry Bagrets,
Steven Shenker and Douglas Stanford are thanked for useful discussions.}

\appendix

\section{Derivation of the \texorpdfstring{\boldmath $\sigma$}{sigma}-model}
\label{app:A}
In this section we derive the nonlinear $\sigma$-model for the SYK model,
see~\cite{Verbaarschot1984} for the complex SYK model, and~\cite{Altland-ml-2017eao} for the
Majorana SYK model. Our derivation follows a different route using
the Fierz transformation, see also~\cite{Liu-ml-2016rdi}.
Since $[\gamma_c, H]=0$
we have the Hamiltonian in the block-diagonal form
\begin{equation}
H = \begin{pmatrix}
H^{U}& \\& H^L
\end{pmatrix} = \sum_{\alpha} J_\alpha \Gamma_\alpha = \begin{pmatrix}
\sum_{\alpha} J_\alpha \Gamma_\alpha^{U}& \\& \sum_{\alpha} J_\alpha \Gamma_\alpha ^L
\end{pmatrix}
\end{equation}
in the chiral basis where
\begin{equation}
\gamma_c = \text{diag}(\underbrace{1,\ldots,1}_{2^{N/2-1}}, \underbrace{-1,\ldots,-1}_{2^{N/2-1}}).
\end{equation}
Taking into account the unitary symmetries as required for universal
RMT correlations, we
focus on the upper block and define the generating function as
\begin{align}
Z(x,y)  & = \left \langle  {\det}^{-n}\left(x+H^U\right) {\det}^{-n}\left(y+H^{U}\right) \right \rangle \nonumber\\
 & = \left \langle  \int \mathcal{D}\phi\mathcal{D}\phi^* e^{i\sum_k\phi_{1k}^*\cdot(x +H^U)\cdot\phi_{1k}-i\sum_k\phi_{2k}^*\cdot(y +H^U)\cdot\phi_{2k}}\right \rangle,
\end{align}
where $k$ is the replica index. We should think of $\phi$ as also having an implicit index of the $D=2^{N/2-1}$ dimensional Hilbert space (the index of the block Hamiltonian $H^U$) and ``$\ \cdot\ $'' indicates summation over the Hilbert space index. The integrals are convergent
because ${\rm Im}(x)=i\epsilon$ and  ${\rm Im}(y)=-i\epsilon$. We also introduce the sign factor
\begin{equation}
\eta_p=(-1)^{p+1},
\end{equation}
which will be inserted in the definition of the partition function such that the integrals are convergent.
The correlation function of two resolvents is given by
\begin{equation}
\left \langle G(x) G(y) \right \rangle = \lim_{n\to 0}\frac 1{n^2} \frac 1{D^2}
\frac d{dx}\frac d {dy} Z(x,y).
  \end{equation}
Recall that the disordered average is over Gaussian variables $J_\alpha$ with variance $v^2$, so after averaging we obtain
\begin{equation}
Z(x,y) =   \int \mathcal{D}\phi e^{i\sum_{p,k}z_{pk}\eta_p\phi_{pk}^* \cdot \phi_{pk}
  -\frac{v^2}2 \sum_\alpha \left(\sum_{p,k}\eta_p\phi^*_{pk}\cdot \left(\Gamma^U_\alpha\right) \cdot \phi_{pk}\right)^2},
\label{initial}
\end{equation}
  where $p=1,2$, $k=1,2,\ldots, n$ and $z_{pk}$ can be viewed as the diagonal entries of
  \begin{equation}
  z :=\text{diag}(\overbrace{x,\cdots,x}^n,\overbrace{y,\cdots,y}^n)
  \end{equation}
  defined in the replica space.
 Next we apply the Fierz transformation
  \begin{equation}
  \left(\Gamma_\alpha\right)_{ij} \left(\Gamma_\alpha\right)_{kl} = 2^{-N}\sum_\mu \Tr(\Gamma_\alpha X_\mu\Gamma_\alpha X_\mu)
  \left(X_{\mu}\right)_{il} \left(X_\mu\right)_{kj},
  \label{Fierz}
  \end{equation}
  where the $X_\mu$'s are defined in equation~\eqref{eqn:matrixCompleteness}. Restricting this equation to the left-upper block, we get
  \begin{equation}\label{eqn:fierzUpperhalf}
   \left(\Gamma^U_\alpha\right)_{ij} \left(\Gamma^U_\alpha\right)_{kl} = 2^{1-N}\sum_{\mu, |\mu| \text{ even}}\Tr(\Gamma^U_\alpha X^U_\mu\Gamma^U_\alpha X^U_\mu)
\left( X^U_{\mu}\right)_{il} \left( X^U_\mu\right)_{kj}.
  \end{equation}
We summed over $\mu$'s with even string length because only a product of even number of Dirac matrices has a non-zero left-upper block, and we   denote the left-upper block of $X_\mu$ by ${X}^U_{\mu}$.  We also used that
\begin{equation}
\Tr(\Gamma_\alpha X_\mu\Gamma_\alpha X_\mu)=2\Tr\left(\Gamma^U_\alpha X^U_\mu\Gamma^U_\alpha X^U_\mu\right).
\end{equation}
  We can further reduce the number of generators in equation~\eqref{eqn:fierzUpperhalf} by half by noticing that generators of $X_\mu$ with $|\mu|>N/2$ are related to $|\mu|<N/2$ one-to-one by multiplication of $\gamma_c$, and half of $X_\mu$'s with $|\mu|=N/2$ is related to the other half in the same way. On the other hand $\gamma_c$ is the identity matrix when restricted to upper block, so within the upper block the above-mentioned pairs are identical. This allows us to reduce the number of generators to $2^{N-2}=D^2$, and
   \begin{equation}\label{eqn:fierzUpperhalfFinal}
   \left(\Gamma^U_\alpha\right)_{ij} \left(\Gamma^U_\alpha\right)_{kl} = \frac{1}{D^2}\sum_\mu{}^{'}\Tr(\Gamma^U_\alpha X^U_\mu\Gamma^U_\alpha X^U_\mu)
\left( X^U_{\mu}\right)_{il} \left( X^U_\mu\right)_{kj},
  \end{equation}
where $\sum{}^{'}_\mu$ denotes the sum over $\mu$'s with even $|\mu|$ and $|\mu|<N/2$ if $N/2$ is odd, and also over half of $\mu$'s with $|\mu|=N/2$ if $N/2$ is even. After the Fierz transformation we  thus obtain the generating function
\begin{equation}
Z(x,y) =   \int \mathcal{D}\phi\mathcal{D}\phi^* e^{i\sum_{p,k} \eta_p z_{pk}\phi_{pk}^*\cdot\phi_{pk}
  -\frac{\sigma^2}2  \frac 1D\sum_{pkql} \eta_p\eta_q\sum{}^{'}_\mu T_{\mu}\phi^*_{pk}\cdot X^U_\mu \cdot \phi_{ql}
  \phi^*_{ql}\cdot X^U_\mu\cdot \phi_{pk}},
\label{after-hub}
   \end{equation}
    where $\sigma^2 =\binom{N}{q}v^2$ is the multi-particle variance and
   \begin{equation}
   T_\mu = \frac 1D \binom{N}{q}^{-1}\sum_\alpha  \Tr (\Gamma^U_\alpha X^U_\mu \Gamma^U_\alpha X^U_\mu).
   \end{equation}

   Now we use that
   \begin{equation}
   \int d a_\mu e^{-\frac 1{2\sigma^2  T_\mu}(a_\mu - i\sigma^2 T_\mu \frac 1{\sqrt {D}} \phi^*\cdot X^U_\mu \cdot \phi)^2}
   = {\rm constant }.
   \end{equation}
   This results in
\begin{equation}
Z(x,y) =   \int \mathcal{D}a_\mu\mathcal{D}\phi\mathcal{D}\phi^* e^{i \sum_{pk}\eta_p z_{pk}\phi_{pk}^*\cdot\phi_{pk}}
  e^{-\sum{}^{'}_\mu \frac 1{2\sigma^2 T_\mu} \tr a_\mu^2  + \sum_{pkql}\eta_p\eta_q\sum{}^{'}_\mu a_\mu^{pk;ql}i\frac{1}{\sqrt D}
    \phi_{pk}^* \cdot X^U_\mu \cdot \phi_{ql}}.
   \end{equation}
   We can now perform the Gaussian integral over $\phi$ and $\phi^*$ resulting in the partition function
\begin{equation}
Z(x,y) =   \int \mathcal{D}a_\mu
e^{-\sum{}^{'}_\mu \frac 1{2\sigma^2 T_\mu} \tr a_\mu^2  - \Tr \log(z +\sum{}^{'}_\mu \frac{1}{\sqrt D}a_\mu X^U_\mu)}.
\label{ab-gen}
  \end{equation}
 A rescaling of the integration variable $a_\mu \to \sigma a_\mu$ gives equation~\eqref{twogen}. This is the partition function obtained by Altland and Bagrets~\cite{Altland-ml-2017eao}.

  Note that the $\mu=0$ term in this sum is exactly the GUE result. This derivation raises an important concern.
  The $\mu=0$ term in~(\ref{after-hub}) consists of $4n^2 D^2{N\choose q}$
  terms of the form    $\phi^{i*}_{pk}  \phi_{ql}^i \phi^{j*}_{ql} \phi_{pk}^j$
each with weight $\sigma^2/2D$.
  However, the expression before the Fierz transformation~(\ref{initial}) has only
  $ 4n^2 D {N/2 \choose q}$  such terms from the diagonal $\Gamma^\alpha$ (there are ${N/2 \choose q} $ of them)
and $4n^2 D \left[{N\choose q} -{N/2\choose q}\right] $ such terms from the off-diagonal $\Gamma^\alpha$,
each with weight $\sigma^2/2$. Note that all $\Gamma^\alpha$ have only one nonzero matrix element in each
column and in each row in the representation we are working in.
So in total we have $4 n^2 D {N \choose q}$ terms of this form in equation~(\ref{initial})
  which is an exponentially smaller number than in equation~(\ref{after-hub}), and most terms of the form $\phi^{i*}_{pk}  \phi_{ql}^i \phi^{j*}_{ql} \phi_{pk}^j$ in~(\ref{after-hub})
   are actually canceled by the other terms obtained after the Fierz transformation. Indeed, the
  dominance of the GUE contribution has to break down for $q=2$, where the eigenvalues are uncorrelated. Yet, there
  is no qualitative difference between the $\sigma$-model for $q=2$ and $q=4$.

  For $N \mod 8 =0$ the $\gamma$ matrices are real and the spectral correlations are in the universality class of the Gaussian Orthogonal Ensemble (GOE)\@. In this case, the term in~(\ref{initial})
  \begin{equation}
  \phi^*_{pk}\cdot \left(\Gamma^U_\alpha\right) \cdot \phi_{pk}
  \end{equation}
  can be written as
  \begin{equation}
\frac 12 \left (  \phi^*_{pk}\cdot \left(\Gamma^U_\alpha\right) \cdot \phi_{pk}+\phi_{pk}\cdot \left(\Gamma^U_\alpha\right) \cdot \phi^*_{pk}\right).
  \end{equation}
  After applying the Fierz transformation, we obtain additional terms of the form
  \begin{equation}
\sum_{pkql}\sum{}^{'}_\mu T_{\mu}\phi^*_{pk}\cdot X^U_\mu \cdot \phi^*_{ql}
\phi_{ql}\cdot X^U_\mu\cdot \phi_{pk},
\end{equation}
which is the so-called Cooperon contribution to the GOE result. Note that in order to get a $\beta$-independent spectral
density for the Wigner-Dyson ensemble, we have to scale the variance of the random matrix Hamiltonian as $1/\beta$.
For the SYK model we do
not have such rescaling and we expect an additional $\sqrt \beta$ dependence in the saddle-point equation and the corresponding
resolvent and semi-circular spectral density.

  \section{Some combinatorial identities}
\label{app:one}
  One can easily prove the identity
 \begin{equation}
 \frac 1{D^2} \sum_{|\mu|=0}^N {N\choose |\mu|} T_{\mu}^2 =  4 {N\choose q}^{-1}.
   \end{equation}
For arbitrary $p$ the sum behaves as
\begin{equation}
\frac 1{D^2} \sum_{|\mu|=0}^N {N\choose |\mu|}  T_{\mu}^p \sim N^{-qp/2}.
\end{equation}
If  $\nu$ is summed over with $| \nu |$ fixed we have
\begin{equation}
\frac 1D  \sum_{ \nu , |\nu| {\rm fixed}} \Tr X_{\mu} X_{\nu} X_{\mu} X_{\nu}=
\sum_s (-1)^s {N\choose |\mu|} {N-|\mu| \choose |\nu|-s}{|\mu| \choose s}
    \end{equation}
    Using this we obtain the identity
     \begin{equation}
     \frac 1{D^3}  \sum_{|\mu|=0}^N {N\choose |\mu|} \sum_{|\nu|=0}^N {N\choose |\nu|} \sum_\nu T_\mu^p T_\nu
  \Tr X_\mu X_\nu X_\mu X_\nu = 4 \eta^p,
     \label{ident}
     \end{equation}
     which can also be shown by applying the Fierz transformation to $T_\nu X_\nu X_\nu$.
To obtain the sixth moment from the $\sigma$-model calculation we need the identity
\begin{align}
& \sum_{m_1=0}^N \sum_{m_2=0}^N \sum_{s=0}^{m_1}
             {N \choose m_1} {m_1\choose s} {N-m_1 \choose  m_2-s}(-1)^s
             T_{m_1} T_{m_2} T_{m_1+m_2-2s}
               \nonumber\\
                & =  4 {N \choose q}^{-2} \sum_{k=0}^q\sum_{m=0}^q (-1)^{q-k-m}
                   {N-2k\choose q-m} {2k \choose m} {N-q \choose k}{q\choose k}.
                   \nonumber\\
                  & \equiv   4 T_6.
\end{align}
Here $T_6$ denotes the value of the single-trace diagram with three chords all intersecting with each other, see the footnote near the end of section~\ref{subsec:onepoint}.

\section{Replica limit of the GUE partition function}\label{app:GUEreplicaLimit}
\subsection{One-point function}
\label{app:replicaGUE}
The GUE partition function for $n$ replicas is given by
\begin{equation}
Z(z) = \int d\sigma \frac 1{\det^N(z+\sigma)} e^{-\frac N2 \sigma^2},
\end{equation}
where the integral is over $n\times n$ Hermitian matrices $\sigma$.
Using that the replica limit of the partition function is equal to one, we
find that the resolvent is given by
\begin{equation}
G(z) = -\lim_{n\to 0} \frac 1n\frac 1N \frac d{dz} \log Z(z)
=\lim_{n\to 0} \frac 1n \int d\tilde \sigma   \tr \tilde \sigma
  e^{-\frac N2\tilde \sigma^2- N\tr \log (z+\tilde\sigma)},
  \end{equation}
  where we have expressed the derivative with respect $z$ in terms of a
  derivative with respect to the $\tilde \sigma_{kk}$ and partially integrated
 these variables.
  We evaluate the resolvent in powers of $1/z$ and in powers of $1/N$.
  The saddle point equation is given by
  \begin{equation}
  \tilde \sigma + \frac 1{\tilde \sigma+z} = 0.
  \end{equation}
  Expanding around the physical solution $\bar \sigma$ that asymptotes
  as $1/z$ for large $z$,
  \begin{equation}
  \tilde  \sigma = \bar \sigma +\sigma,
  \end{equation}
and  using the saddle point equation we obtain the expansion
  {\rdmathspace[h]
\begin{align}
  G(z)  & =   \bar G(z) +\lim_{n\to 0} \frac 1n  \int d\sigma \tr \sigma
  e^{-\frac N2 (1-\bar \sigma^2)\sigma^2 + \sum_{k=3}^\infty \frac 1k \tr\bar \sigma^k \sigma^k}
\nonumber\\
 & = \bar G(z) +
\lim_{n\to 0}\frac 1n\!\left (\!
    \frac N3\left \langle \tr \sigma \tr (\bar \sigma\sigma)^3 \right \rangle
  +  \frac N5\left \langle\tr \sigma  \tr (\bar \sigma \sigma)^5 \right \rangle
  +  \frac N7\left \langle \tr \sigma \tr (\bar \sigma \sigma)^7 \right \rangle
  +  \frac N9\left \langle \tr \sigma \tr (\bar \sigma \sigma)^9 \right \rangle
   \right . \nonumber \\
& \quad
  +  \frac {N^2}{12}\left \langle \tr \sigma \tr (\bar \sigma \sigma)^3 \tr (\bar \sigma \sigma)^4 \right \rangle
  +  \frac {N^2}{18}\left \langle \tr \sigma \tr (\bar \sigma \sigma)^3 \tr (\bar \sigma \sigma)^6 \right \rangle
  +  \frac {N^2}{20}\left \langle \tr \sigma \tr (\bar \sigma \sigma)^4 \tr (\bar \sigma \sigma)^5 \right \rangle\nonumber \\
& \quad
\left .   +  \frac{N^3}{162}\left \langle \tr \sigma (\tr (\bar \sigma \sigma)^3)^3  \right \rangle \right )+ \cdots,
\end{align}}\relax
where the expectation values are given by the sum over all corresponding Wick contractions.
The terms with an odd number of factors of $\sigma$ vanish and we have not included them in the above.
One can easily show that the following terms vanish in the replica limit
\begin{align}
 \lim_{n\to 0}\frac 1n \left \langle \tr \sigma\tr \sigma^3 \right \rangle & =0, \nonumber\\
  \lim_{n\to 0}\frac 1n   \left \langle \tr \sigma\tr \sigma^7 \right \rangle & =0, \nonumber\\
  \lim_{n\to 0}\frac 1n   \left \langle \tr \sigma \tr \sigma^3\tr \sigma^6 \right \rangle & =0, \nonumber\\
  \lim_{n\to 0}\frac 1n   \left \langle\tr \sigma  \tr \sigma^4\tr \sigma^5 \right \rangle & =0.
\end{align}
This leaves us with the following nonvanishing contributions
\begin{align}
A_1  & = \lim_{n\to 0}\frac 1n  \frac N5 \left \langle \tr \sigma \tr (\bar \sigma\sigma)^5 \right \rangle =
\frac 1{N^2} \frac {\bar\sigma^5}{(1-\bar\sigma^2)^3} = \frac 1{N^2}\left [\frac 1{z^5}+
8 \frac 1{z^7} + 47 \frac 1{z^9}+\cdots\right ],\nonumber\\
A_2 & = \lim_{n\to 0}\frac 1n \frac N9 \left \langle\tr \sigma  \tr (\bar \sigma \sigma)^9 \right \rangle =
\frac {21}{N^4} \frac {\bar\sigma^9}{(1-\bar\sigma^2)^5} = \frac {21}{N^4} \frac 1{z^9},\nonumber\\
A_3 & = \lim_{n\to 0}\frac 1n\frac {N^2}{12} \left \langle \tr\sigma \tr( \bar \sigma^3 \sigma)^3 \Tr (\bar \sigma \sigma)^4 \right\rangle = \frac 1{N^2} \frac {\bar\sigma^7}{(1-\bar\sigma^2)^4} = \frac 1{N^2}\left [\frac 1{z^7}+
11 \frac 1{z^9} +\cdots\right ],\nonumber\\
A_4 & = \lim_{n\to 0}\frac 1n\frac {N^2}{12}  \left \langle \tr \sigma \tr (\bar \sigma\sigma)^3 \Tr (\bar \sigma \sigma)^4 \right\rangle = \frac 1{N^2} \frac {\bar\sigma^7}{(1-\bar\sigma^2)^3} = \frac 1{N^2}\left [\frac 1{z^7}+
11 \frac 1{z^9} +\cdots\right ],\nonumber\\
A_6 & = \lim_{n\to 0}\frac 1n\frac {N^3}{162} \left \langle \tr   \sigma (\tr (\bar \sigma \sigma)^3)^3\right\rangle
= \frac 1{N^2} \frac {\bar\sigma^9}{(1-\bar\sigma^2)^5} = \frac 1{N^2}\left [
 \frac 1{z^9} +\cdots\right ].
\label{gue-one}
\end{align}
A few comments on the evaluation of the contractions are in order.
In total $9\times 105$ different diagrams contribute to $\langle\tr \sigma \tr \sigma^9\rangle$
but only $9\times 21$ of them do not vanish in the replica limit. The expression for $A_3$ and $A_4$ correspond to two different
contraction patterns, namely,
 \begin{equation}
\contraction[1ex]{\tr} {\sigma}{\tr}{\sigma}
\contraction[2ex]{\tr \sigma \tr \sigma}{\sigma}{\sigma \tr }{\sigma}
\contraction[3ex]{\tr \sigma \tr \sigma\sigma}{\sigma}{\tr \sigma\sigma}{\sigma}
\contraction[4ex]{\tr \sigma \tr \sigma\sigma\sigma \tr \sigma}{\sigma}{\sigma}{\sigma}
\tr \sigma \tr \sigma\sigma\sigma \tr \sigma\sigma\sigma\sigma
\end{equation}
and
\begin{equation}
\contraction[1ex]{\tr}{ \sigma} {\tr \sigma\sigma\sigma \tr}{\sigma}
\contraction[2ex]{\tr \sigma \tr}{\sigma} {\sigma\sigma \tr \sigma}{\sigma}
  \contraction[3ex]{\tr \sigma \tr \sigma}{\sigma} {\sigma\tr \sigma\sigma}{\sigma}
    \contraction[4ex]{\tr \sigma \tr \sigma\sigma}{\sigma}{\tr \sigma\sigma\sigma}{\sigma}
\tr \sigma \tr \sigma\sigma\sigma \tr \sigma\sigma\sigma\sigma.
\end{equation}

This result~(\ref{gue-one}) for the moments is in agreement with the general
formula obtained by Mehta~\cite{Mehta-2004} (see also~\cite{Witte-ml-2013cea})
\begin{equation}
M_{2p} = (2p-1)!! \sum_{j=0}^p{p\choose j}{N\choose j+1} 2^j,
\end{equation}
by comparing to $M_{2p}/(N M_2^p)$.

\subsection{Two-point function}
\label{app:replicaGUETwoPoint}
The generating function for the two-point function of the GUE is given by
\begin{equation}
Z(z) = \int d\sigma \frac 1{\det^N(z+\sigma)} e^{-\frac N2 \sigma^2},
\end{equation}
with $\sigma$ a Hermitian $2n\times 2n$ matrix and $z=\underbrace{(x,\cdots,x}_n,\underbrace{y,\cdots,y}_n)$.
The corresponding  blocks of the $\sigma$ matrix are referred to as the 11 block, the 12 block, the 21
block and the 22 block.
For $x=y$ and both with the same infinitesimal increment,
this becomes the generating function for the one-point function but now with $2n$ replicas. Note that
for the two-point function, $x$ and $y$ have an opposite infinitesimal imaginary part, while for the one-point
function, the imaginary parts of all $z$ have the same sign.

The two-point correlation function is given by
\begin{align}
C(x,y)  & =  \left . \lim_{n\to 0}\frac 1{n^2N^2} \frac d{dx}\frac d{dy} \log Z(z)\right |_{\rm connected},\nonumber \\
 & =  \lim_{n\to 0}\frac 1{n^2}\int d\sigma
\tr  P_{11}\sigma \tr P_{22} \sigma \frac 1{\det^N(z+\sigma)} e^{-\frac N2 \sigma^2}
\label{cor}
\end{align}
with $P_{kk}$ the projection on the $kk$ block.
The differentiation gives other contributions but they do not contribute to the connected
two-point function. We evaluate the integral by a saddle point approximation.
The saddle point equation is given by
\begin{equation}
\frac 1{z+\sigma} + \sigma = 0.
\end{equation}
In a 12 block notation, the solution with $\sigma_{12} =0$ has the replica-diagonal form
\begin{equation}
\bar \sigma = \left (\begin{array}{cc} \bar \sigma (x) & 0 \\0& \bar \sigma(y) \end{array} \right ).
\end{equation}
The propagators and the vertices follow from the expansion of the logarithm
\begin{equation}
-\tr \log (z+\sigma) = \sum_{k=1}^\infty \frac 1k (\bar \sigma \sigma)^k.
\end{equation}
This results in the following quadratic part of the action
\begin{equation}
-\frac N2 \left [ (1-\bar \sigma(x)^2 ) \tr\sigma_{11}^2   +(1-\bar \sigma(y)^2 ) \Tr\sigma_{22}^2
  + 2(1-\bar \sigma(x)\bar\sigma(y) ) \tr\sigma_{12}\sigma_{21}   \right ].
\end{equation}

To leading order in $1/N^2$ two diagrams contribute to the two-point function:
\begin{equation}
\lim_{n\to 0} \frac 1{n^2N^2} \left \langle \tr P_{11} \sigma \tr P_{22} \sigma \frac {1}{4} \tr (\bar\sigma \sigma)^4  \right \rangle,
\end{equation}
and
\begin{equation}
\lim_{n\to 0} \frac 1{n^2N^2} \left \langle \tr P_{11} \sigma \tr P_{22} \sigma \frac {1}{18} \tr^2 (\bar\sigma \sigma)^3  \right \rangle.
\end{equation}
The connected part of the first diagram is given by
\begin{align}
&\lim_{n\to 0} \frac 1{n^2N^2}\left \langle \tr \sigma_{11} \tr \sigma_{22} \tr \sigma_{11} \sigma_{12}\sigma_{22}\sigma_{21}\right\rangle_c
\nonumber\\
 & = \frac 1{N^2}\frac{(\bar \sigma(x) \bar \sigma(y))^2}{(1-\bar\sigma(x) \bar\sigma(y))(1-\bar \sigma^2(x))(1-\bar \sigma^2(x))}
\nonumber\\
 & =  \frac 1{N^2} \left( \frac 1{x^2y^2}+\frac 1{x^3y^3}+\cdots \right ).
\end{align}
The connect part of the second diagram can be evaluated as
\begin{align}
&\lim_{n\to 0} \frac 1{n^2N^2}\left \langle \tr \sigma_{11} \tr \sigma_{22} \tr \sigma_{11} \sigma_{12}\sigma_{21} \tr \sigma_{22}\sigma_{21}\sigma_{12}\right \rangle_c
\nonumber\\
 & = \frac 1{N^2}\frac{(\bar \sigma(x) \bar \sigma(y))^3}{(1-\bar\sigma(x) \bar\sigma(y))^2(1-\bar \sigma^2(x))(1-\bar  \sigma^2(x))}\nonumber\\
 & =  \frac 1{N^2}\left(  \frac  1{x^3y^3} +\cdots\right ).
\end{align}
The sum of the two contributions is equal to
\begin{align}
 & = \frac 1{N^2}\frac{(\bar \sigma(x) \bar \sigma(y))^2}{(1-\bar\sigma(x) \bar\sigma(y))^2(1-\bar \sigma^2(x))(1-\bar  \sigma^2(x))}\nonumber\\
 & =  \frac 1{N^2}\left(  \frac  1{x^2y^2}+\frac  2{x^3y^3} +\cdots\right ),
\end{align}
which gives the correct result for two-point correlator to order $1/N^2$ (see~\cite{Verbaarschot1984}) and
the $M_{1,1}$ and $M_{2,2}$ moments. Note that in the normalization of this appendix,
$\bar\sigma(x) \bar\sigma(y)\to 1 $ for $x\to y$, so that the two-point function behaves as
$1/(N^2(x-y)^2)$ in this limit.

\section{Illustration of~(\ref{eqn:generalDescriptionStarts})--(\ref{eqn:generalDescriptionEnds}) by 3-cross-linked examples}\label{app:mom3}

To illustrate the general prescription for the evaluation double-trace diagrams (see section \ref{sec:dtrAndChord})
we work out two double-trace diagrams with three cross-links in this appendix.

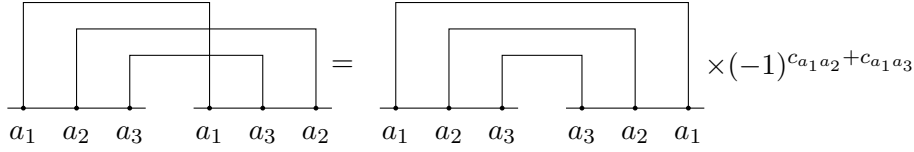
\begin{figure}
\centering

\begin{tikzpicture}[scale=0.7]
\draw[fill=black] (0,0) circle (1pt);
\draw[fill=black] (1,0) circle (1pt);
\draw[fill=black] (2,0) circle (1pt);
\draw[fill=black] (3.5,0) circle (1pt);
\draw[fill=black] (4.5,0) circle (1pt);
\draw[fill=black] (5.5,0) circle (1pt);

\draw[] (0,0)--(0,2)--(3.5,2)--(3.5,0);
\draw[] (1,0)--(1,1.5)--(5.5,1.5)--(5.5,0);
\draw[] (2,0)--(2,1)--(4.5,1)--(4.5,0);

\node at (0,-0.5) {$a_1$};
\node at (1,-0.5) {$a_2$};
\node at (2,-0.5) {$a_3$};
\node at (3.5,-0.5) {$a_1$};
\node at (4.5,-0.5) {$a_3$};
\node at (5.5,-0.5) {$a_2$};
\draw (-0.3,0)--(2.3,0);
\draw (3.2,0)--(5.8,0);
\node at (6,0.8) {$=$};

\draw[fill=black] (7,0) circle (1pt);
\draw[fill=black] (8,0) circle (1pt);
\draw[fill=black] (9,0) circle (1pt);
\draw[fill=black] (10.5,0) circle (1pt);
\draw[fill=black] (11.5,0) circle (1pt);
\draw[fill=black] (12.5,0) circle (1pt);

\draw[] (7,0)--(7,2)--(12.5,2)--(12.5,0);
\draw[] (8,0)--(8,1.5)--(11.5,1.5)--(11.5,0);
\draw[] (9,0)--(9,1)--(10.5,1)--(10.5,0);
\draw (6.7,0)--(9.3,0);
\draw (10.2,0)--(12.8,0);
\node at (14.8,0.8) {$\times (-1)^{c_{a_1a_2}+c_{a_1a_3}}$};
\node at (7,-0.5) {$a_1$};
\node at (8,-0.5) {$a_2$};
\node at (9,-0.5) {$a_3$};
\node at (10.5,-0.5) {$a_3$};
\node at (11.5,-0.5) {$a_2$};
\node at (12.5,-0.5) {$a_1$};

\end{tikzpicture}
\caption{Shuffling a double trace to its canonical ordering (subscripts not summed over).\label{fig:shuffleOrder}}
\end{figure}

The starting point is the fact that $2^{-N/2}\Tr (\Gamma_{a_1}\Gamma_{a_2}\ldots\Gamma_{a_m})$ can only be $0$ or $\pm 1$. This implies
\begin{equation}
  \label{eqn:canonicalOrder}
2^{-N}\Tr (\Gamma_{a_1}\Gamma_{a_2}\ldots\Gamma_{a_m})\Tr (\Gamma_{a_m}\Gamma_{a_{m-1}}\ldots\Gamma_{a_1})=2^{-N}|\Tr (\Gamma_{a_1}\Gamma_{a_2}\ldots\Gamma_{a_m})|^2 = 0 \text{ or } 1.
\end{equation}
This particular simplicity motivates us to shuffle the $\Gamma$'s to the above ``canonical'' ordering. The shuffling will introduce phase factors due to the relation
\begin{equation}
\Gamma_\alpha \Gamma_\beta = (-1)^{q+c_{\alpha\beta}}\Gamma_\beta\Gamma_\alpha,
\end{equation}
where $c_{\alpha\beta}=|\alpha \cap \beta|$ is the number of common elements in sets $\alpha$ and $\beta$. For example let us consider the contraction $\Tr(\Gamma_{a_1}\Gamma_{a_2}\Gamma_{a_3})\Tr(\Gamma_{a_1}\Gamma_{a_3}\Gamma_{a_2})$, the shuffling (see figure~\ref{fig:shuffleOrder}) will introduce a phase factor of $(-1)^{c_{a_1 a_2}+c_{a_1a_3}}$, and hence
\begin{equation}\label{eqn:threeGamma}
\sum_{a_1,a_2,a_3}\Tr(\Gamma_{a_1}\Gamma_{a_2}\Gamma_{a_3})\Tr(\Gamma_{a_1}\Gamma_{a_3}\Gamma_{a_2})=\sum_{a_1,a_2,a_3} (-1)^{c_{a_1 a_2}+c_{a_1a_3}} |\Tr(\Gamma_{a_1}\Gamma_{a_2}\Gamma_{a_3})|^2.
\end{equation}
In fact in this particular case we can get rid of the phase factor by cyclically permuting the $\Gamma$'s in the second trace, so we must have
\begin{equation*}
\sum_{a_1,a_2,a_3} (-1)^{c_{a_1 a_2}+c_{a_1a_3}} |\Tr(\Gamma_{a_1}\Gamma_{a_2}\Gamma_{a_3})|^2
= \sum_{a_1,a_2,a_3}  |\Tr(\Gamma_{a_1}\Gamma_{a_2}\Gamma_{a_3})|^2.
\end{equation*}
However we want to illustrate a general point beyond this simple example, so we keep the phase factors in our discussion. We see in general each intersection among the chords introduces a phase factor of $(-1)^{q+c_{\alpha\beta}}$. We would still like to get rid of the trace in our equation~\eqref{eqn:threeGamma}  in favor of a purely combinatorial term, so that it can be effectively handled by computers. The key question is, when is $2^{-N}|\Tr(\Gamma_{a_1}\Gamma_{a_2}\Gamma_{a_3})|^2$ equal to $0$ and when is it equal to $1$?

Recall that each $\Gamma$ is a product of $q$ different Dirac matrices. The necessary and sufficient condition for $\Tr(\Gamma_{a_1}\Gamma_{a_2}\Gamma_{a_3})$ to be nonvanishing is that every one of the $N$ Dirac matrices occurs exactly even number (including zero) of times in the totality of  $\Gamma_{a_1}, \Gamma_{a_2}$ and $\Gamma_{a_3}$.

This suggests that a useful perspective will be provided by the $d$-variables: $d_{a_{i_1} a_{i_2}\ldots a_{i_k}}$ is the number of elements common and only common to the sets $a_{i_1}, a_{i_2},\ldots , a_{i_k}$ (naturally, $i_1,\ldots,i_k$ are all different from each other in this definition).
In the case of three index sets ($k=3$), we have the $d$-variables
$\{d_{a_1a_2},d_{a_1a_3},d_{a_2a_3},d_{a_1a_2a_3}\}$. There are also $\{d_{a_1}, d_{a_2}, d_{a_3}\}$ which count the number of subscripts that occur exactly once in only $a_1, a_2$ or $a_3$, but they are not independent variables due to the constraint that each index set has $q$ elements.\footnote{For example see equation~\eqref{eqn:onceExample33}.} Both the $c$-variables ($c_{a_{i_1}a_{i_2}}$) and the $d$-variables have played an important role in calculating single-trace contractions, and the relations between them were discussed in~\cite{Garcia-Garcia-ml-2018kvh}. Here we cite one diagram that makes their relation clear, see figure~\ref{fig:indexVenn}:
\begin{equation}
\begin{split}
c_{a_1 a_2} &= d_{a_1a_2}+d_{a_1a_2a_3};\\
c_{a_1 a_3} &= d_{a_1a_3}+d_{a_1a_2a_3};\\
c_{a_2 a_3} &= d_{a_2a_3}+d_{a_1a_2a_3}.
\end{split}
\end{equation}
In general,
\begin{equation}
c_{a_{i_1}a_{i_2}}= d_{a_{i_1}a_{i_2}}+ \sum_{i_3\notin \{i_1,i_2\}}d_{a_{i_1}a_{i_2}a_{i_3}}+\sum_{i_3,i_4\notin \{i_1,i_2\}}d_{a_{i_1}a_{i_2}a_{i_3}a_{i_4}}+\cdots+ \sum_{i_3,\ldots, i_k\notin \{i_1,i_2\}}d_{a_{i_1}a_{i_2}a_{i_3}\ldots a_{i_k}}.
\label{c-def}
\end{equation}
We come back to the case of $|\Tr(\Gamma_{a_1}\Gamma_{a_2}\Gamma_{a_3})|^2$. Here $d_{a_1a_2}$ counts the number of Dirac matrices of which the subscripts appear exactly in sets $a_1$ and $a_2$, no less and no more. So there the Dirac matrices appear exactly twice in the totality of $\Gamma_{a_1}, \Gamma_{a_2}$ and $\Gamma_{a_3}$.
\begin{figure}
\centering
\resizebox*{1\textwidth}{!}{%
\begin{tikzpicture}
      \begin{scope}
    \clip \secondcircle;
    \fill[red] \thirdcircle;
      \end{scope}
    \draw (-2.5,-2)--(-2.5,3.75)--(4.25,3.75)--(4.25,-2)--(-2.5,-2);
      \draw \firstcircle (-0.25,-0.25) node {$a_1$};
      \draw \secondcircle (2.25,-0.25) node {$a_2$};
      \draw \thirdcircle (0.85,2.25) node  {$a_3$};
\draw (1,-2.5) node {Red region has cardinality $c_{a_2a_3}$};
 \begin{scope}
    \clip \fifthcircle;
    \fill[red] \sixthcircle;
      \end{scope}
      \begin{scope}
      \clip \fourthcircle;
    \clip \fifthcircle;
    \fill[white] \sixthcircle;
      \end{scope}
    \draw (6,-2)--(6,3.75)--(12.75,3.75)--(12.75,-2)--(6,-2);
      \draw \fourthcircle (8.25,-0.25) node {$a_1$};
      \draw \fifthcircle (10.75,-0.25) node {$a_2$};
      \draw \sixthcircle (9.35,2.25) node  {$a_3$};
      \draw (9.5,-2.5) node {Red region has cardinality $d_{a_2a_3}$};
    \end{tikzpicture}}\relax
\caption{Venn diagrams with three index sets. Each index set is represented by a circle, containing $q$ elements. The box is the set of all possible values an index can take, which has cardinality $N$. The box is partitioned into eight regions. (Taken from~\cite{Garcia-Garcia-ml-2018kvh}.)\label{fig:indexVenn}}
\end{figure}\relax
The same can be said about $d_{a_1a_3}$ and $d_{a_2a_3}$. The total number of Dirac matrices that appear exactly twice in the totality of $\Gamma_{a_1}, \Gamma_{a_2}, \Gamma_{a_3}$ is thus
\begin{equation}
d_2 = d_{a_1a_2}+d_{a_1a_3}+d_{a_2a_3}.
\end{equation}

 On the other hand $d_{a_1a_2a_3}$ counts the number of Dirac matrices that appear exactly three times in the totality of $\Gamma_{a_1}, \Gamma_{a_2}, \Gamma_{a_3}$. Hence if $d_{a_1a_2a_3}\neq 0$, $|\Tr(\Gamma_{a_1}\Gamma_{a_2}\Gamma_{a_3})|^2=0$. We also want to know how many Dirac matrices appear exactly once:
\begin{equation}\label{eqn:onceExample33}
\begin{split}
\text{Exactly once in $a_1:$}&\ d_{a_1}=q-d_{a_1a_2}-d_{a_1a_3}-d_{a_1a_2a_3};\\
\text{Exactly once in $a_2:$}&\ d_{a_2}=q-d_{a_1a_2}-d_{a_2a_3}-d_{a_1a_2a_3};\\
\text{Exactly once in $a_3:$}&\ d_{a_3}=q-d_{a_1a_3}-d_{a_2a_3}-d_{a_1a_2a_3}.
\end{split}
\end{equation}
And  $|\Tr(\Gamma_{a_1}\Gamma_{a_2}\Gamma_{a_3})|^2=0$ also if any of the $\{d_{a_1},d_{a_2},d_{a_3}\}$ is nonzero. Synthesizing everything discussed so far, we arrive at the formula
\begin{equation}\label{eqn:example33}
\begin{split}
&2^{-N}\sum_{a_1,a_2,a_3}\Tr(\Gamma_{a_1}\Gamma_{a_2}\Gamma_{a_3})\Tr(\Gamma_{a_1}\Gamma_{a_3}\Gamma_{a_2})\\
&=2^{-N}\sum_{a_1,a_2,a_3} (-1)^{c_{a_1 a_2}+c_{a_1a_3}} |\Tr(\Gamma_{a_1}\Gamma_{a_2}\Gamma_{a_3})|^2\\
&= \sum_{d_{a_1a_2}=0}^q\sum_{d_{a_1a_3}=0}^q\sum_{d_{a_2a_3}=0}^q (-1)^{d_{a_1 a_2}+d_{a_1a_3}} \frac{N!}{(N-3q+d_2)!d_{a_1a_2}!d_{a_1a_3}!d_{a_2a_3}!}\\&\quad \qquad\qquad\qquad\times\delta(q-d_{a_1a_2}-d_{a_1a_3})\delta(q-d_{a_1a_2}-d_{a_2a_3})\delta(q-d_{a_1a_3}-d_{a_2a_3}).
\end{split}
\end{equation}

\enlargethispage*{\baselineskip}\relax

Note for the second equality we replaced the sum over index sets $a_1, a_2, a_3$ by a sum over numbers $d_{a_1a_2}, d_{a_1a_3}, d_{a_2a_3}$. In principle there is also a sum over $d_{a_1a_2a_3}$, but we have argued only the $d_{a_1a_2a_3}=0$ case contributes. It is clear that in this case there is only one scenario where the three Kronecker $\delta$ constraint is satisfied:
\begin{equation}
d_{a_1a_2}= d_{a_1a_3}= d_{a_2a_3}=\frac{q}{2},
\end{equation}
which implies
\begin{equation}
2^{-N}\sum_{a_1,a_2,a_3} (-1)^{c_{a_1 a_2}+c_{a_1a_3}} |\Tr(\Gamma_{a_1}\Gamma_{a_2}\Gamma_{a_3})|^2=
\begin{cases}0 \ \,\quad\qquad\qquad\qquad \text{for odd }q, \\
\frac{N!}{\left(N-\frac{3q}{2}\right)!\left(\left(\frac{q}{2}\right)!\right)^3}\qquad \text{for even }q.
\end{cases}
\end{equation}
In more general cases no such drastic simplification can be expected.

\pagebreak

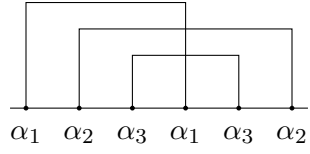
\begin{figure}
\centering
\begin{tikzpicture}[scale=0.7]
\draw[fill=black] (0,0) circle (1pt);
\draw[fill=black] (1,0) circle (1pt);
\draw[fill=black] (2,0) circle (1pt);
\draw[fill=black] (3,0) circle (1pt);
\draw[fill=black] (4,0) circle (1pt);
\draw[fill=black] (5,0) circle (1pt);

\draw[] (0,0)--(0,2)--(3,2)--(3,0);
\draw[] (1,0)--(1,1.5)--(5,1.5)--(5,0);
\draw[] (2,0)--(2,1)--(4,1)--(4,0);

\draw[] (-0.3,0)--(5.3,0);
\node at (0,-0.5) {$\alpha_1$};
\node at (1,-0.5) {$\alpha_2$};
\node at (2,-0.5) {$\alpha_3$};
\node at (3,-0.5) {$\alpha_1$};
\node at (4,-0.5) {$\alpha_3$};
\node at (5,-0.5) {$\alpha_2$};
\end{tikzpicture}
\caption{The single trace chord diagram that has the same intersection structure as that of figure~\ref{fig:shuffleOrder}, note here all chords are attached to a single backbone as opposed to two in figure~\ref{fig:shuffleOrder}.\label{fig:singleTrace33}}
\end{figure}

It is instructive to contrast equation~\eqref{eqn:example33} with the single trace contraction that has the same intersection structure (figure~\ref{fig:singleTrace33}), using the results of~\cite{Garcia-Garcia-ml-2018kvh}:
\begin{align}\label{eqn:singleTrace33}
&2^{-N}\sum_{\alpha_1,\alpha_2,\alpha_3}\Tr(\Gamma_{\alpha_1}\Gamma_{\alpha_2}\Gamma_{\alpha_3}\Gamma_{\alpha_1}\Gamma_{\alpha_3}\Gamma_{\alpha_2})\nonumber\\
&=\sum_{\alpha_1,\alpha_2,\alpha_3} (-1)^{c_{\alpha_1 \alpha_2}+c_{\alpha_1\alpha_3}}\nonumber\\
&= \sum_{d_{\alpha_1\alpha_2\alpha_3}}\sum_{d_{\alpha_1\alpha_2}}\sum_{d_{\alpha_1\alpha_3}}\sum_{d_{\alpha_2\alpha_3}} (-1)^{d_{\alpha_1 \alpha_2}+d_{\alpha_1\alpha_3}} \frac{N!}{(N-3q+d_2+2d_{\alpha_1\alpha_2\alpha_3})!d_{\alpha_1\alpha_2}!d_{\alpha_1\alpha_3}!d_{\alpha_2\alpha_3}!}\nonumber\\&\quad \qquad\qquad\times\frac{1}{d_{\alpha_1\alpha_2\alpha_3}!}\frac{1}{(q-d_{\alpha_1\alpha_2}-d_{\alpha_1\alpha_3}-d_{\alpha_1\alpha_2\alpha_3})!}\frac{1}{(q-d_{\alpha_1\alpha_2}-d_{\alpha_2\alpha_3}-d_{\alpha_1\alpha_2\alpha_3})!}\nonumber\\
&\quad \qquad\qquad\times\frac{1}{(q-d_{\alpha_1\alpha_3}-d_{\alpha_2\alpha_3}-d_{\alpha_1\alpha_2\alpha_3})!}.
\end{align}
How about cases with both single-trace links and cross links? Let us consider the example in figure~\ref{fig:mixed331}, we would need to calculate
\begin{equation}\label{eqn:mixed331}
\begin{split}
&2^{-N}\sum_{a_1,a_2,a_3,\beta}\Tr(\Gamma_{a_1}\Gamma_{a_2}\Gamma_{a_3})\Tr(\Gamma_{a_1}\Gamma_{a_3}\Gamma_\beta\Gamma_{a_2}\Gamma_\beta)\\
&=2^{-N}\sum_{a_1,a_2,a_3,\beta} (-1)^{q+c_{a_1 a_2}+c_{a_1a_3}+c_{a_2\beta}} |\Tr(\Gamma_{a_1}\Gamma_{a_2}\Gamma_{a_3})|^2.
\end{split}
\end{equation}
Now the $d$-variables are
\begin{equation}\label{eqn:mixed331dvars}
\{d_{a_1a_2}, d_{a_1a_3}, d_{a_2a_3}, d_{a_1\beta},d_{a_2\beta}, d_{a_3\beta}, d_{a_1a_2a_3},d_{a_1a_2\beta}, d_{a_1a_3\beta}, d_{a_2a_3\beta}, d_{a_1a_2a_3\beta}\}.
\end{equation}
However the constraint of the sum on the right-hand side is still given by the Kronecker deltas arising from $|\Tr(\Gamma_{a_1}\Gamma_{a_2}\Gamma_{a_3})|^2$, just as in equation~\eqref{eqn:example33}, so we may say similar things about the conditions on the contributing summands:
\begin{figure}
\centering
\begin{tikzpicture}[scale=0.7]
\draw[fill=black] (0,0) circle (1pt);
\draw[fill=black] (1,0) circle (1pt);
\draw[fill=black] (2,0) circle (1pt);
\draw[fill=black] (3.5,0) circle (1pt);
\draw[fill=black] (4.5,0) circle (1pt);
\draw[fill=black] (5.5,0) circle (1pt);
\draw[fill=black] (6.5,0) circle (1pt);
\draw[fill=black] (7.5,0) circle (1pt);

\draw[] (0,0)--(0,2)--(3.5,2)--(3.5,0);
\draw[] (1,0)--(1,1.5)--(6.5,1.5)--(6.5,0);
\draw[] (2,0)--(2,1)--(4.5,1)--(4.5,0);
\draw[] (5.5,0)--(5.5,1)--(7.5,1)--(7.5,0);
\node at (0,-0.5) {$a_1$};
\node at (1,-0.5) {$a_2$};
\node at (2,-0.5) {$a_3$};
\node at (3.5,-0.5) {$a_1$};
\node at (4.5,-0.5) {$a_3$};
\node at (5.5,-0.5) {$\beta$};
\node at (6.5,-0.5) {$a_2$};
\node at (7.5,-0.5) {$\beta$};

\draw[] (-0.3,0)--(2.3,0);
\draw[] (3.2,0)--(7.8,0);
\end{tikzpicture}
\caption{A chord diagram with three cross links and one single-trace link.\label{fig:mixed331}}
\end{figure}
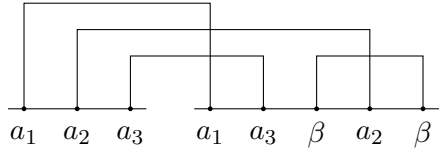\relax
the traces with  Dirac matrices whose subscripts appear odd number of times in the totality of $a_1, a_2, a_3$ are vanishing. This means
\[d_{a_1\beta},d_{a_2\beta}, d_{a_3\beta}, d_{a_1a_2a_3}, d_{a_1a_2a_3\beta}\]
must all be zero. So the list~\eqref{eqn:mixed331dvars} is shortened to
\begin{equation}
\{d_{a_1a_2}, d_{a_1a_3}, d_{a_2a_3}, d_{a_1a_2\beta}, d_{a_1a_3\beta}, d_{a_2a_3\beta}\}.
\end{equation}
We still need to take into account the Dirac matrices whose subscripts only appear once exclusively in either $a_1$, $a_2$ or $a_3$, and they are counted respectively by
\begin{align}
\begin{split}
  d_{a_1}& =q-d_{a_1a_2}-d_{a_1a_3}-d_{a_1a_2\beta}-d_{a_1a_3\beta}, \\
  d_{a_2}& =q-d_{a_1a_2}-d_{a_2a_3}-d_{a_1a_2\beta}-d_{a_2a_3\beta},\\
  d_{a_3}& =q-d_{a_1a_3}-d_{a_2a_3}-d_{a_2a_3\beta}-d_{a_1a_3\beta}.
 \end{split}
\end{align}
So these must also be zero. We also have
\begin{equation}
\begin{split}
c_{a_1a_2} &= d_{a_1a_2}+d_{a_1a_2\beta},\\
c_{a_1a_3} &= d_{a_1a_3}+d_{a_1a_3\beta},\\
c_{a_2\beta} &= d_{a_1a_2\beta}+d_{a_2a_3\beta}.
\end{split}
\end{equation}
We conclude then that equation~\eqref{eqn:mixed331} is equal to
\begin{equation}\label{eqn:example331combinatorics1}
\begin{split}
&\sum_{d_{a_1a_2}}\sum_{d_{a_1a_3}}\sum_{d_{a_2a_3}}\sum_{d_{a_1a_2\beta}}\sum_{ d_{a_1a_3\beta}}\sum_{d_{a_2a_3\beta}}(-1)^{q+d_{a_1a_2}+d_{a_1a_3}+d_{a_1a_3\beta}+d_{a_2a_3\beta}}\\
&\qquad\qquad \times \frac{N!}{(N-4q+d_2+2d_3)!}\frac{1}{d_{a_1a_2}! d_{a_1a_3}! d_{a_2a_3}! d_{a_1a_2\beta}! d_{a_1a_3\beta}! d_{a_2a_3\beta}!}\\
&\qquad\qquad \times \delta(q-d_{a_1a_2}-d_{a_1a_3}-d_{a_1a_2\beta}-d_{a_1a_3\beta})\\
&\qquad\qquad \times\delta(q-d_{a_1a_2}-d_{a_2a_3}-d_{a_1a_2\beta}-d_{a_2a_3\beta})\\
&\qquad\qquad\times \delta(q-d_{a_1a_3}-d_{a_2a_3}-d_{a_2a_3\beta}-d_{a_1a_3\beta}),
\end{split}
\end{equation}
where
\begin{equation}\label{eqn:example331combinatorics2}
\begin{split}
d_2 &= d_{a_1a_2}+d_{a_1a_3}+d_{a_2a_3},\\ d_3 &= d_{a_1a_2\beta}+ d_{a_1a_3\beta}+ d_{a_2a_3\beta}.
\end{split}
\end{equation}
These examples enable us to see how the general situation can be handled, which is summarized in equations~\eqref{eqn:generalDescriptionStarts}--\eqref{eqn:generalDescriptionEnds}.

\section{Low-order double-trace moments}\label{sec:lowOrderMoments}
Using the properties of double traces discussed so far, we can sort out\footnote{The sorting is mostly based on the dihedral action property~\ref{item:dihedral}  discussed in section~\ref{sec:properties}. There are a few groups of diagrams which are not related by the dihedral action but they have the same intersection structures, hence the same~values.} the following few nontrivial low-order double-trace moments:
\begin{align}
\tilde{M}_{3,3} &= 3\left(1+(-1)^{q(q-1)/2}\right)t_{123|321},\\
\tilde{M}_{4,4} &= \tilde M_4^2+8\binom{N}{q}^{-1}\tilde M_4^2+8t_{1234|4321}+16t_{1234|3421},\\
\tilde{M}_{3,5} &= 15(1+\eta)\left(1+(-1)^{q(q-1)/2}\right)t_{123|321},\\
\tilde M_{4,6} &= \tilde M_4 \tilde M_6+12\binom{N}{q}^{-1}\tilde M_4\tilde M_6+6(1+\eta)\left(8t_{1234|4321}+16t_{1234|3421}\right)\nonumber \\
& \quad +24\left(t_{1234|43\alpha21\alpha}+t_{1234|34\alpha21\alpha}+t_{1234|42\alpha31\alpha}\right),\\
\tilde M_{5,5} &= 75\left(1+(-1)^{q(q-1)/2}\right)(1+\eta)^2t_{123|321}+5\left(1+(-1)^{q(q-1)/2}\right)\nonumber \\
& \quad \qquad\quad\times\left(t_{12345|54321}+5t_{12345|45321}+5t_{12345|35421}+t_{12345|42531}\right),\\
\tilde M_{6,6} &= \tilde M_6^2+18\binom{N}{q}^{-1}\tilde M_6^2+36(1+\eta)^2\left(8t_{1234|4321}+16t_{1234|3421}\right)\nonumber \\
& \quad +36\times8(1+\eta)\left(t_{1234|43\alpha21\alpha}+t_{1234|34\alpha21\alpha}+t_{1234|42\alpha31\alpha}\right)\nonumber \\
& \quad +36\left(t_{\alpha12\alpha34|43\beta21\beta}+t_{\alpha12\alpha34|34\beta21\beta}+t_{\alpha12\alpha34|42\beta31\beta}\right.\nonumber \\
& \quad \qquad\qquad\qquad\qquad\quad\left.+2t_{\alpha12\alpha34|24\beta31\beta}+t_{\alpha12\alpha34| 23\beta41\beta}\right)\nonumber \\
& \quad +12\left(t_{123456|654321}+6t_{123456|564321}+12t_{123456|465321}\right.\nonumber \\
& \quad \left. \qquad\qquad+3t_{123456|456321}+9t_{123456|563421}+12t_{123456|536421}\right.\nonumber \\
& \quad \left.\qquad\qquad\qquad+12t_{123456|356421}+2t_{123456|436521}+3t_{123456|462531}\right).
\end{align}
The $t$-variables with subscripts denote chord diagrams, whose meaning should be evident from the two examples in figure~\ref{fig:tvarDef}.
\begin{figure}
\centering
\begin{tikzpicture}[scale=0.7]
\draw[fill=black] (1,0) circle (1pt);
\draw[fill=black] (2,0) circle (1pt);
\draw[fill=black] (3,0) circle (1pt);
\draw[fill=black] (4,0) circle (1pt);
\draw[fill=black] (5.5,0) circle (1pt);
\draw[fill=black] (6.5,0) circle (1pt);
\draw[fill=black] (7.5,0) circle (1pt);
\draw[fill=black] (8.5,0) circle (1pt);
\node at (1,-.5) {$1$};
\node at (2,-.5) {$2$};
\node at (3,-.5) {$3$};
\node at (4,-.5) {$4$};
\node at (5.5,-.5) {$3$};
\node at (6.5,-.5) {$4$};
\node at (7.5,-.5) {$2$};
\node at (8.5,-.5) {$1$};
\node at (4.75,-1.5) {$t_{1234|3421}$};
\draw[] (0.7,0)--(4.3,0);
\draw[] (5.2,0)--(8.8,0);
\draw[] (1,0) -- (1,2)--(8.5,2)--(8.5,0);
\draw[] (2,0)--(2,1.5)--(7.5,1.5)--(7.5,0);
\draw[] (3,0)--(3,1)--(5.5,1)--(5.5,0);
\draw[] (4,0)--(4,0.5)--(6.5,0.5)--(6.5,0);

\end{tikzpicture}
\hspace{1cm}
\begin{tikzpicture}[scale=0.7]
\draw[fill=black] (1,0) circle (1pt);
\draw[fill=black] (2,0) circle (1pt);
\draw[fill=black] (3,0) circle (1pt);
\draw[fill=black] (4,0) circle (1pt);
\draw[fill=black] (5.5,0) circle (1pt);
\draw[fill=black] (6.5,0) circle (1pt);
\draw[fill=black] (7.5,0) circle (1pt);
\draw[fill=black] (8.5,0) circle (1pt);
\draw[fill=black] (9.5,0) circle (1pt);
\draw[fill=black] (10.5,0) circle (1pt);
\node at (1,-.5) {$1$};
\node at (2,-.5) {$2$};
\node at (3,-.5) {$3$};
\node at (4,-.5) {$4$};
\node at (5.5,-.5) {$3$};
\node at (6.5,-.5) {$4$};
\node at (7.5,-.5) {$\alpha$};
\node at (8.5,-.5) {$2$};
\node at (9.5,-.5) {$1$};
\node at (10.5,-.5) {$\alpha$};

\node at (5.5,-1.5) {$t_{1234|34\alpha21\alpha}$};

\draw[] (0.7,0)--(4.3,0);
\draw[] (5.2,0)--(10.8,0);
\draw[] (1,0) -- (1,2)--(9.5,2)--(9.5,0);
\draw[] (2,0) -- (2,1.5)--(8.5,1.5)--(8.5,0);
\draw[] (3,0) -- (3,1)--(5.5,1)--(5.5,0);
\draw[] (4,0) -- (4,0.5)--(6.5,.5)--(6.5,0);
\draw[] (7.5,0) -- (7.5,0.5)--(10.5,.5)--(10.5,0);
\end{tikzpicture}
\caption{Definition of the $t$-variables.\label{fig:tvarDef}}
\end{figure}
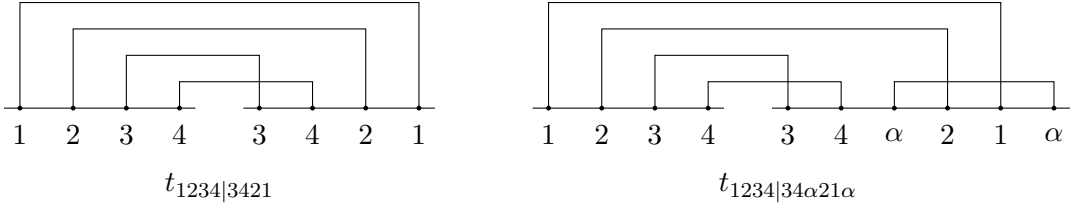\relax
We are interested in the case of $N=32$ and $q=4$ for comparison with the numerical results. Every term in these low-order double-trace moments can be calculated using the formulas developed in section~\ref{sec:dtrAndChord}. We only
present the final values of these terms in the next subsection of this appendix.

\boldmath
\subsection{Values for low-order double-trace contractions (\texorpdfstring{$N=32, q=4$}{N = 32, q = 4})}
\unboldmath\label{appen:dtrContractionVals}
In this appendix we list the explicit values of the terms that appeared in the low-order double -trace moments. Some of the contributing terms are purely single-trace ($\eta, M_4, M_6$), and are thoroughly discussed in~\cite{Garcia-Garcia-ml-2018kvh}:
\begin{equation}
\begin{split}
\eta&=1191/4495,\\
\tilde M_4 &= 10181/4495,\\
\tilde M_6 & =137227959/20205025.
\end{split}
\end{equation}
The rest of the terms are the connected double-trace contractions that can be calculated by the method laid out in section~\ref{sec:dtrAndChord}:
\begin{equation}
\begin{split}
t_{123|321}&=567/323280400,\\
t_{1234|4321}&=7168277/23250326368000,\\
t_{1234|3421}&=-555243/23250326368000,\\
t_{1234|43\alpha21\alpha}&=5928671067/104510217024160000,\\
t_{1234|34\alpha21\alpha}&=518354267/104510217024160000,\\
t_{1234|42\alpha31\alpha}&=-711893157/104510217024160000,\\
t_{12345|54321}&=8116630557/104510217024160000,\\
t_{12345|45321}&=-462480963/104510217024160000,\\
t_{12345|35421}&=160002717/104510217024160000,\\
t_{12345|42531}&=-6933507/104510217024160000,\\
t_{\alpha12\alpha34|43\beta21\beta}&=7519992447797/469773425523599200000,\\
t_{\alpha12\alpha34|34\beta21\beta}&=520655935797/469773425523599200000,\\
t_{\alpha12\alpha34|42\beta31\beta}&=-368559700939/469773425523599200000,\\
t_{\alpha12\alpha34|24\beta31\beta}&=120146887221/469773425523599200000,\\
t_{\alpha12\alpha34|23\beta41\beta}&=4043901987381/469773425523599200000,\\
t_{123456|654321}&=6135081997081/234886712761799600000,\\
t_{123456|564321}&=-60665945079/234886712761799600000,\\
t_{123456|465321}&=-687382239/234886712761799600000,\\
t_{123456|456321}&=31234591/7576990734251600000,\\
t_{123456|563421}&=89686592361/234886712761799600000,\\
t_{123456|536421}&=3173529/234886712761799600000,\\
t_{123456|356421}&=-19709513199/234886712761799600000,\\
t_{123456|436521}&=5189715001/234886712761799600000,\\
t_{123456|462531}&=589707393/234886712761799600000.
\end{split}
\end{equation}

\pagebreak

\end{document}